\documentclass[manuscript,revised]{geophysics}

\usepackage[utf8]{inputenc}
\usepackage{amsmath, amssymb, amsfonts, amsbsy, latexsym, amsthm}
\usepackage[english]{babel}
\usepackage{doi}
\usepackage{algorithm}
\usepackage{algorithmic}
\usepackage{diagbox}
\usepackage{color}
\usepackage{graphicx}
\usepackage{epstopdf}
\usepackage{subfigure}
\usepackage{grffile} 
\usepackage{multirow}
\usepackage[toc,page]{appendix}
\usepackage{ulem}
\usepackage{cancel}

\usepackage{marginnote}

\usepackage{url}
\usepackage{hyperref}
\hypersetup{
colorlinks,%
citecolor=black,%
filecolor=black,%
linkcolor=black,%
urlcolor=black
}

\title{Deep learning  for denoising}

\address{
\footnotemark[1]Center of Geophysics, School of Mathematics and Artificial Intelligence Laboratory,\\
Harbin Institute of Technology,\\
Harbin, 150001, China
}

\author
{Siwei Yu\footnotemark[1], Jianwei Ma\footnotemark[1] and Wenlong Wang\footnotemark[1]
}

\righthead{DL  for denoising}

\begin{document}

\maketitle

\begin{abstract}

Compared with traditional seismic noise attenuation algorithms that depend on signal models and their corresponding prior assumptions, removing noise with a deep neural network is trained based on a large training set, where the inputs are the raw datasets and the corresponding outputs are the desired clean data. After the completion of training, the deep learning method achieves adaptive denoising with no requirements of (i) accurate modelings of the signal and noise, or (ii) optimal parameters tuning. We call this intelligent denoising. We use a convolutional neural network as the basic tool for deep learning.  {In random and linear noise attenuation, the training set is generated with artificially added noise. In the multiple attenuation step, the training set is generated with  acoustic wave equation.} Stochastic gradient descent is used to solve the optimal parameters for the convolutional neural network. The runtime of deep learning on a graphics processing unit for denoising has the same order as the $f-x$ deconvolution method.  Synthetic and field results show the potential applications of deep learning in automatic attenuation of random noise (with unknown variance), linear noise, and multiples.

\end{abstract}

\newpage

\section{Introduction}

In seismic {data acquisition}, the geophones {record} reflected seismic signals as well as random and coherence noises. Random noise is caused by environmental interferences. {Coherent noise, such as linear noise (ground-roll waves), and multiple reflections, are  generated by the sources.} Noises lead to undesired artifacts in seismic migration and inversion; therefore, noise attenuation must be applied before {subsequent} seismic processing steps.

Traditional random noise attenuation methods are typically based on filtering techniques. Generally, noise is assumed as Gaussian distributed while the data can satisfy different assumptions, such as linear events (\citealp{Spitz1991Seismic,Naghizadeh2012Seismic}), sparsity  \citep{Zhang2003Physical,Hennenfent2006Seismic,Fomel2013Seislet,yu2015interpolation}, and low-rank assumptions \citep{trickett2008f,Kreimer2012A},  such that the data can be distinguished from the noise with specially designed  {algorithms}. 

Coherent noise attenuation methods are based on two technologies:  filtering and prediction. The ground roll travel{s} along the land surface, with high amplitude, low frequency, and low velocity. Shallow reflections may be masked by strong ground roll, which must be removed first. For removing ground-roll, filtering methods based on the combinations of  {its} frequency and velocity properties \citep{corso2003seismic,zhang2010analysis,Liu2013Seismic} have been proposed,  as well as prediction methods based on  model-driven techniques \citep{yarham2006curvelet} and data-driven techniques \citep{herman2006predictive}.
     
Multiples are recorded signals that are scattered or reflected more than once. Multiples cannot be migrated to the correct positions with most of state-of-the-art imaging methods (e.g., reverse-time migration) \citep{weglein16}. Many algorithms have been proposed based on two properties of multiples \citep{berkhout06}: the moveout differences between multiples and primaries \citep{hampson86, herrmann2000aliased, trad03}, and the predictability of multiples \citep{robinson57, verschuur92,berkhout2012seismic}.

Although many {methods} are proposed for random and coherent noise attenuation, the algorithms still encounter two bottlenecks: inaccurate assumptions and improper parameter settings. The seismic data models are still approximations of the field data. For example, the sparse-transform-based methods assume that seismic data can be represented sparsely with a specially designed transform; however, this assumption is often invalid for field data. Adaptive dictionary learning methods \citep{yu2015interpolation} train an adaptive sparse transform from the dataset, rather than using a pre-designed transform.  However, dictionary learning is still based on the sparsity assumption. Furthermore, the fine-tuning of parameters by experience is required for good denoising quality. For example, the noise level is unknown in field data random noise attenuation, such that repeated numerical tests with different variances are required, causing low efficiency. Noise estimation is used for reducing labor and uncertainty \citep{Liu2013Single}, which is still an approximate procedure.

Can we have a uniform framework for general noise attenuation? Rather than establishing the model and estimating the parameters in advance, we introduce deep learning (DL), an advanced machine learning method as an alternative seismic noise attenuation method with little prior knowledge of the data or noise.  DL achieves “intelligent denoising” by exploiting the hidden relationship between corrupted data and clean data from {a large} amount of existing datasets. 

Before describing DL in detail, we first introduce its origin: machine learning (ML). ML algorithms are designed to learn the features and relationships hidden in large numbers of datasets automatically. ML is used primarily in the regression, prediction, and classification of large datasets, such as facial recognition \citep{rowley1998neural} and medical diagnosis \citep{kononenko2001machine}. 

ML has also been used in seismic exploration. \citet{zhang2014machine} proposed using a kernel-regularized least squares (KRLS) \citep{evgeniou2000regularization} method for fault detection from seismic records. The authors used toy velocity models to generate records and set the velocity models and records as inputs and outputs of a network, respectively. They used KRLS to construct and optimize the network and obtain meaningful results. \citet{Jia2017What} proposed using supported vector regression (SVR) \citep{cortes1995supportvector} for seismic interpolation. The authors used linearly interpolated data and original data as the inputs and outputs, respectively, and they used SVR to obtain the relation between the inputs and outputs. They claimed that no assumptions are imposed on the data and that no parameter tuning is required for interpolation. 
The class of methods called artificial neural networks have been widely explored in seismic data processing and interpretation, such as time-to-depth conversion \citep{roth1994neural}, event picking \citep{Glinsky1996Automatic}, tomography \citep{doi:10.1046/j.1365-246x.1999.00835.x}, parameters determination and pattern detection \citep{Huang2006Neural}, facies-classification \citep{Ross2017A}, detecting faults \citep{huang2017a,zhao2018fault,guo2018new,wu2018convolutional,wu2019faultseg3d}, channels \citep{pham2019automatic} and salt bodies \citep{shi2018automatic},  among others. 

Given the rapid development of computer hardware, especially graphic processing units (GPUs), deep neural network has been a popular topic since 2010, such as the deep belief network \citep{hinton2006a}, stacked autoencoder \citep{vincent2010stacked}, and deep convolutional neural network (CNN) \citep{lecun1998gradientbased}. CNN's use a shared local convolutional filter bank designed for images, which contains many fewer parameters compared to a fully connected multilayer neural network (FCNN). The FCNN encounters computational and storage problems owing to the massive parameters when the network becomes deeper, or the size of the inputs becomes larger. The FCNN ignores the structure of the input entirely. Seismic data have a strong local structure: neighboring samples are highly correlated. CNN's utilize the correlation using a shared local convolutional filter, thus avoiding the use of numerous parameters. 

\citet{lecun1998gradientbased} proved that CNN's with fewer parameters provided superior classification results on the MNIST set compared to the FCNN. The CNN was developed rapidly since 2010 for image classification and segmentation, such as VGGNet \citep{Simonyan2014Very}, and AlexNet \citep{krizhevsky2012imagenet}. The CNN was also used in image denoising \citep{Jain2008Natural,Zhang2017Beyond} and super-resolution \citep{Dong2014Learning,7961250}. 
A CNN with 17 convolutional layers was used by \citet{Zhang2017Beyond} for image denoising.
The authors used noise as the {outputs} rather than clean data, i.e., residual learning. They claimed that residual learning accelerated the training process as well as improved the denoising performance. {Compared to dictionary learning with single-layer decomposition,} DL (if not specified, DL refers to deep learning with CNN herein) with deeper layers enables the exploration of rich structures in seismic data training set with different levels of abstraction.

The applications of DL in image processing provide new ideas and techniques to geophysicists. We herein propose using DL for seismic noise attenuation. Our contributions are as follows: (i) we construct three training sets for synthetic noise attenuation and two training sets for field noise attenuation, (ii) our methods show better automation  and higher quality in random noise attenuation compared to traditional methods,  and (iii) we present an indirect visualization of the trained convolutional filters and discuss the hyperparameter tuning of the CNN. The second part introduces the theory, including the design of the CNN, optimization algorithm, inverting the CNN for visualization, and transfer learning for field data training. The third part provides the preparation of the training datasets and presents the numerical tests on synthetic and field datasets. Subsequently, we discuss the performance, parameters, and other aspects of DL. Finally, we conclude the paper.

\section{Method}

The seismic data acquisition model is
\begin{equation}\label{eq:data}
\mathbf{y = x + n}
\end{equation}
where ${\mathbf{x}}$ is the clean seismic data, ${\mathbf{y}}$ is the contaminated data, and ${\mathbf{n}}$ is the noise. The basic idea of ML based on a NN is to establish a relationship between ${\mathbf{x}}$ and ${\mathbf{y}}$ using the following formula:
\begin{equation}\label{eq:cnn}
{\mathbf{x = \text{Net}(y;\Theta)}}
\end{equation}
where Net stands for the NN architecture, which is equivalent to a denoising operator, and ${\mathbf{\Theta=\{W,b\}}}$ contains network parameters, including the weight matrix ${\mathbf{W}}$ and bias ${\mathbf{b}}$. In specific applications, the residual is used as the output:
\begin{equation}\label{eq:rcnn}
{\mathbf{y - x = \text{R}(y;\Theta)}}
\end{equation}
where R stands for residual learning. In this section, we  introduce the network architectures, the optimization of ${\mathbf{\Theta}}$, {inverting CNN and transfer learning}. 

\subsection{Network architecture and optimization}

\begin{figure*}
    \centering

    \subfigure[]{\label{fig:nn}
    \includegraphics[width=0.8\textwidth]{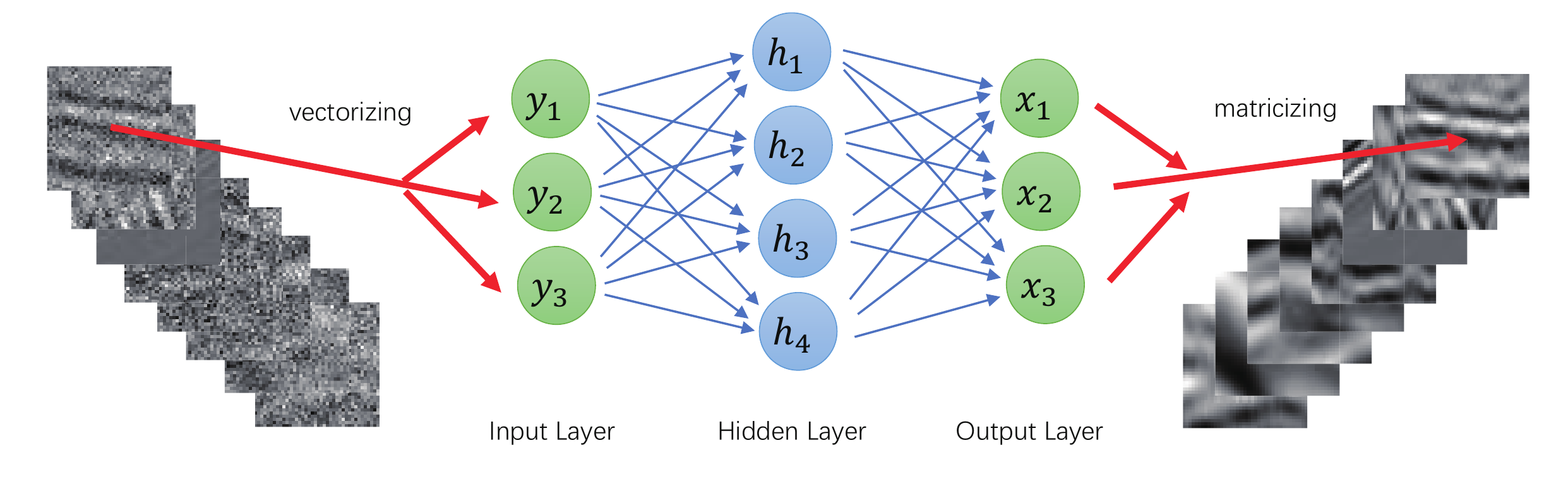}}
    \subfigure[]{\label{fig:cnn}
    \includegraphics[width=0.8\textwidth]{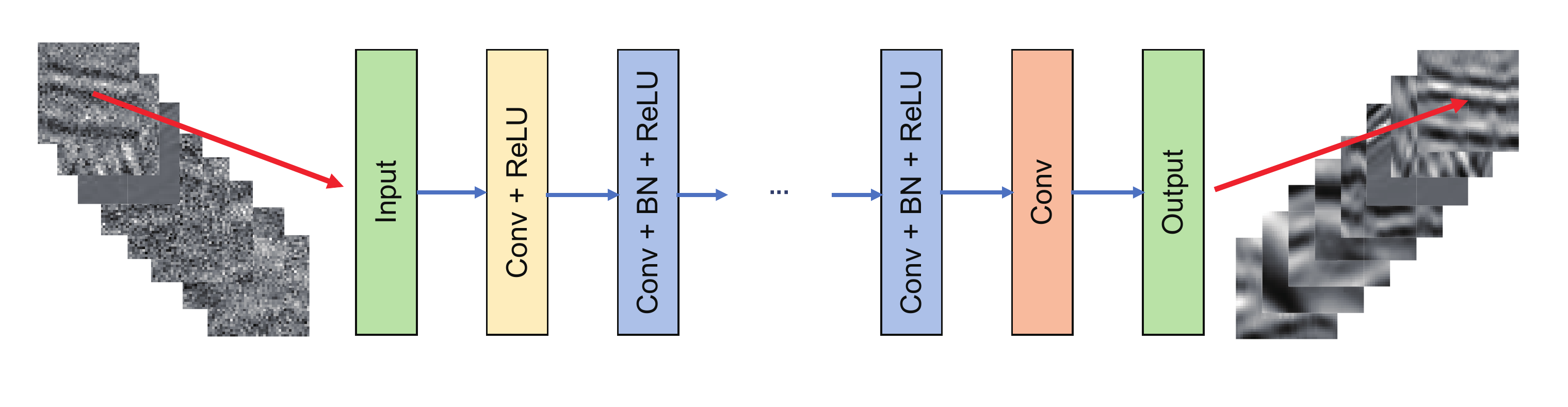}}
    \caption{(a) Sketch of a FCNN with one hidden layer. Vectorizing  implies turning  matrices into vectors. Matricizing implies turning vectors into matrices. The number of elements in input/hidden/output layer is randomly selected for simple illustration. (b) Sketch of a deep CNN architecture.}
    \label{fig:nns}
\end{figure*}

In this section, we begin by introducing a simple NN, namely a fully-connected NN (FCNN) with {a} single hidden layer. First, we present {the} formula of the FCNN:
\begin{eqnarray}\label{eq:nn}
{\mathbf{\text{R}(y;\Theta) = W_2\text{f}(W_1y+b_1)+b_2}}
\end{eqnarray}
where ${\mathbf{\Theta = \{W_1, W_2, b_1, b_2\}}}$ are the weight matrices and biases of the hidden layer and output layer, respectively, {and} $f(\cdot)$ introduces the nonlinearity, such as the rectified linear units (ReLUs), defined as $\max(0,\cdot)$. Figure \ref{fig:nn} shows the sketch of the FCNN with one hidden layer. Here, “fully connected” implies that every two nodes in adjacent layers are connected. 

The CNN uses convolutional filters and it is equivalent to the FCNN where the elements in the weight matrices are shared across different local windows. ${\mathbf{\text{R}(y;\Theta)}}$ in the CNN is expressed explicitly as
\begin{eqnarray}\label{eq:cnn}
\mathbf{\text{R}(y;\Theta)} &=&   \mathbf W_M\ast\mathbf  a_{M-1}+ \mathbf b_{M} \nonumber\\
& \cdots & \nonumber\\
\mathbf a_m &=&  \text{ReLU}\cdot \text{BN}\cdot (\mathbf W_m\ast \mathbf  a_{m-1}+ \mathbf b_m) \nonumber\\
& \cdots & \nonumber\\
\mathbf a_1 &=&  \text{ReLU}\cdot (\mathbf W_1\ast \mathbf y + \mathbf b_1)
\end{eqnarray}
where $M$ is the number of convolutional layers, $\mathbf W_m, m\in(1,\cdots,M)$ are convolutional filters, $\mathbf b_m, m\in(1,\cdots,M)$ are the biases, `$*$' stands for the convolutional operator. and $\mathbf a$ is the intermediate output, named activation.  The definitions of ReLU  and BN (Batch Normalization) are shown in Table \ref{tab:layers}.   The sketch of the CNN network architecture  $\mathbf{\text{R}(y;\Theta)}$ is shown in Figure \ref{fig:cnn}.

\begin{table}[h]
\center
\begin{tabular}{|c|c|c|c|c|}
  \hline
    Layers & Description          & Definition \\
    \hline
    Conv   & Convolutional layer  & ${\mathbf{a = W * x + b}}$\\
    \hline
    BN     & Batch normalization  & normalization, scale and shift\\
    \hline
    ReLU     & Rectified linear units  &$\max(0,\cdot)$\\
    \hline
\end{tabular}
\caption{Definition of different layers.  }\label{tab:layers}

\end{table}

\begin{figure*}
    \centering

    \subfigure[]{\label{fig:p-conv}
    \includegraphics[width=0.6\textwidth]{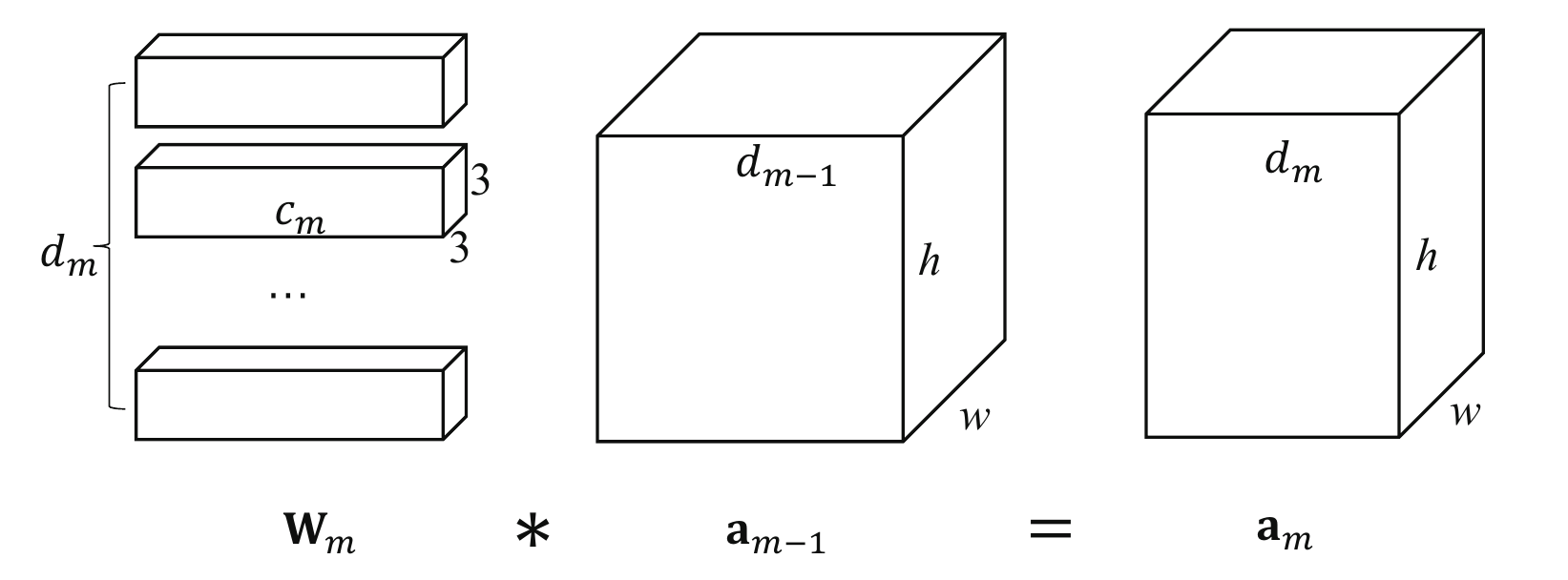}}
    \subfigure[]{\label{fig:rf}
    \includegraphics[width=0.5\textwidth]{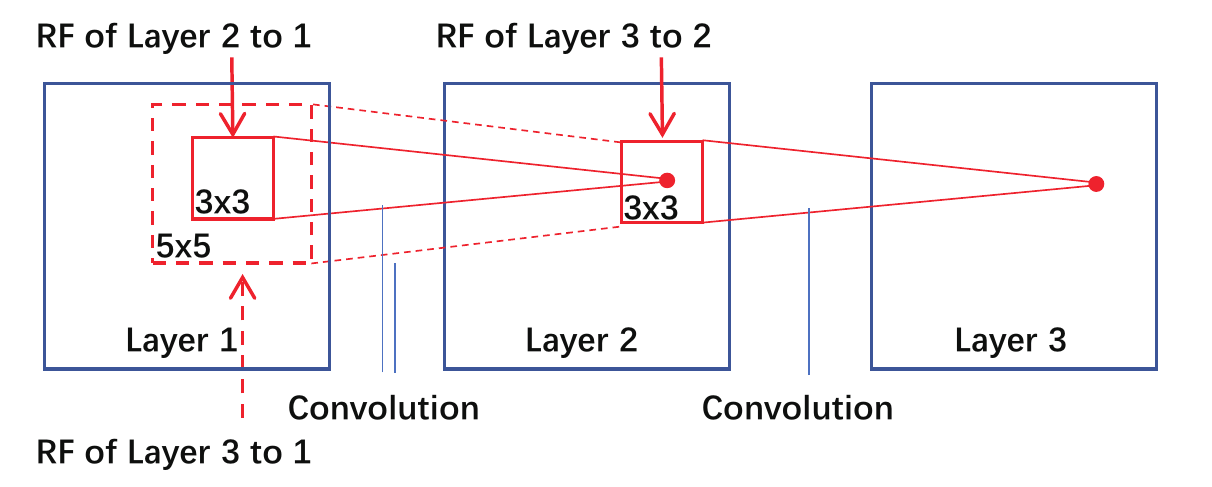}}
    \subfigure[]{\label{fig:p-relu}
    \includegraphics[width=0.4\textwidth]{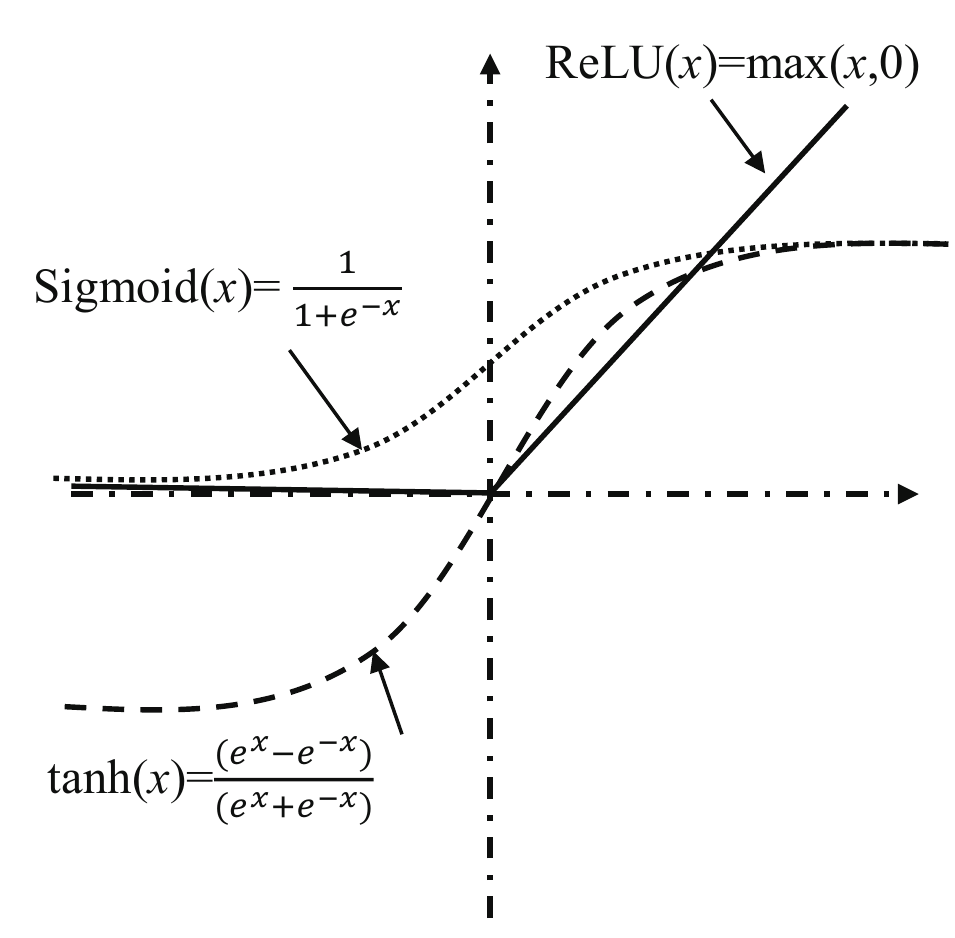}}
    \caption{(a) Convolution operation. (b) Receptive field (RF) for convolutional filter size of 3$\times$3.  (c) ReLU, sigmoid and tanh operators.}
    \label{fig:sketch}
\end{figure*}

In the CNN, $\mathbf W_m, \mathbf b_m$, and $\mathbf a_m$ are written in the tensor form.  $\mathbf W_m\in \text R^{p\times p\times c_m \times d_m}, \mathbf a_m, \mathbf b_m \in \text R^{h\times w\times d_m}$, where $h$ and $w$ are the dimensions of the input, $p\times p$ is the size of the convolutional filter, $c_m = d_{m-1}$ is the number of channels in layer $m$, and $d_m$ is the number of convolutional filters in layer $m$. Each three-dimensional (3-D) convolutional filter is applied on the 3-D input tensor to produce one output channel, as shown in Figure \ref{fig:p-conv}.
A convolutional filter causes each unit in a layer to receive inputs from a set of units located in a small neighborhood in the previous layer, as shown in Figure \ref{fig:rf}. The region of the neighborhood is named the local receptive field \citep{lecun1998gradientbased}. 

A traditional method of introducing nonlinearity $f$ to a network is to use the sigmoid or tanh function. However, in gradient descent, these nonlinearities are much slower than ReLU, because the gradient is small in the nonlinear regions \citep{krizhevsky2012imagenet}. The comparison among sigmoid, tanh, and ReLU is shown in Figure \ref{fig:p-relu}. BN is introduced by \citet{ioffe2015batch} and used to accelerate training.

To optimize $\mathbf\Theta$ in equation \ref{eq:rcnn}, a loss function  is defined as
\begin{equation}\label{eq:loss}
l({\mathbf{\Theta}})=\frac{1}{2N} \sum_{i=1}^N \text D\left(\text{R}({\mathbf{y}}_i;{\mathbf{\Theta}}),{\mathbf{y}}_i-{\mathbf{x}}_i\right)
\end{equation}
where $\{({\mathbf{y}}_i,{\mathbf{x}}_i)\}_{i=1}^N$ are $N$ training pairs. {D measures the discrepancy between the desired output and the network output, which in our case is simply chosen as the mean square error, i.e.,  D$(\mathbf x, \mathbf y)=\|\mathbf x - \mathbf y\|_F^2$, where $\|\cdot\|_F$ stands for the Frobenious norm.} Because $N$ is extremely large, computing the gradient of $l({\mathbf{\Theta}})$ numerically is impractical. Therefore, a  mini-batch stochastic gradient descent (SGD) \citep{Yann1998Efficient} is used to minimize $l({\mathbf{\Theta}})$. In every iteration, only a small subset of $\{({\mathbf{y}}_i,{\mathbf{x}}_i)\}_{i=1}^N$ is used to approximate the gradient. One pass through the whole training set is defined as an epoch. The training samples are first shuffled into a random order, and subsequently chosen sequentially in mini-batches to ensure a whole pass. 

A tradeoff exists for the batch size in SGD. A small batch size allows for frequent updates per epoch, better use of GPU memory, and acceleration of convergence. However, a smaller batch size uses totally random gradients, leading to low efficiency. Smaller batch sizes also do not utilize the acceleration of parallel matrix--matrix products. According to \citet{bengio2012practical}, the sweet region of the batch size is between 1 and a few hundred.  We select the batch size according to \citep{Zhang2017Beyond} and under the restriction of GPU memory in different types of noise attenuation.

A tradeoff also exists for the number of epochs of SGD. If the number of epochs is small, a satisfactory solution would not be obtained. However, too large a number will cause overfitting. We decide the number of epochs by observing the training curve of the validation loss function (loss function of the validation set). If the loss function appears stable, we end the training after 10--20 more epochs. This empirical principle is called early stopping, which also avoids strong overfitting \citep{bengio2012practical}.

\subsection{Inverting a CNN for indirect visualization of convolutional filters}

In the dictionary learning method \citep{yu2015interpolation}, we wish to observe the images of the dictionaries for a better understanding of their operations. In DL, after the network is trained, we also wish to observe the convolutional filters, especially in different layers. However, the filter size is 3$\times$3, which is extremely small for visualization. Therefore, an inverting CNN \citep{mahendran2015understanding} is used to visualize the filters in the data domain by backpropagating the corresponding activations to the first layer. 

The inverting CNN firstly sets the activation $\mathbf{a}_{m,d}$ to a given $\mathbf{a}_0$  and other activations to zeros, which  corresponds to $\mathbf W_{m,d}$, i.e.,  the $d$th filter  in  the $m$th hidden layer. Subsequently, we obtain input $\mathbf y$  in the data domain,   which contributes most to $\mathbf{a}_{m,d}$. Therefore, $\mathbf y$ will contain the information from $\mathbf W_{m,d}$, which is an indirect visualization method. We can compute $\mathbf y$, which activates $\mathbf a_{m,d}$ the most by solving
\begin{equation}\label{eq:loss_inv}
\min_{\mathbf y} \|\mathbf a_{m,d}(\mathbf y)-\mathbf a_0\|_\text{F}^2+\lambda \Lambda(\mathbf y)
\end{equation}
where $\lambda$ is a balancing parameter, $\mathbf{\Lambda(y)}$ is a regularization term, such as the Tikhonov regularization or total variation, in case of multi solutions. The elements in $\mathbf{a}_0$ can be set as ones. The optimization problem (\ref{eq:loss_inv}) can be solved by the gradient descent method. The details of parameter setting and implementation are referred to \citep{mahendran2015understanding}.

\subsection{Transfer learning}

The trained network can also be used as an initialization for learning a new network, i.e., transfer learning \citep{donahue2014decaf}. Fine tuning a network with transfer learning is typically much faster and easier than training a network with randomly initialized weights from scratch. With a smaller number of training samples, we can promptly transfer the learned features to a new task, such as from random noise attenuation to coherent noise attenuation, or a new dataset, such as from a synthetic dataset to a real dataset. This work will focus on the latter, i.e., real dataset random noise attenuation. The relationship between the transferability of features, and the distance between different tasks and different datasets are discussed by \citet{yosinski2014transferable} and \citet{oquab2014learning}, respectively.

\section{Numerical Results}

In this section, we first introduce how the training set is generated. Subsequently, we use the generated training sets to train three CNN's for the attenuations of random noise, linear noise, and surface-related multiples. Further, we test the CNN on field random noise and scattered ground roll attenuation. The networks are trained on an HP Z840 workstation with one Tesla K40 GPU, 32 Core Xeon CPU, 128GB RAM, and Ubuntu operating system.

\subsection{Dataset preparation}\label{sec:dataprep}

To train an effective network, a good training set for DL should be well labeled, large, and diverse.  {For example,} ImageNet \citep{Deng2009ImageNet} {was} designed for natural images, {and} contains images of human beings, animals, plants, scenes, etc. Seismic data typically consist of events with smoothly varying slopes, which differ completely from the structures of natural images.  However, no open training set for seismic data applications is  available {currently, to the authors’ knowledge.} Three factors render the establishment of the seismic training sets difficult: (i) {private} companies do not share seismic data, (ii) labeling a large number of training samples requires intense human labor, and (iii) certain tasks, such as denoising or inversion, are difficult to be labeled by humans. Therefore, we concentrate on the pseudo-training set herein, implying that the clean dataset is either synthetic or obtained by the existing denoising methods. Three pseudo training sets are first established to prove the {suitability} of DL for seismic data processing. 

A dataset for training a neural network typically consists of three subsets: a training set, a validation set, and a test set. The training set is used to fit the parameters in the network. The validation set is used to tune the hyperparameters. The test set is used to test the performance of the trained network. The training, validation, and test sets are generated with the same rule but are independent of each other. The sizes of the validation and test sets are approximately 25$\%$ of the training set. The training loss, validation loss, and test loss are defined as the average loss in equation \ref{eq:loss} during training, validation, and testing.

\subsubsection{Random noise}

In a random noise situation, we can use clean signals as the {outputs}, and manually add Gaussian noise to the clean signals to serve as the inputs (we use the added noise as the {outputs} for residual learning). Further, the inputs can be split into small patches rather than using the whole sections directly for saving memory, because random noises are locally incoherent with useful data. 

Although the network architecture does not change, the sizes of the intermediate outputs vary according to the input size. Suppose that the network contains 17 convolutional layers with 64 convolutional filters in each layer, and the input size is 1000$\times$1000 (time sample and space sample), and the size of the intermediate outputs after convolutional layers is  $1000\times1000\times64\times17$ (4.05 GB in single precision). If the input is split into patches of size $35\times35$, subsequently the total intermediate size is  $35\times35\times64\times17$ (0.41 MB), which is only 0.01$\%$ of the original one. It is noteworthy that the denoised result is not overlapping of small patches. Although we used small patches in training, the input can be of any sizes in testing. Each unit of one layer is computed by the weighted sum of the units in its corresponding receptive field in the previous layer.

\begin{figure*}
    \centering
    \subfigure[]{\label{fig:03-1}
    \includegraphics[width=0.4\textwidth]{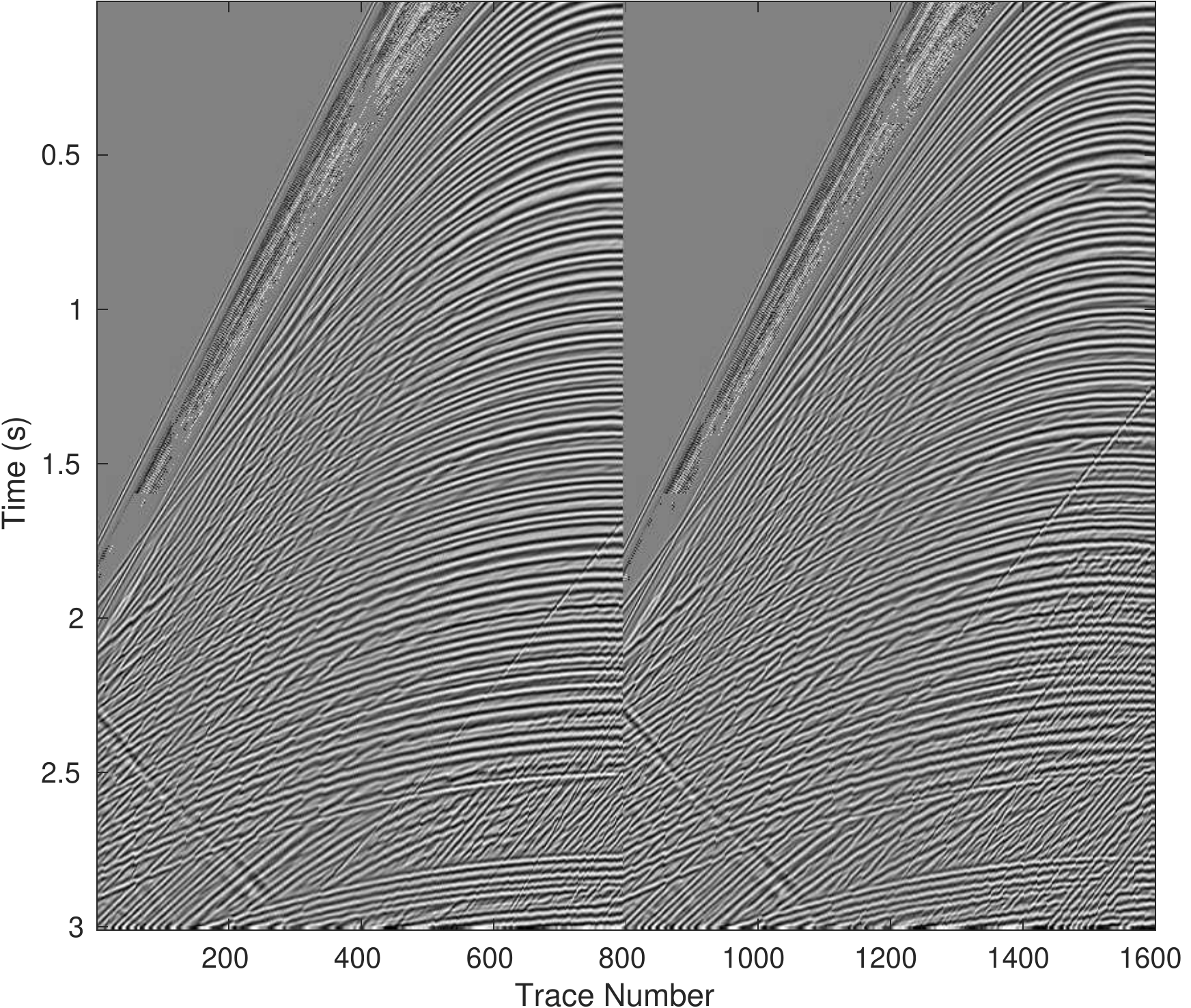}}
    \subfigure[]{\label{fig:03-6}
    \includegraphics[width=0.4\textwidth]{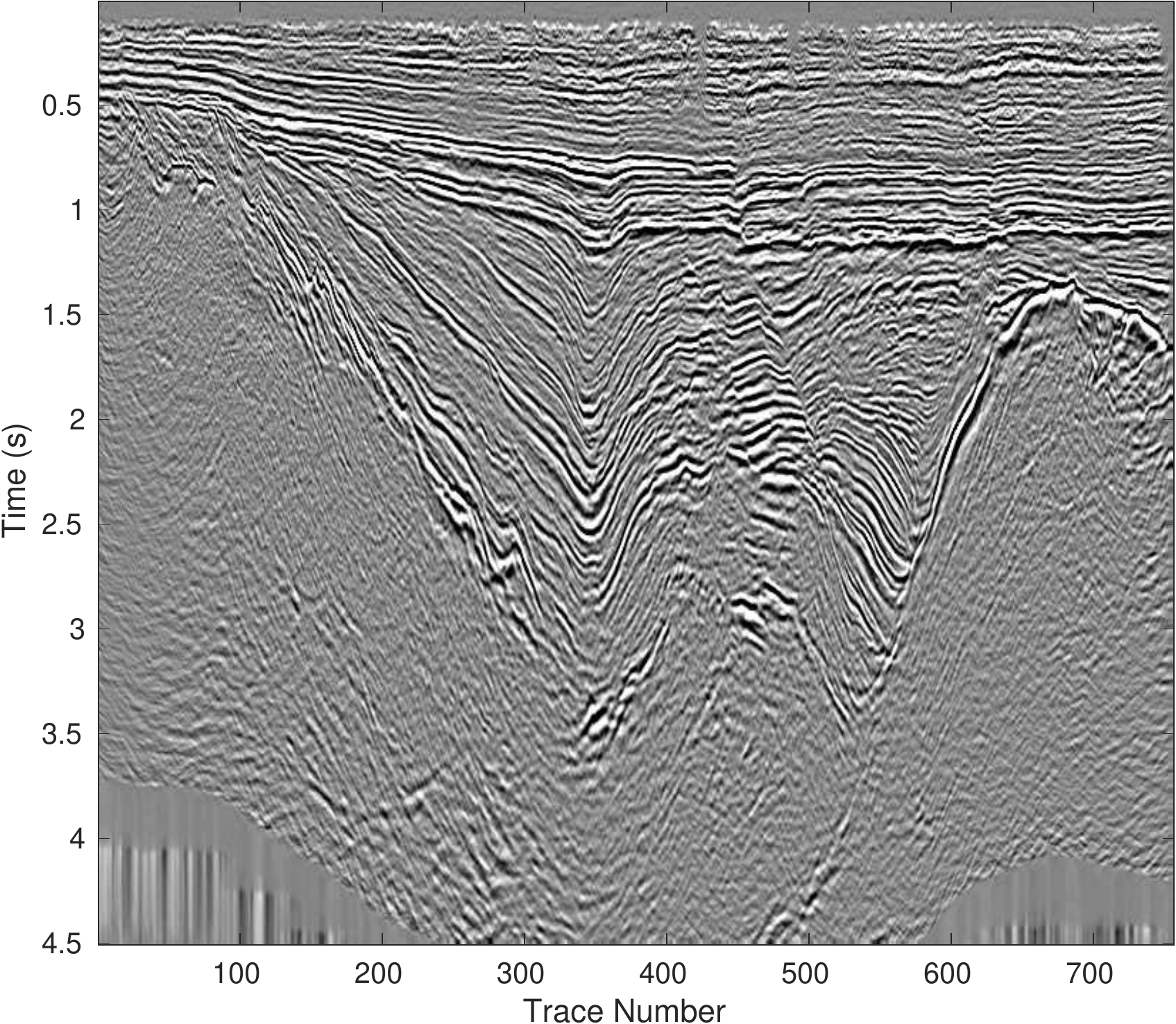}}

    \subfigure[]{\label{fig:04-clean}
    \includegraphics[width=0.4\textwidth]{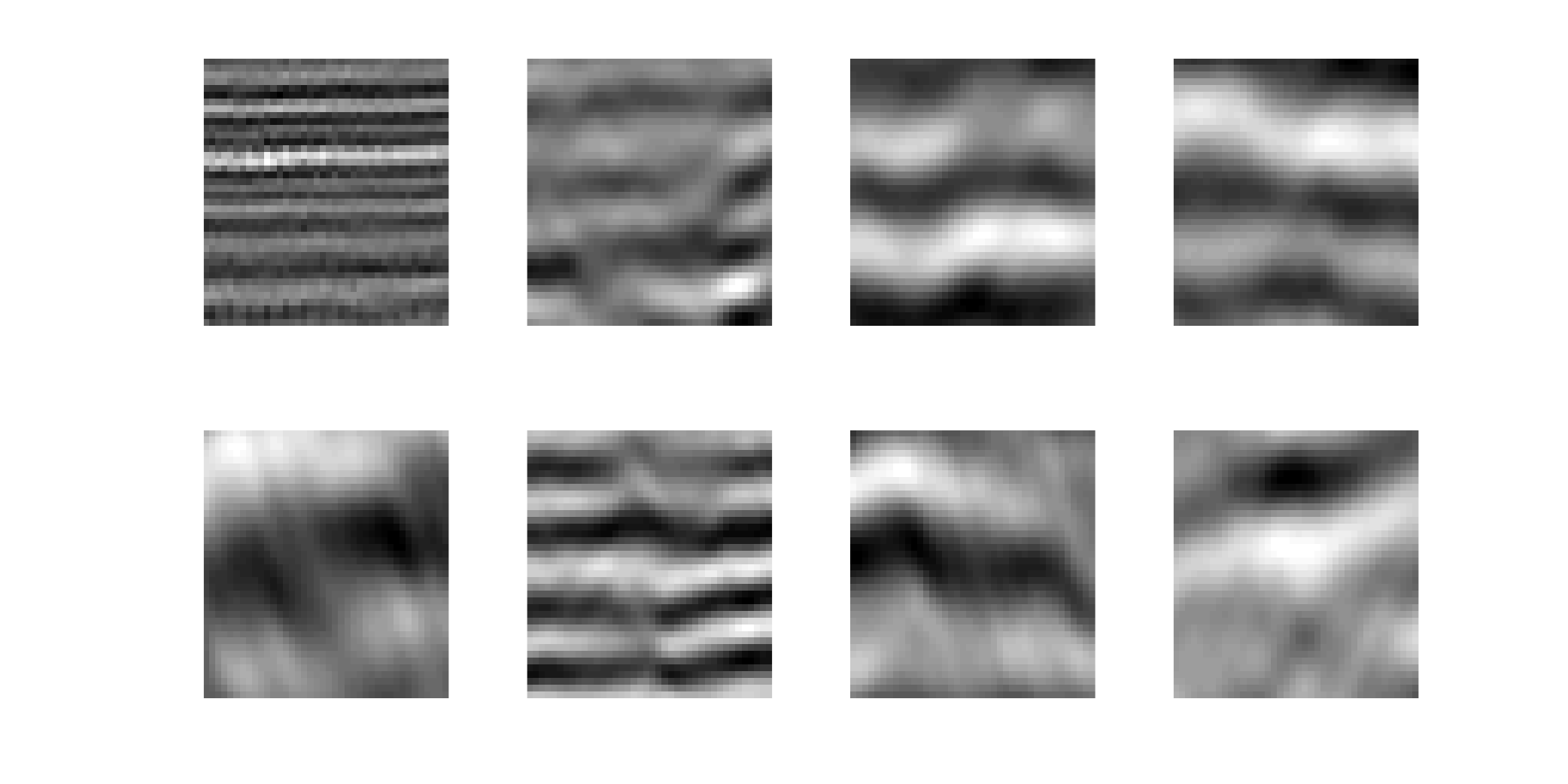}}
    \subfigure[]{\label{fig:04-noisy}
    \includegraphics[width=0.4\textwidth]{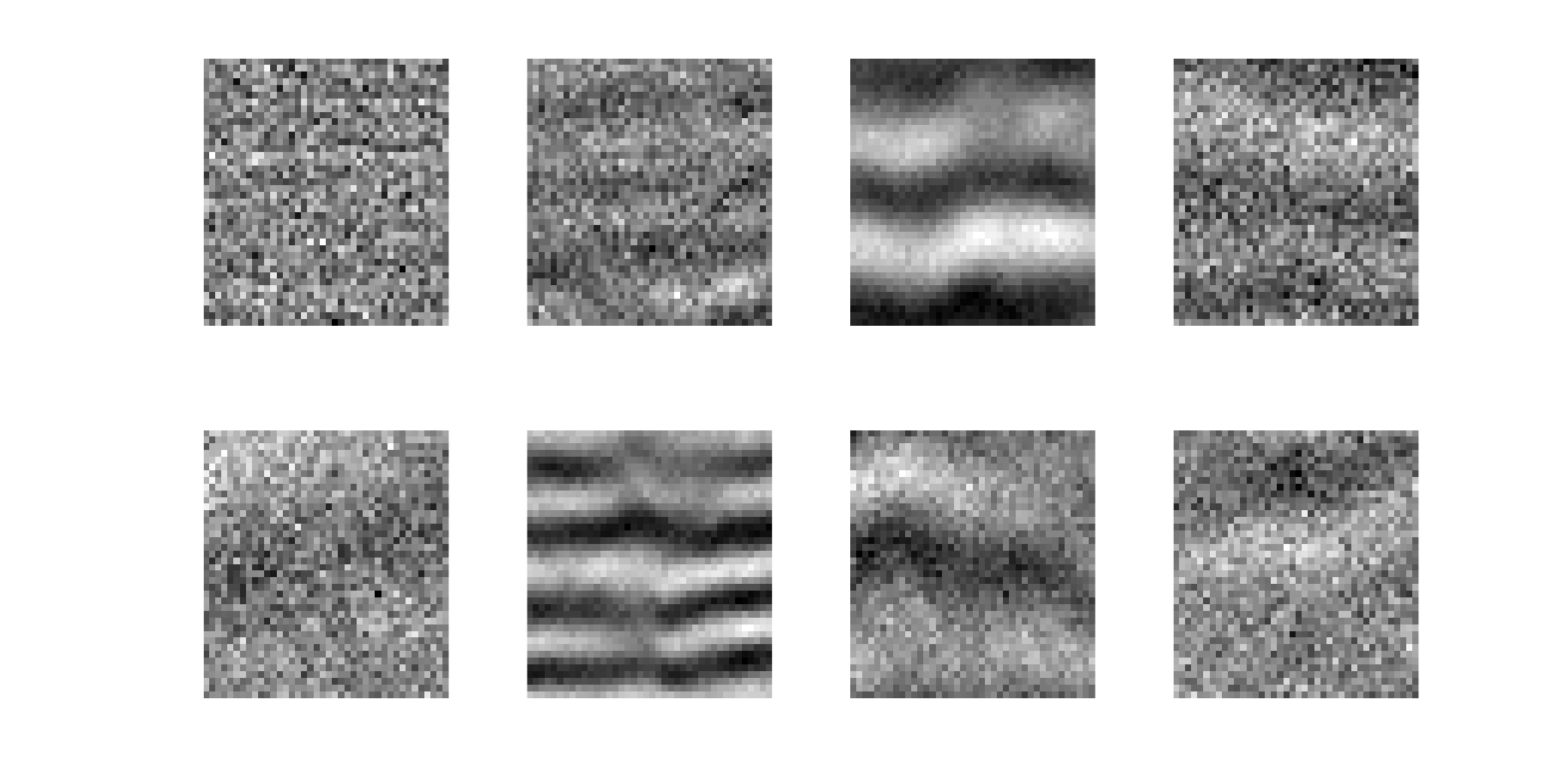}}
    \caption{ (a) and (b) are two datasets from the training datasets. (c) and (d) are eight training samples extracted from the training sets for random noise attenuation. (c) Original data. (d) Noisy data. Note that the noise variances are different.}
    \label{fig:datasets}
\end{figure*}

The training set is downloaded from SEG (Society of Exploration Geophysicists) open datasets (\url{https://wiki.seg.org/wiki/Open_data\#2D_synthetic_seismic_data}). Two of the datasets are shown in Figure \ref{fig:03-1} and Figure \ref{fig:03-6}. To improve the diversity of the training set, we select the pre-stack data, post-stack data, two-dimensional data and 3-D data. In 3-D data, we select sections with intervals of 20 shots, because the adjacent shot sections are similar.

Some portions of the seismic data are almost all zeros, which does not contribute to the training process. We introduce a Monte Carlo strategy \citep{Yu2016Monte} while generating the training set. For each training sample, we compare its variance with a uniformly distributed random number. We maintain the sample if its variance is greater than the random number. Finally, the training set contains approximately 50,000 samples. The size of each training sample is $35\times 35$ by referring to \citet{Zhang2017Beyond}. To achieve blind denoising, we add Gaussian noise with different levels of variances ($\sigma=0-40\%$ of the maximum amplitude, uniformly distributed) to the data. Figure \ref{fig:04-clean} and Figure \ref{fig:04-noisy} show eight examples from the training set. It is noteworthy that the samples contain different types of structures and noise variances. 

\subsubsection{Linear noise}

\begin{figure*}
    \centering
    \subfigure[]{\label{fig:04L-clean}
    \includegraphics[width=0.45\textwidth]{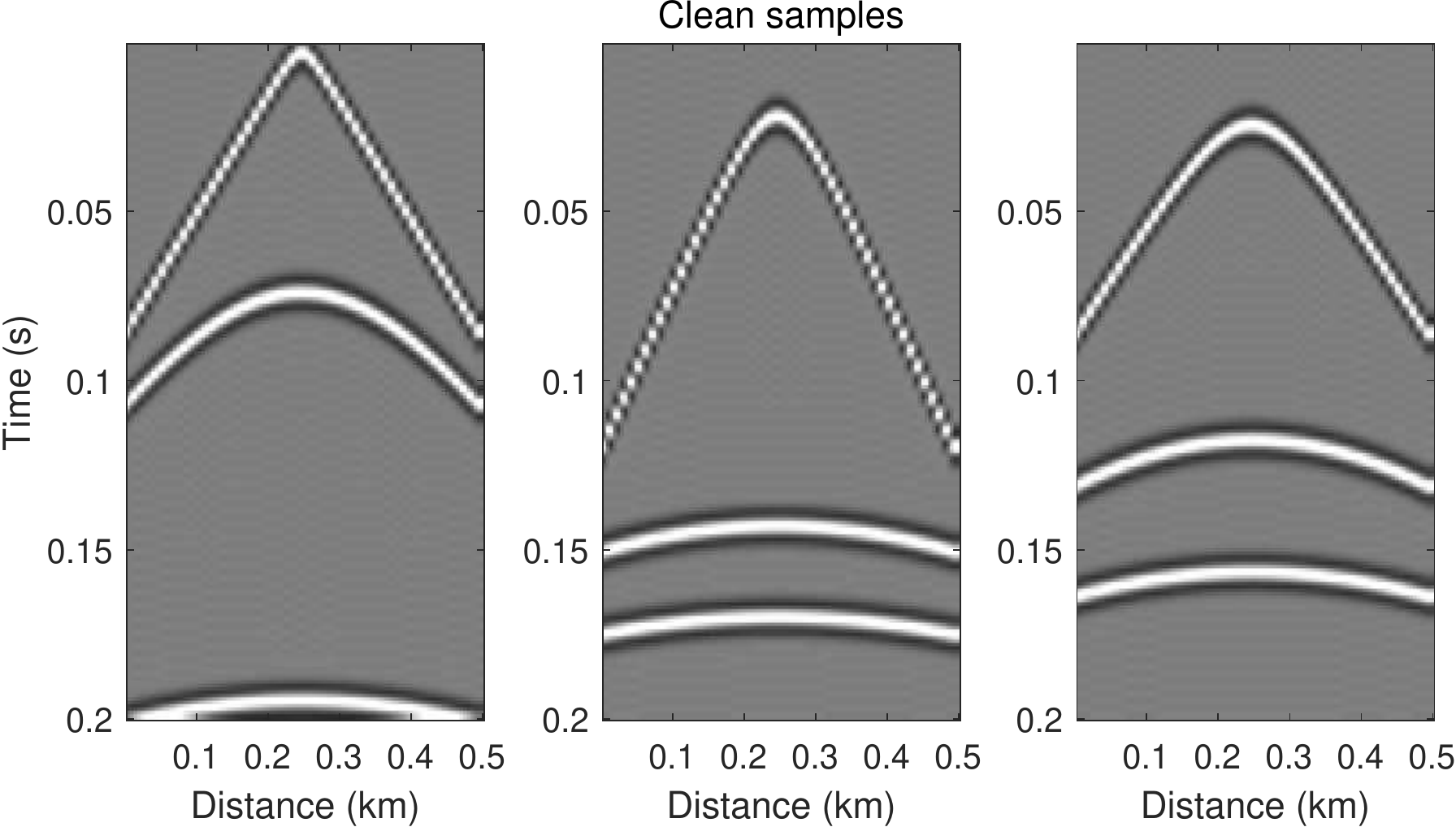}}
    \subfigure[]{\label{fig:04L-noisy}
    \includegraphics[width=0.45\textwidth]{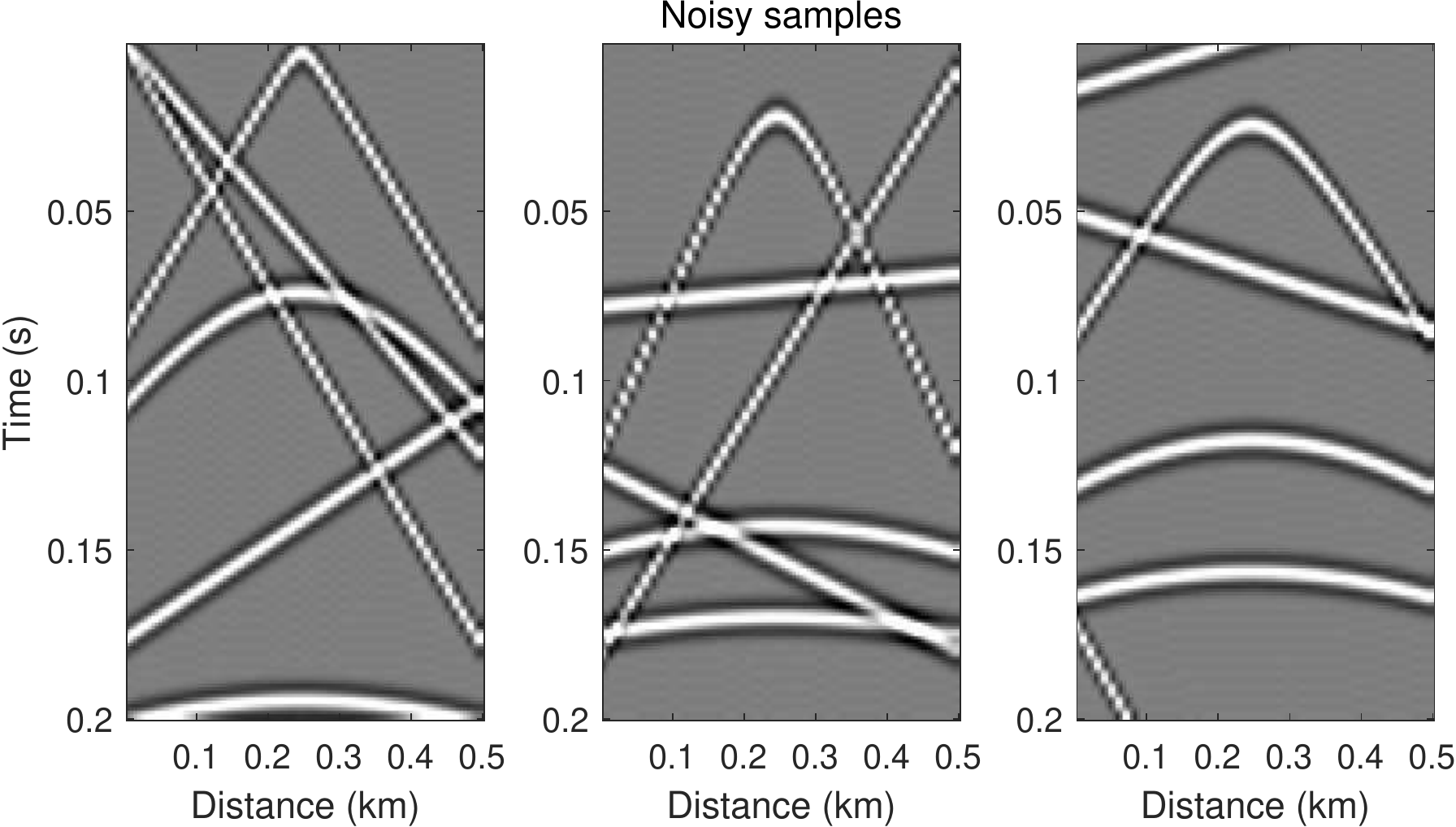}}
    \caption{ Three synthetic training samples for linear noise attenuation. (a) Original data. (b) Data with linear noise. }
    \label{fig:samples-linear}
\end{figure*}

Splitting is not valid in coherent noise attenuation, because the coherent noise is difficult to be distinguished on the local scale. Therefore, the size of the training data is larger than in the random noise situation. The linear noise we consider herein  {are} the events with random time shifts and slopes, rather than ground rolls. The number of linear events is three. The desired signals are three hyperbolic events with random curvatures and random zero offset arrival times. The amplitudes of the desired signals and linear noises are the same. We set the size of the training sample as $100\times50$ and the size of the training set as 8000. Figure \ref{fig:samples-linear} shows three training samples and one test sample with linear noise.

\subsubsection{Multiple}

We generate models of the same size (150 $\times$ 75), containing three interfaces to simulate the water bottom and {various} interfaces. The medium above the first interface is water with a {P-}wave velocity 1500 $\mathrm{m/s}$. The three underlying layers contain P-wave velocities of 2000, 2500, and 3000 $\mathrm{m/s}$, respectively. The size of the training set is  900.

\begin{figure*}
    \centering

    \subfigure[]{\label{fig:vel}
    \includegraphics[width=0.4\textwidth]{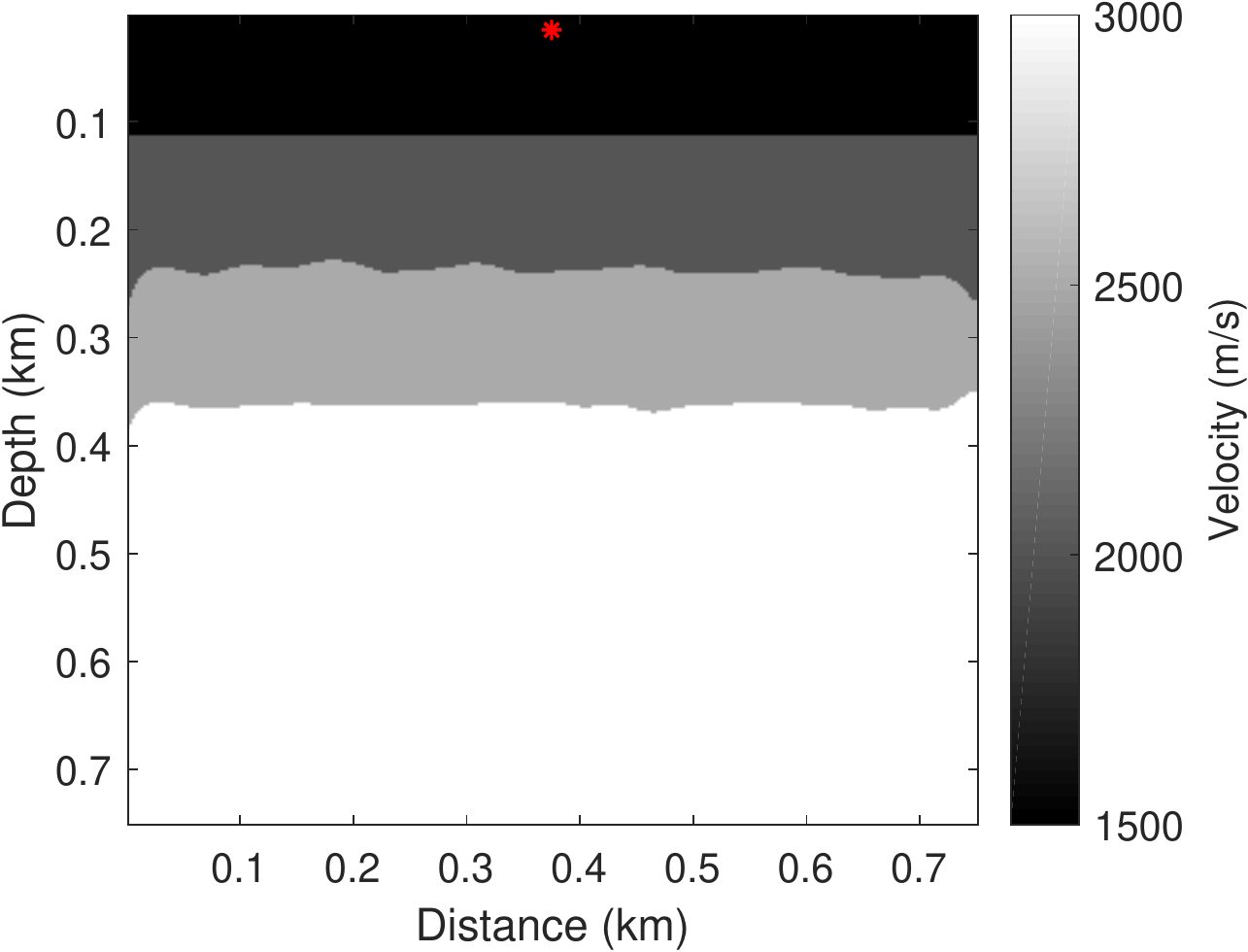}}

    \subfigure[]{\label{fig:04M-noisy}
    \includegraphics[width=0.6\textwidth]{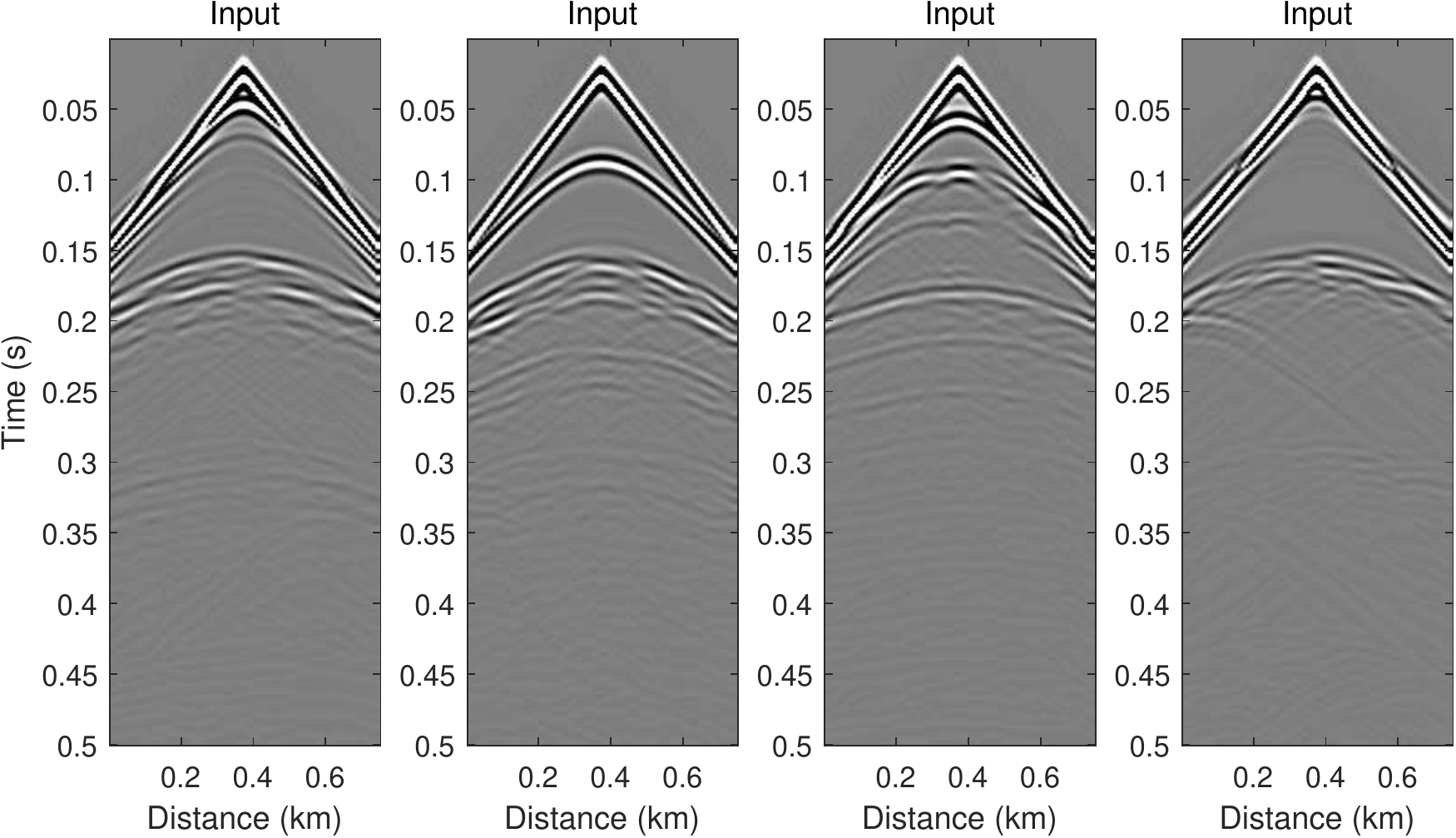}}

    \subfigure[]{\label{fig:04M-clean}
    \includegraphics[width=0.6\textwidth]{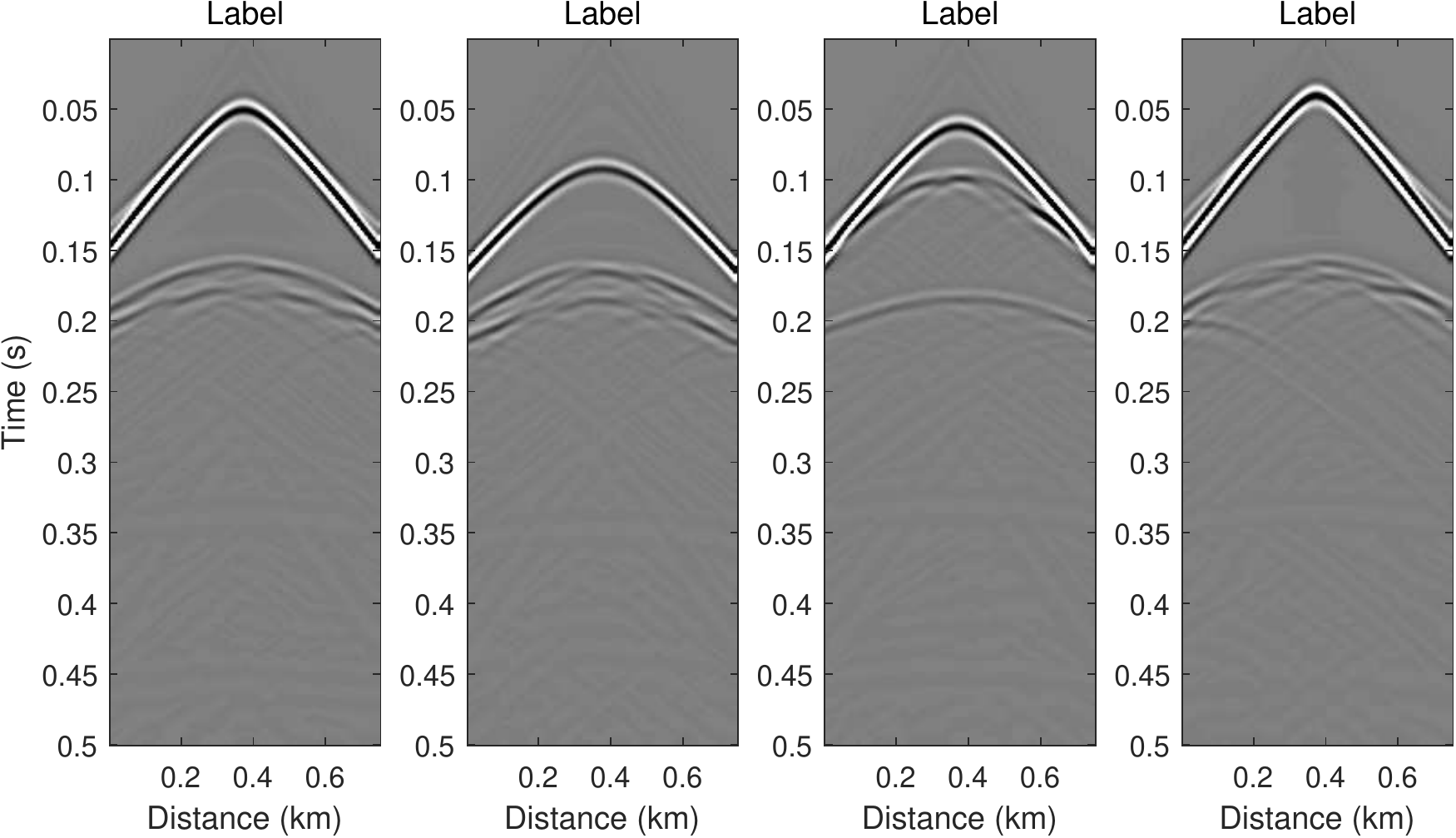}}
    \caption{ Four synthetic training samples for testing DL with seismic data containing multiples. (a) Velocity model for generating a multiple. The star represents the source location. (b) Original data. (c) Data without multiple refection. The events on the top of figures in (b) are caused by the ghost source rather than the first arrivals.}
    \label{fig:samples-multiple}
\end{figure*}

{To obtain the maximum diversity for the training set, each interface is placed randomly with uniform distribution. The first interface contains no dip to produce obvious first-order multiples. The second and third interfaces are dipping, with local slopes chosen randomly (Gaussian distribution followed by average smoothing). Figure \ref{fig:vel} shows a sample from the generated velocity models.} We use {an} eighth order in space, and second order in a time-finite difference algorithm to solve the acoustic wave equation and generate synthetic seismograms on the surface. One fixed source is placed at ($x, z$) = (0.375, {0.005}) km for each model. Additionally, 75 receivers are placed evenly below the surface at {a} depth of 5 m with 10 m spacing. A convolutional perfectly matched layer (CPML) absorbing boundary conditions \citep{komatitsch07} are used on the left, right, and bottom grid edges to reduce unwanted reflections. 

For each {velocity} model, two synthetic forward {models} are performed to generate the seismograms. The first one has a free-surface boundary on the top, and the data generated contains both primary reflections and surface-related multiples, including ghosts and water bottom reverberations. Thus, the generated seismogram can serve as the input of the training set. The other forward modeling has an absorbing zone at the top boundary to avoid generating any surface-related multiples, and the differences between the two generated seismogram are used as the output for training. Figure \ref{fig:04M-noisy} and Figure \ref{fig:04M-clean} show four seismograms with multiple reflections and the corresponding clean signals from the training set. The {first arrivals are removed. The events on the top of Figure \ref{fig:04M-noisy} are caused by the ghost source rather than the first arrivals.} 

\subsection {Dataset tests}
\subsubsection {Random noise attenuation}

In a random noise situation, the size of the convolutional filters in each layer is set to be $3\times 3$, with 64 filters referring to \citep{Simonyan2014Very,Zhang2017Beyond}. The batch size is 128. The intermediate inputs are padded with zeros to maintain the output size.  The number of convolutional layers is 17. The parameters used in SGD are the same as in DnCNN \citep{Zhang2017Beyond}. The number of epochs is set as 50.

\begin{figure*}
    \centering
    \includegraphics[width=0.6\textwidth]{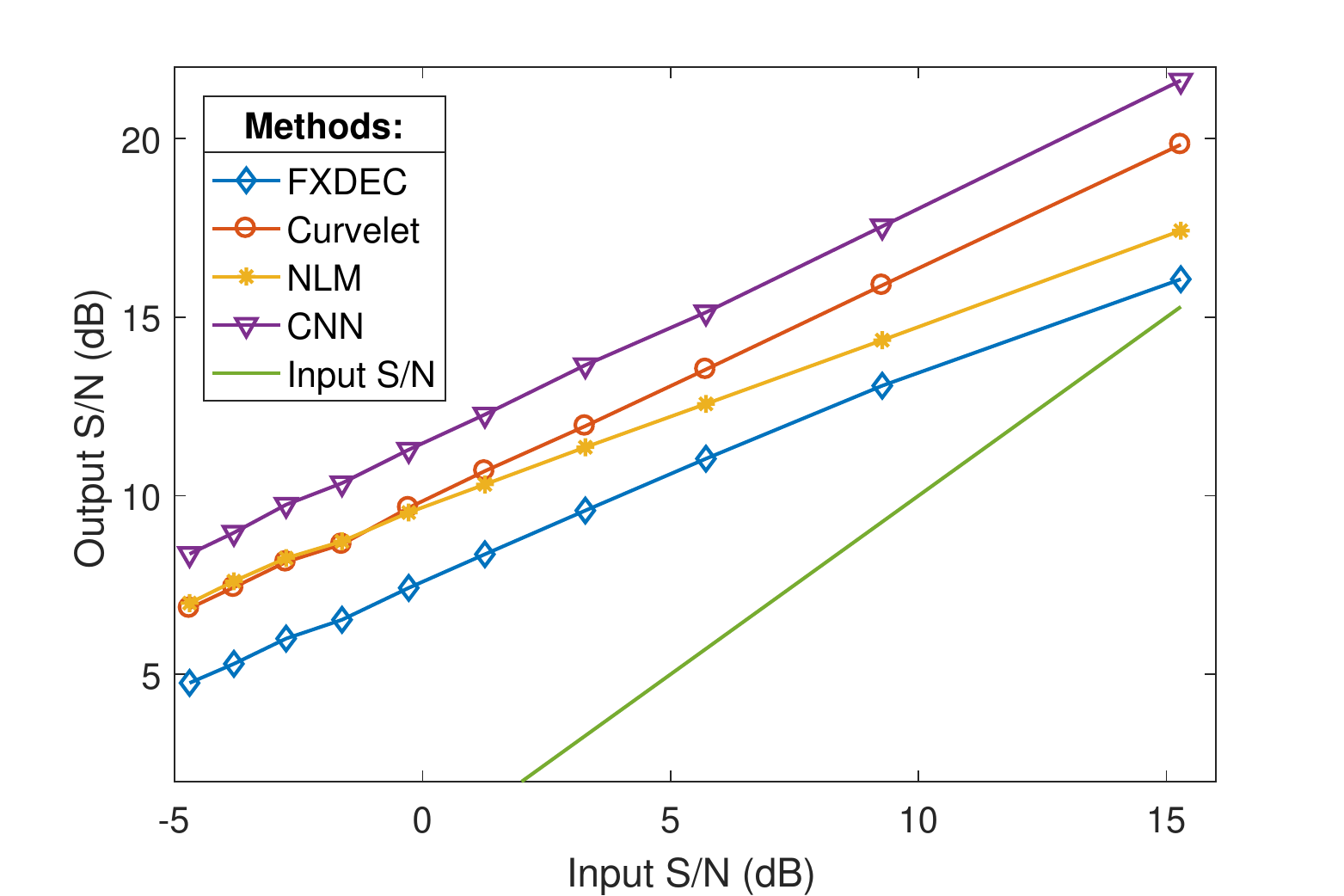}
    \caption{Comparisons of denoising results S/N with different methods. The bottom line indicates output S/N equals input S/N.}
    \label{fig:snr_methods}
\end{figure*}

\begin{figure*}
    \centering
    \includegraphics[width=0.9\textwidth]{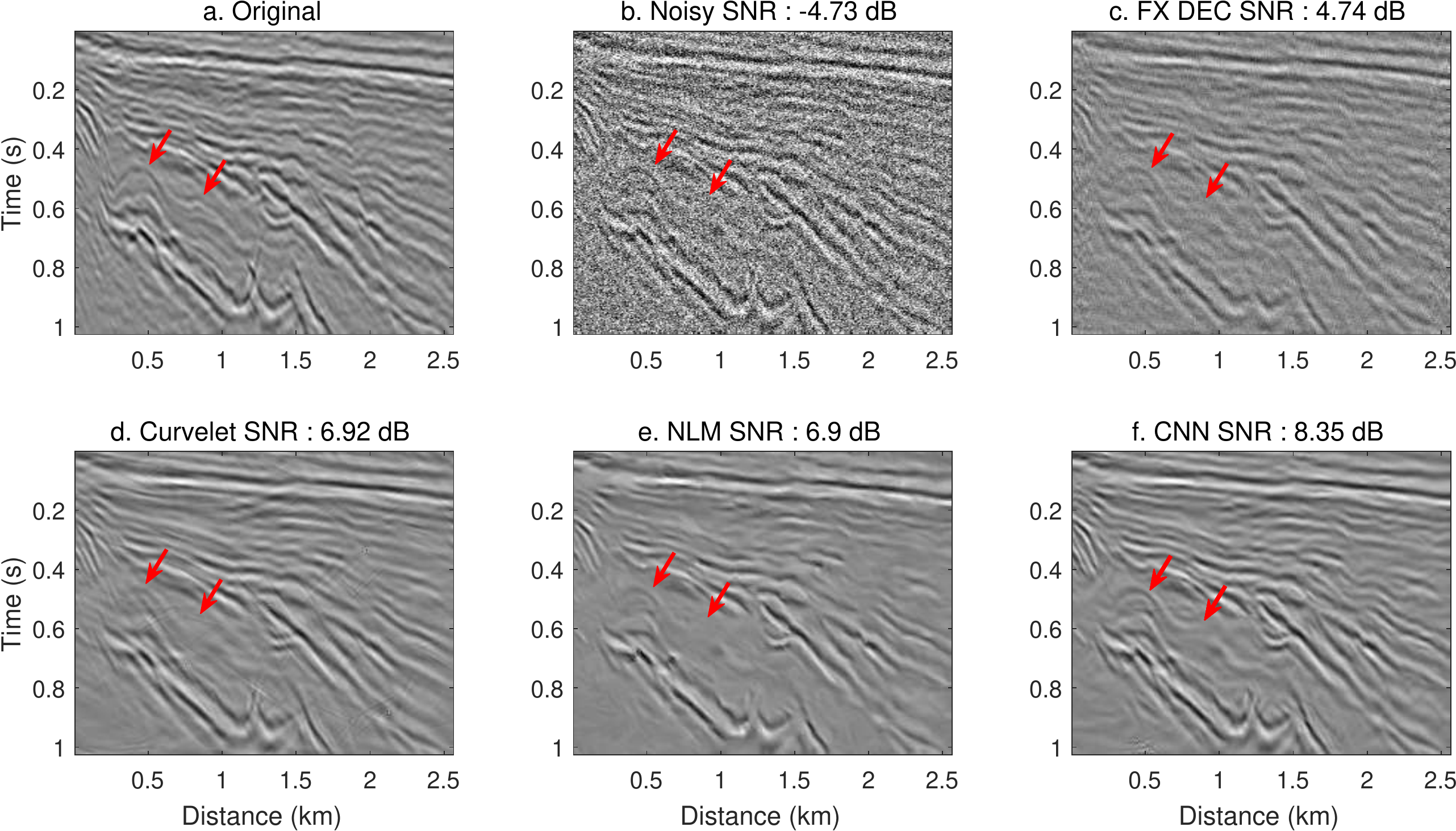}
    \caption{Comparisons of denoising results with different methods for a post-stack dataset.}
    \label{fig:data1}
\end{figure*}

\begin{figure*}
    \centering
    \subfigure[]{\label{fig:rd-denoise-1d}
    \includegraphics[width=0.45\textwidth]{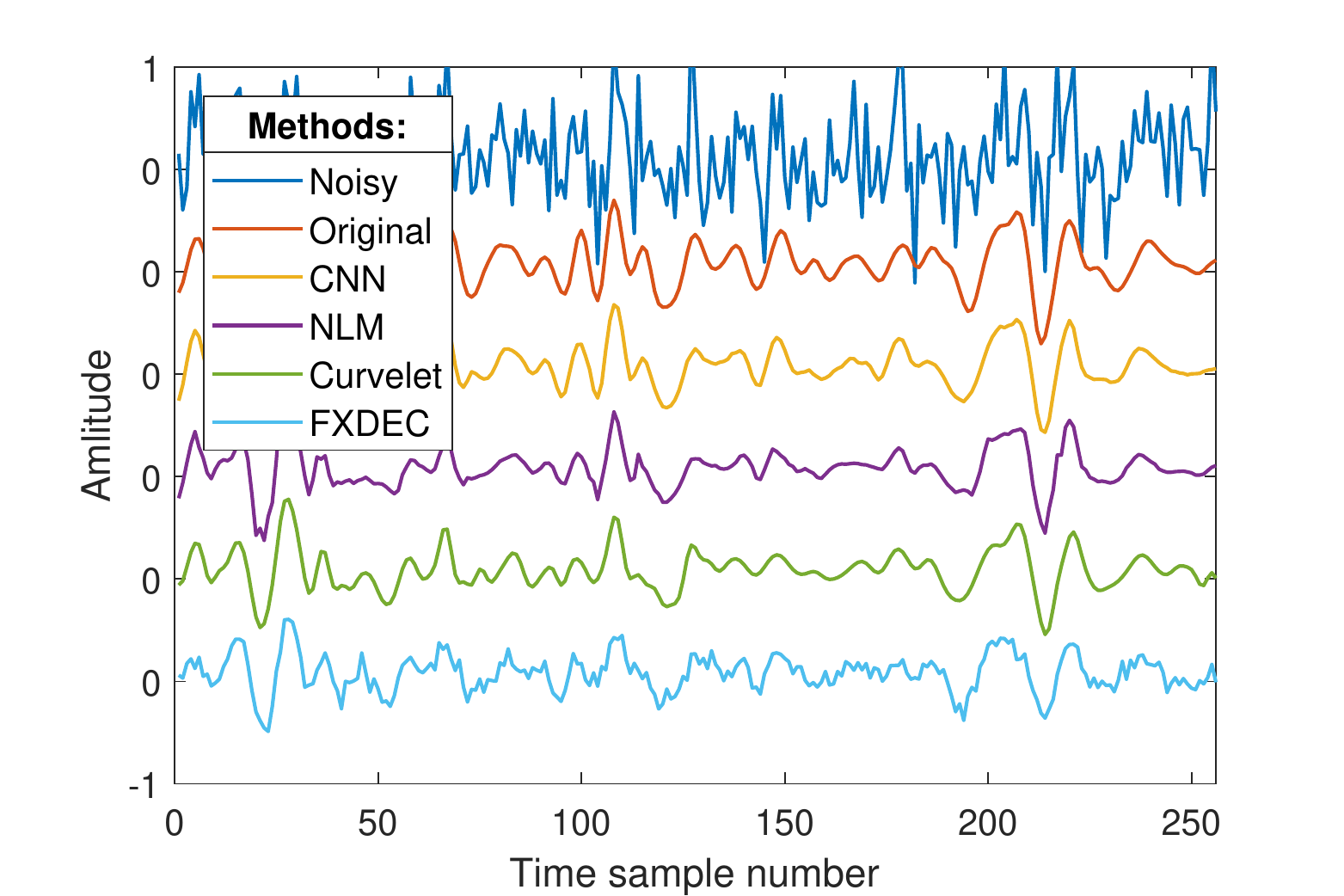}}
    \subfigure[]{\label{fig:rd-error-1d}
    \includegraphics[width=0.45\textwidth]{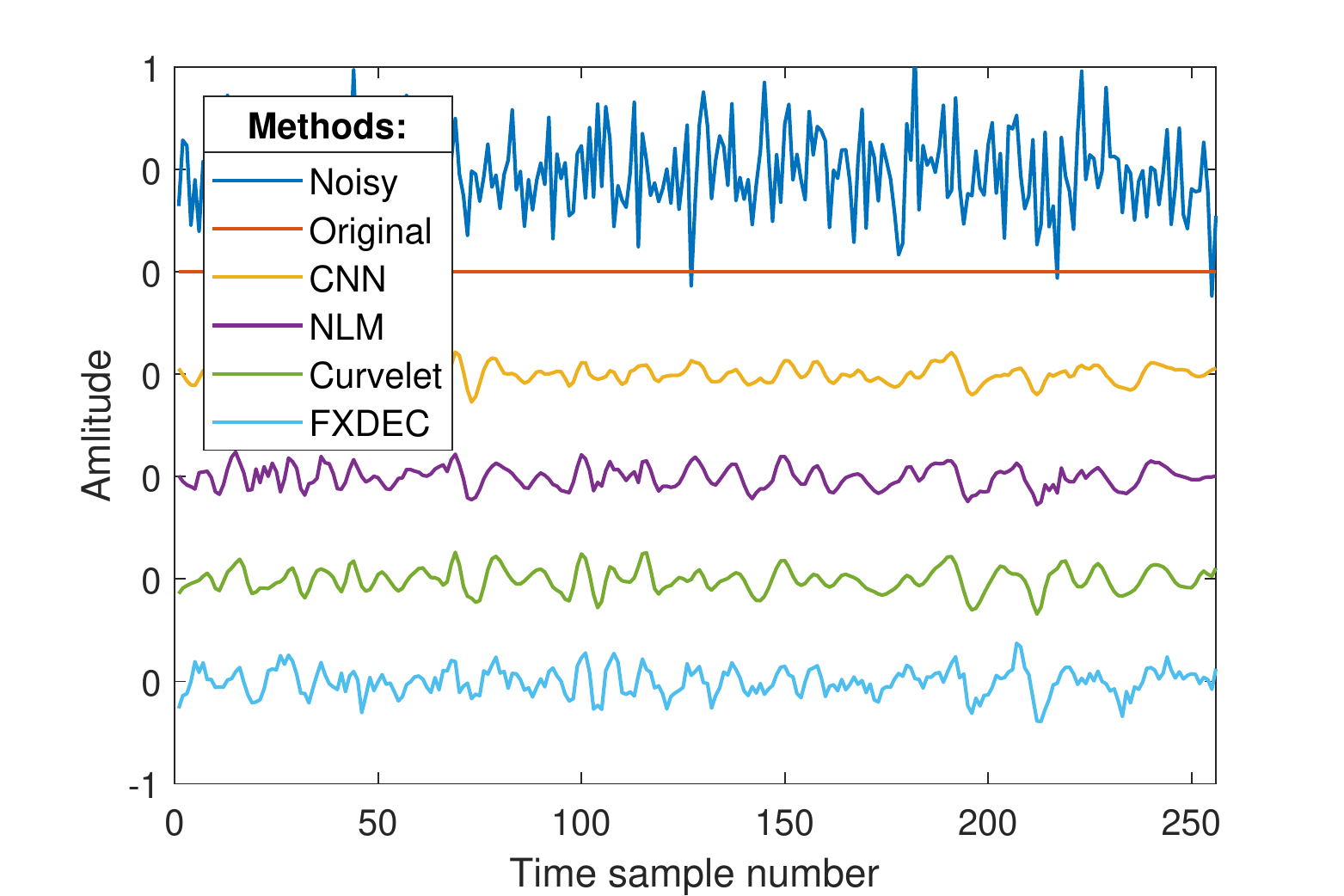}}
    \caption{ The denoised 80th trace (a) and the difference (b) with respect to the original trace. From top to bottom are noisy data, original data, CNN, BM3D, curvelet, and $f-x$ deconvolution methods. The differences are multiplied by 2.}
    \label{fig:rd-1d}
\end{figure*}

We used $f-x$ deconvolution \citep{Canales1984Random}, curvelet \citep{Hennenfent2006Seismic}, and nonlocal mean (NLM) \citep{Bonar2012Denoising} methods for comparison. For these three methods, we tested different parameters for the best denoising quality. This test was performed on the dataset shown in Figure \ref{fig:data1}b. The output S/N versus input S/N are shown in Figure \ref{fig:snr_methods}. {After the network is trained, the parameters need not be tuned for the testing datasets,}  {and} the CNN still achieves the best denoising quality (approximately 2 dB higher than the second) among the tested methods. However, {unlike} the {other} methods, we cannot control the harshness of the CNN manually without tuning the parameters. We can assume that the harshness is controlled adaptively by CNN itself according to the input, which avoids human interventions and achieves intelligent denoising. Figure \ref{fig:data1}  shows the denoised results.  The weak events marked by the arrows are much better preserved with the CNN method than with the other methods, but the diffractions are also smoothed while denoising. In traditional methods, we can tune the parameters to preserve the diffractions along with the noise. In the CNN, datasets with diffractions must be included to train the network. Figure \ref{fig:rd-1d} shows the 80th trace from the denoised result and the corresponding difference concerning the original trace. The differences are multiplied by two. It is clear that CNN preserves the amplitudes best and produces the smallest difference. The training time is approximately nine hours. The elapsed time for the denoising test, shown in Figure \ref{fig:data1}, of each method on the CPU is:  0.08 s ($f-x$ deconvolution), 0.24 s (curvelet), 5.46 s (NLM), and 1.68 s (CNN). The CNN on the GPU requires 0.008 s,  {which makes it feasible for large-scale datasets.}

\begin{figure*}
    \centering
    \subfigure[]{\label{fig:07b-1}
    \includegraphics[width=0.2\textwidth]{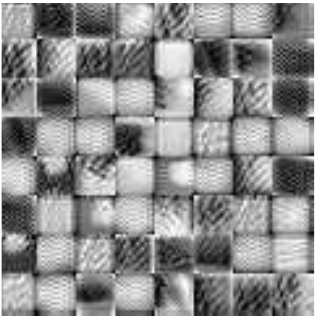}}
    \subfigure[]{\label{fig:07b-2}
    \includegraphics[width=0.3\textwidth]{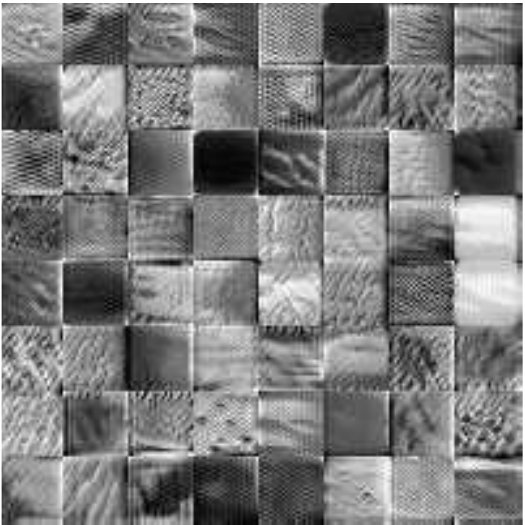}}
    \subfigure[]{\label{fig:07b-3}
    \includegraphics[width=0.4\textwidth]{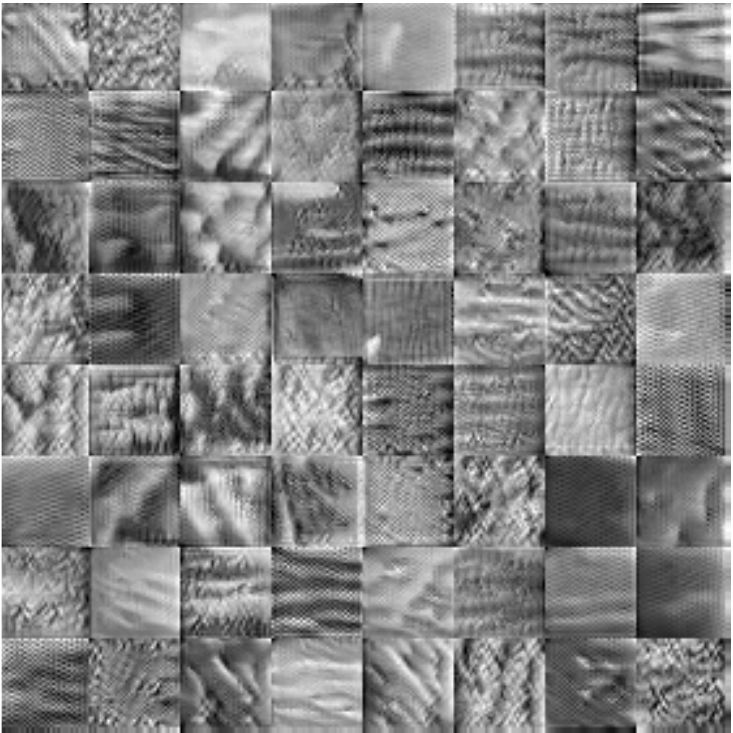}}
    \caption{ (a--c) Visualizations of the intermediate activations of CNN corresponding to each convolutional filter in layers 7, 12, 17  in the image domain. The hierarchical textures in different layers represent how a CNN understands seismic datasets.}
    \label{fig:cnn_layer}
\end{figure*}

As shown in Figure \ref{fig:cnn_layer}, the inverting CNN method \citep{mahendran2015understanding} is used to visualize the activations in the image domain. The activations corresponding to every convolutional filter in layers 7, 12, and 17 are set to ones. In each subfigure, each block represents the input that activates the given activations best. The blocks show obvious texture features in different scales corresponding to different layers. The textures represent how CNN understands seismic datasets, which is currently difficult to explain.

\begin{figure*}
    \centering
    \subfigure[]{\label{fig:05i-1}
    \includegraphics[width=0.3\textwidth]{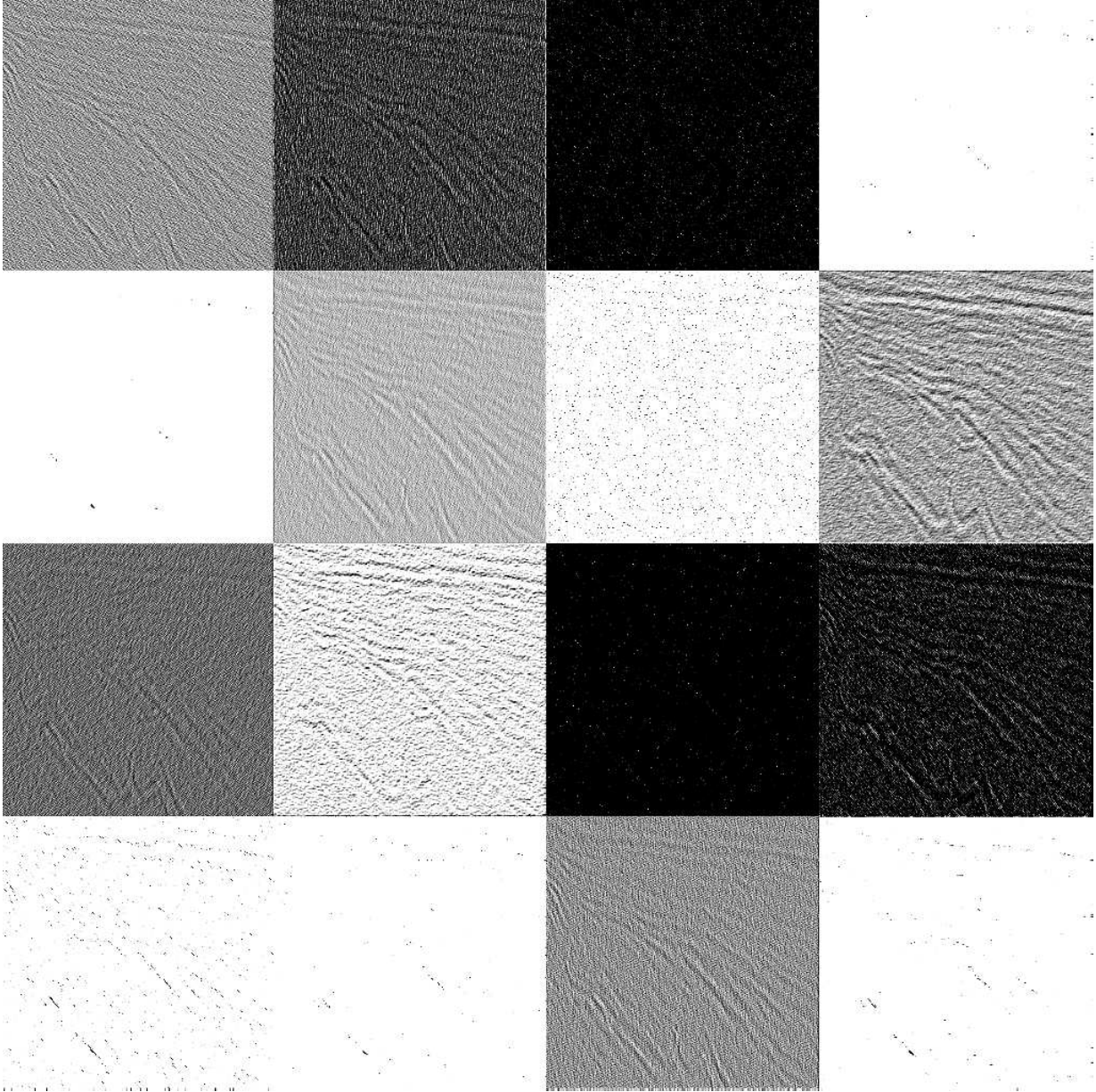}}
    \hfill
    \subfigure[]{\label{fig:05i-2}
    \includegraphics[width=0.3\textwidth]{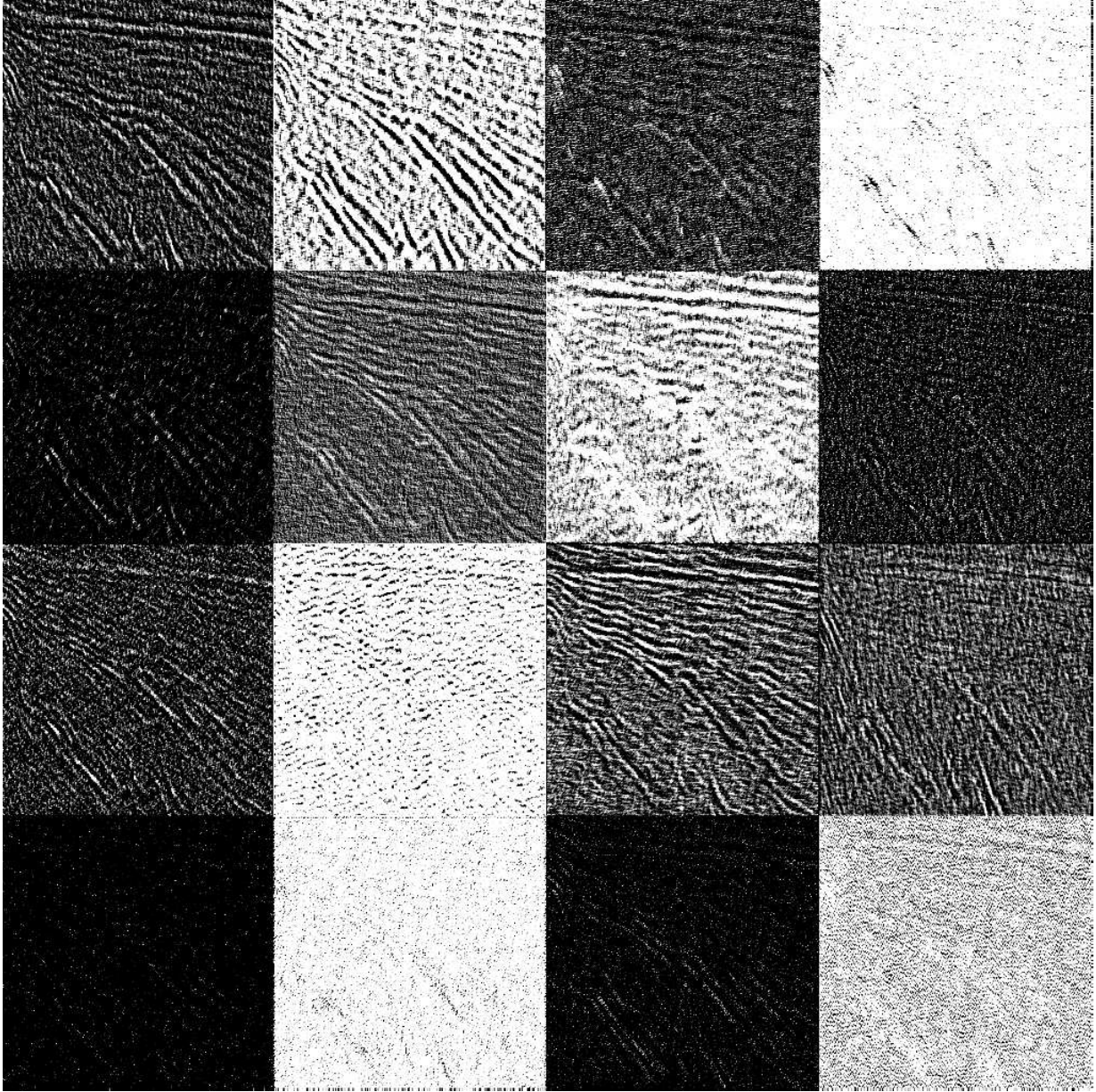}}
    \hfill
    \subfigure[]{\label{fig:05i-3}
    \includegraphics[width=0.3\textwidth]{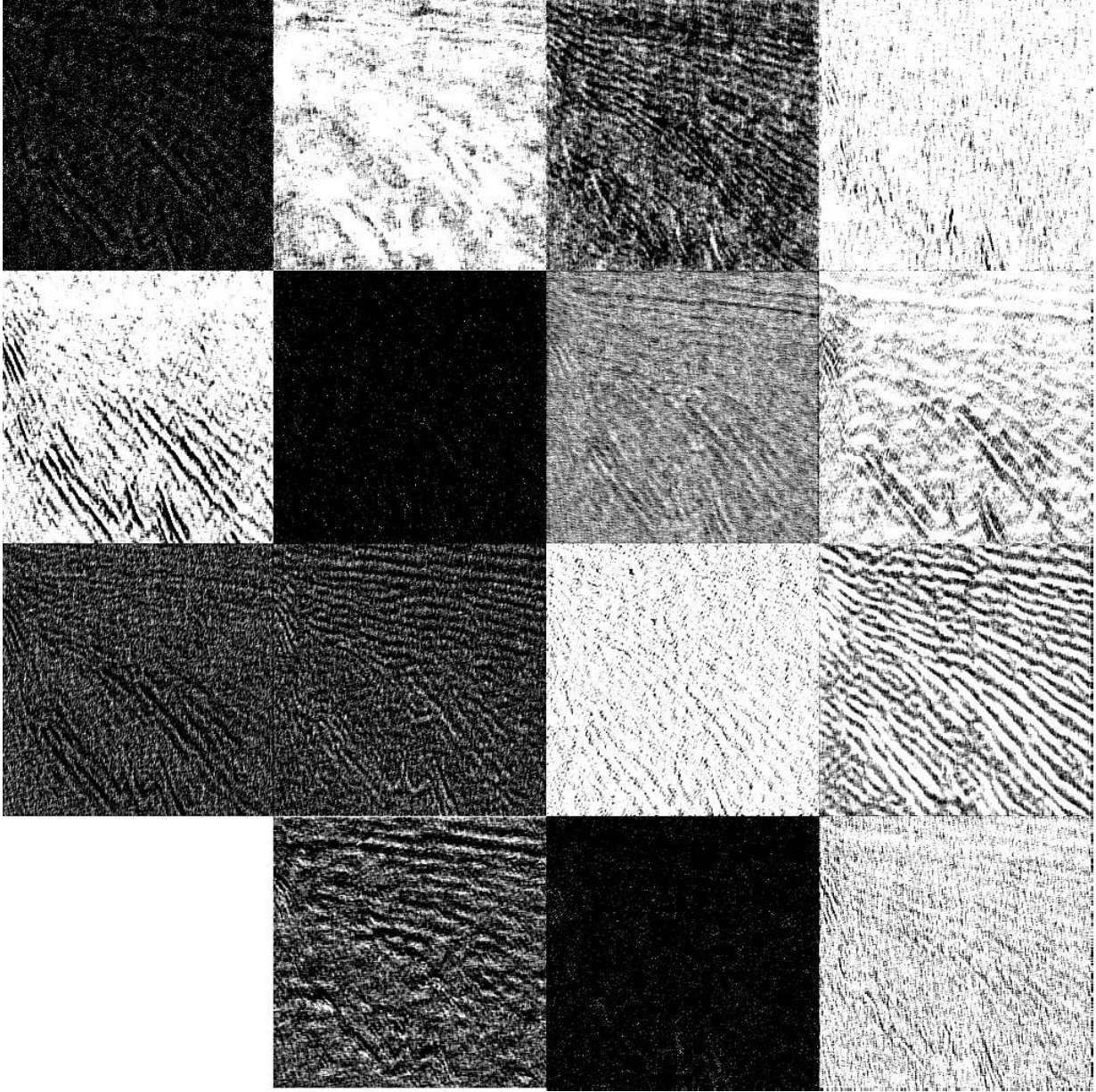}}

    \subfigure[]{\label{fig:05i-4}
    \includegraphics[width=0.3\textwidth]{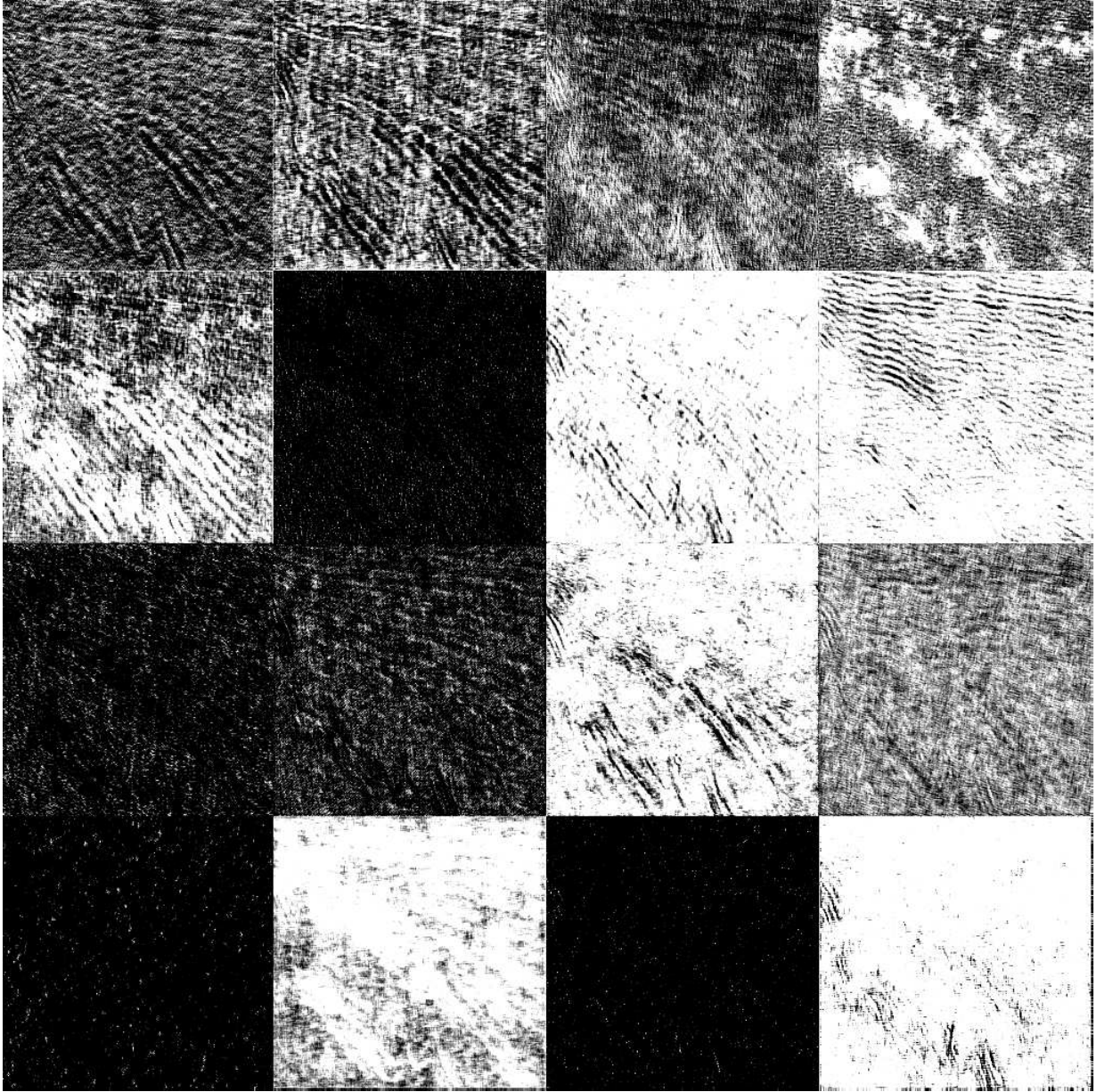}}
    \hfill
    \subfigure[]{\label{fig:05i-5}
    \includegraphics[width=0.3\textwidth]{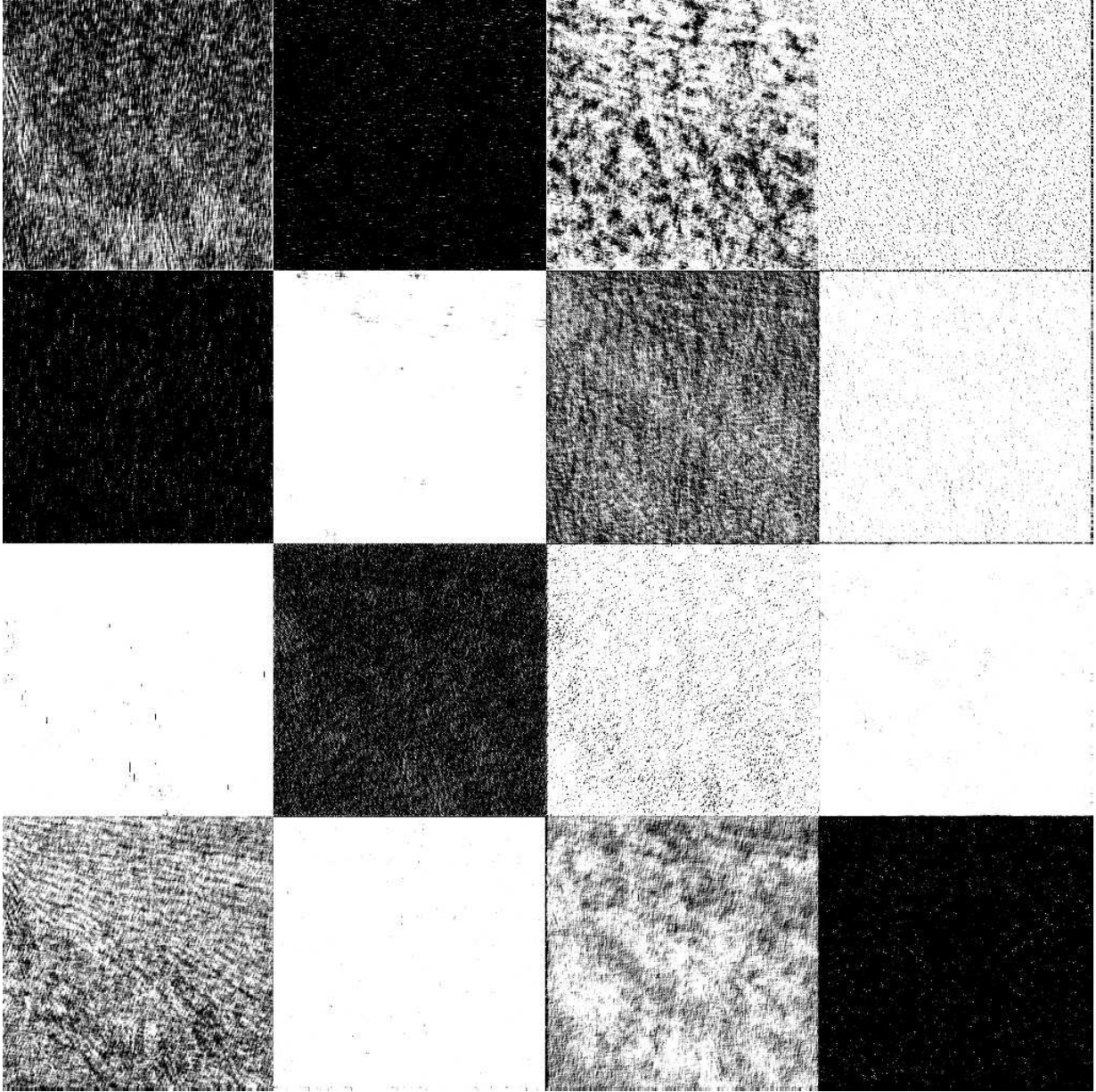}}
    \hfill
    \subfigure[]{\label{fig:05i-6}
    \includegraphics[width=0.3\textwidth]{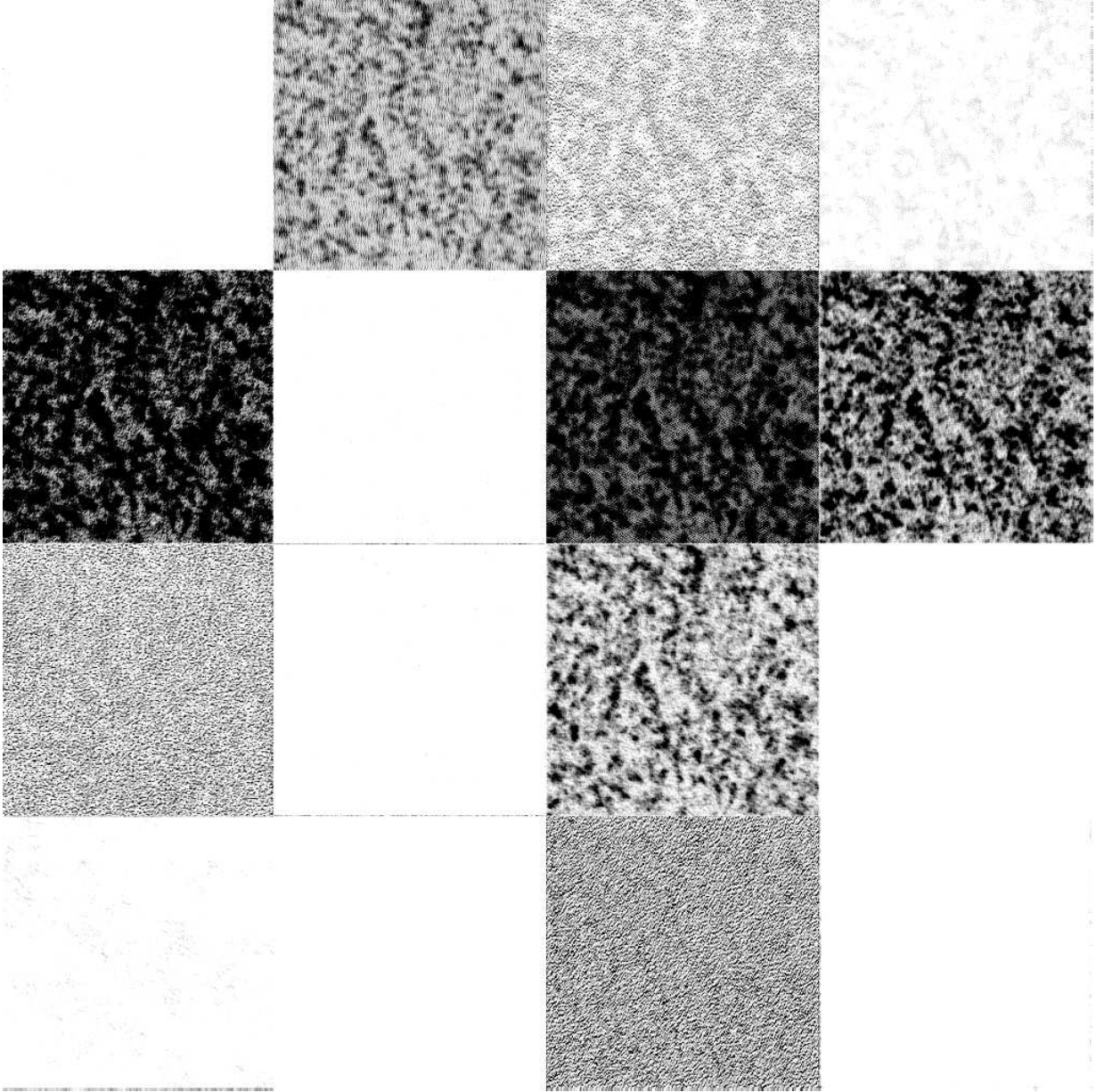}}
    \caption{ (a)--(f) Intermediate outputs of 6 ReLU layers (out of 16). In each subfigure, there are outputs from the first 16 channels (out of 64), with each corresponding to one convolutional filter. The subfigures are sorted according to the direction from input to output. The event and noise are gradually separated. White pixels indicate zeros and gray/black pixels indicate positive values.}
    \label{fig:random_inside}
\end{figure*}

It is challenging to analyze how CNN operates in theory because many layers and nonlinearities exist. From the numerical perspective, we present the intermediate outputs of six ReLU layers (out of 16) in Figure \ref{fig:random_inside}. In each subfigure, outputs are presented from the first 16 channels (out of 64), with each corresponding to one convolutional filter. The subfigures are sorted according to the direction from the input to output. This test is performed on the data in Figure \ref{fig:data1}. We observed that the seismic events are removed gradually and the random noise {is} retained (we use noise as the {outputs}).

\subsubsection{Linear noise attenuation}

\begin{figure*}
    \centering
    \subfigure[]{\label{fig:07-1}
    \includegraphics[width=0.3\textwidth]{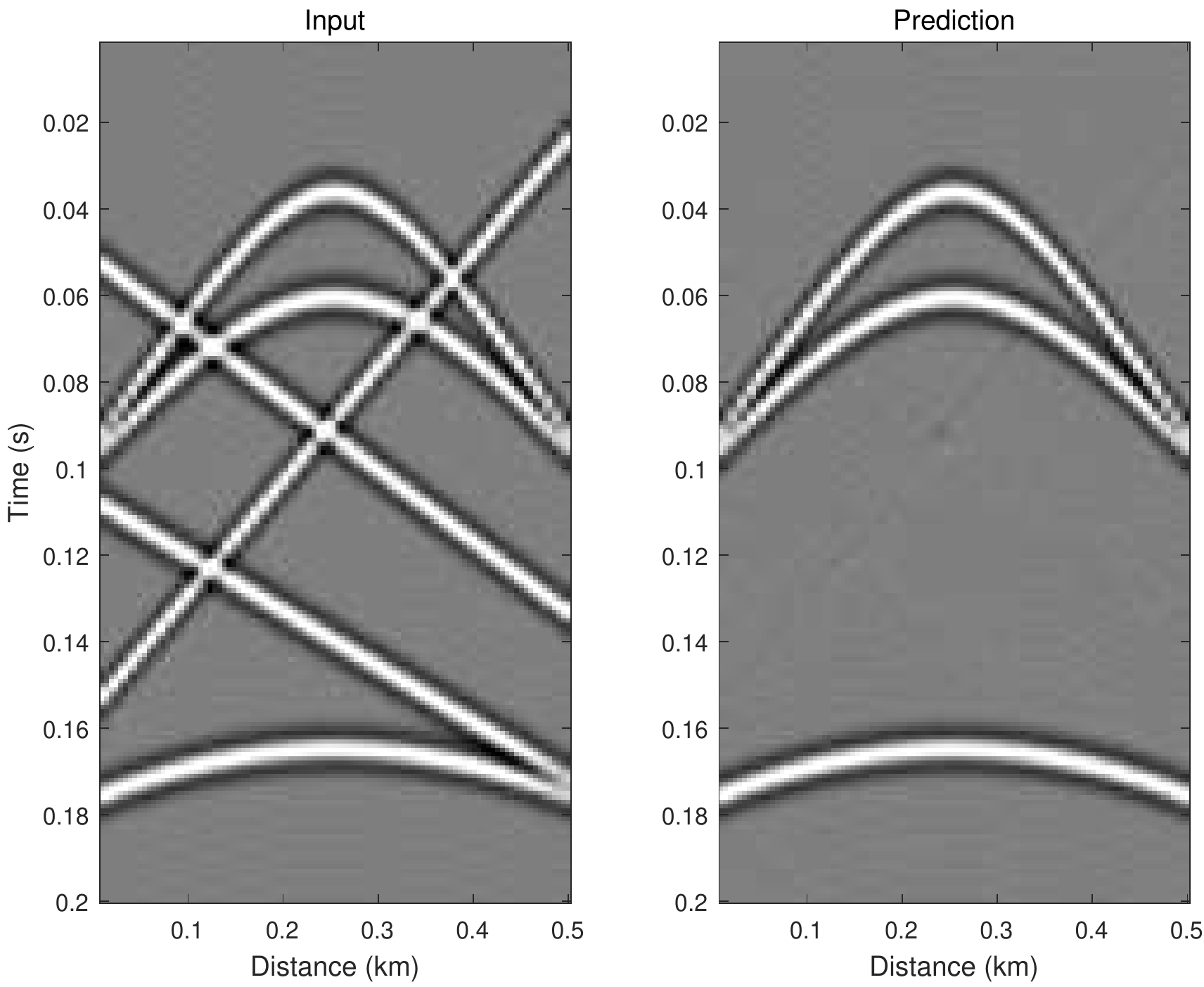}}
    \subfigure[]{\label{fig:07-2}
    \includegraphics[width=0.3\textwidth]{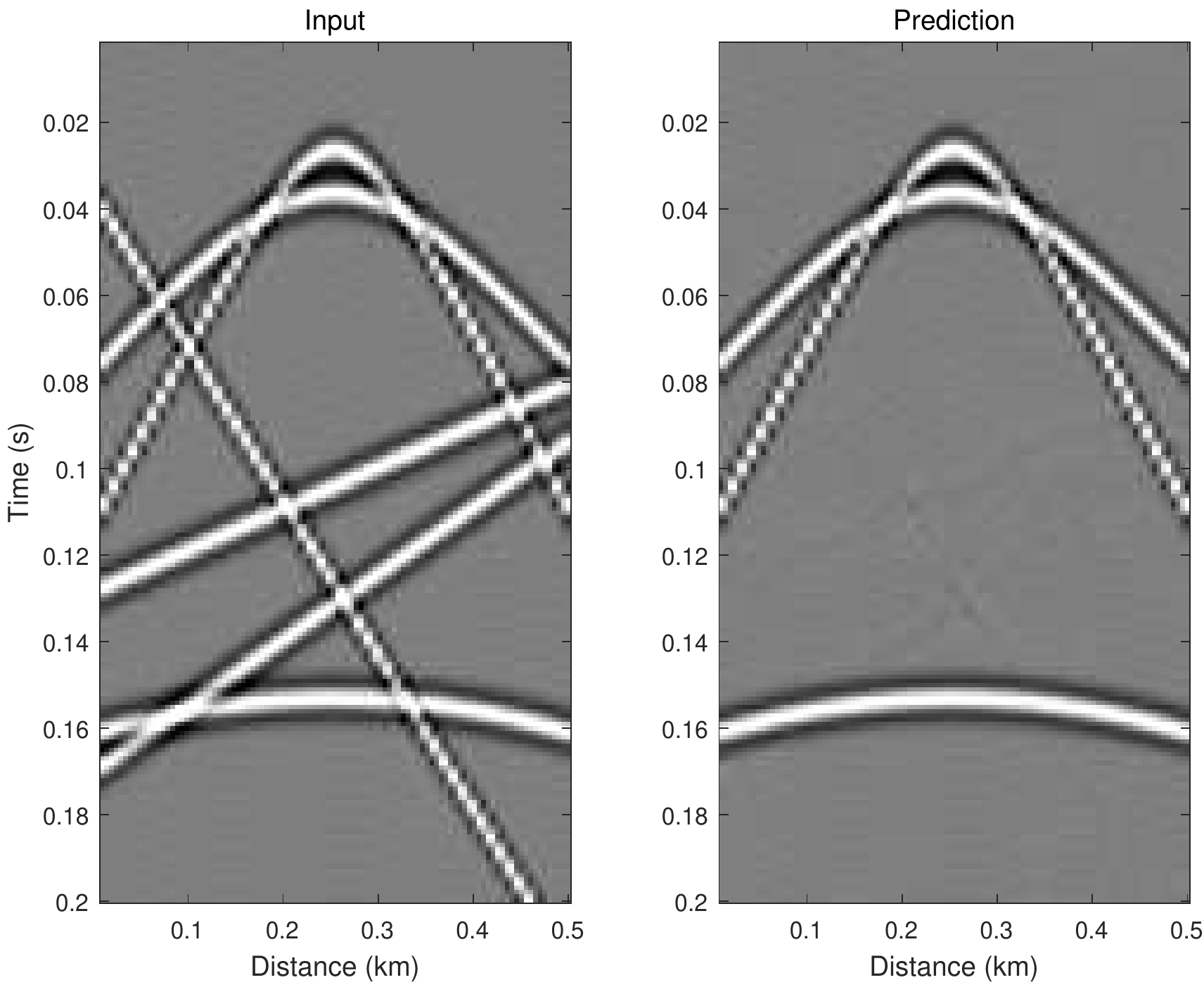}}
    \subfigure[]{\label{fig:07-3}
    \includegraphics[width=0.3\textwidth]{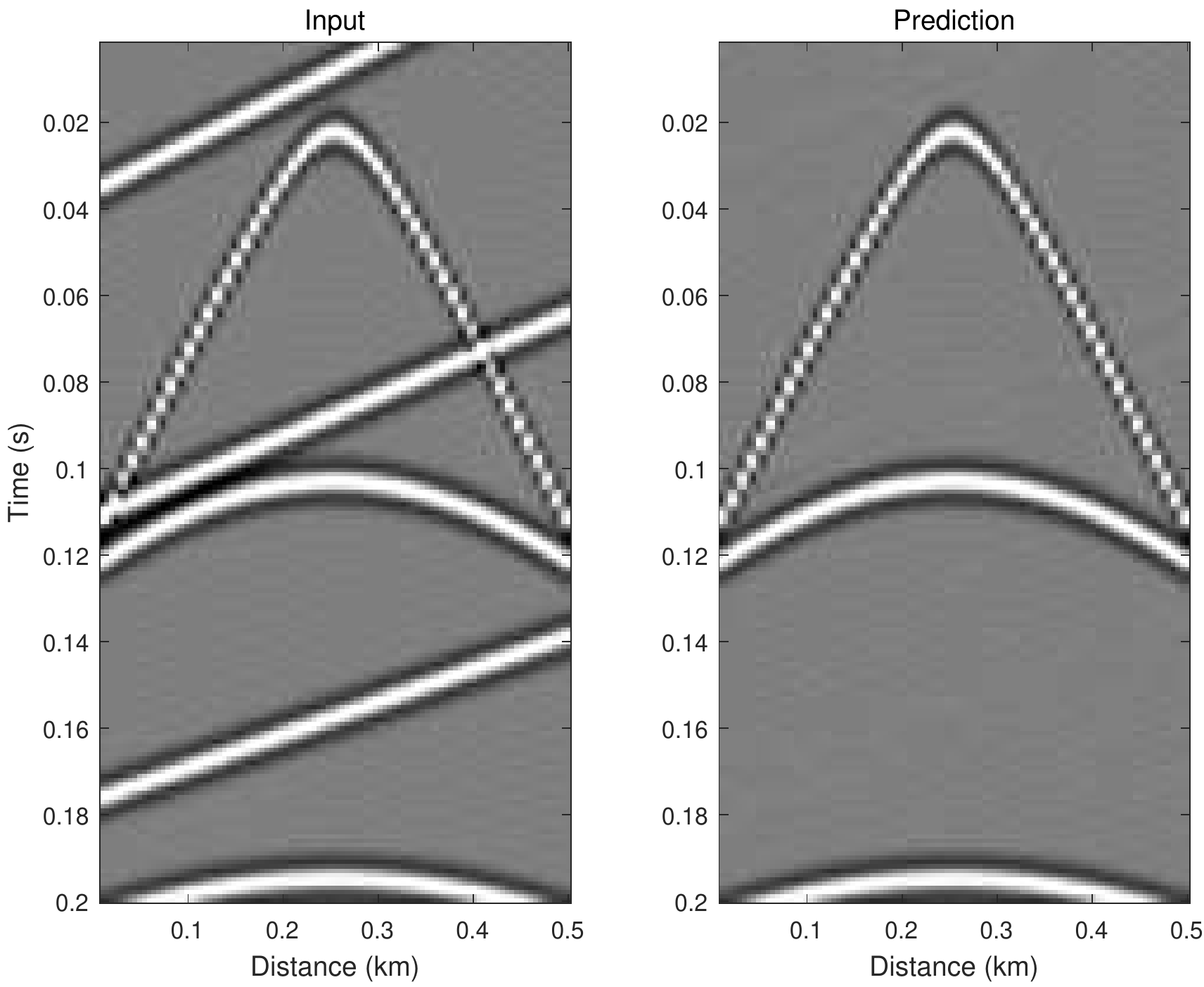}}
    \caption{ Three linear noise attenuation results. In each subfigure, left is the data with linear noise, right is the denoised data.}
    \label{fig:linear_denoise}
\end{figure*}

For the linear noise situation, we used the CNN with the same  {architecture}  as in the random noise situation. The batch size is 64. Figure \ref{fig:linear_denoise} shows three denoising results from the test set. In each subfigure, data with linear noise is on the left, and denoised data is on the right. The linear noise is removed and hardly observed in the prediction section. DL shows the ability to learn knowledge from the training set and to use the knowledge on new data.  The training time is approximately five hours, while the denoising time on the input with size 100$\times 50$ is 0.13 s on the CPU and 0.008 s on the GPU.

\begin{figure*}
    \centering
    \subfigure[]{\label{fig:07i-1}
    \includegraphics[width=0.3\textwidth]{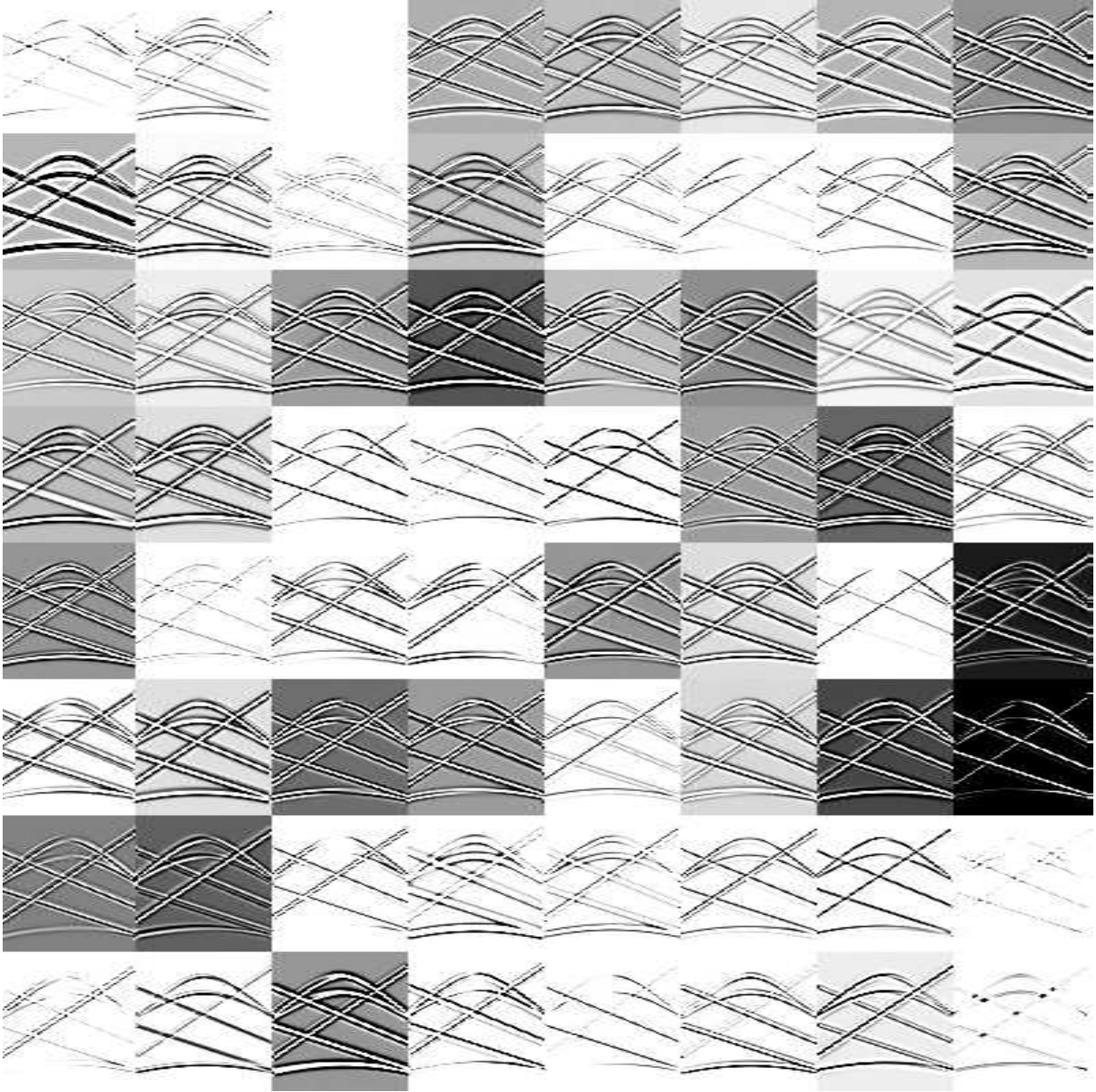}}
    \hfill
    \subfigure[]{\label{fig:07i-2}
    \includegraphics[width=0.3\textwidth]{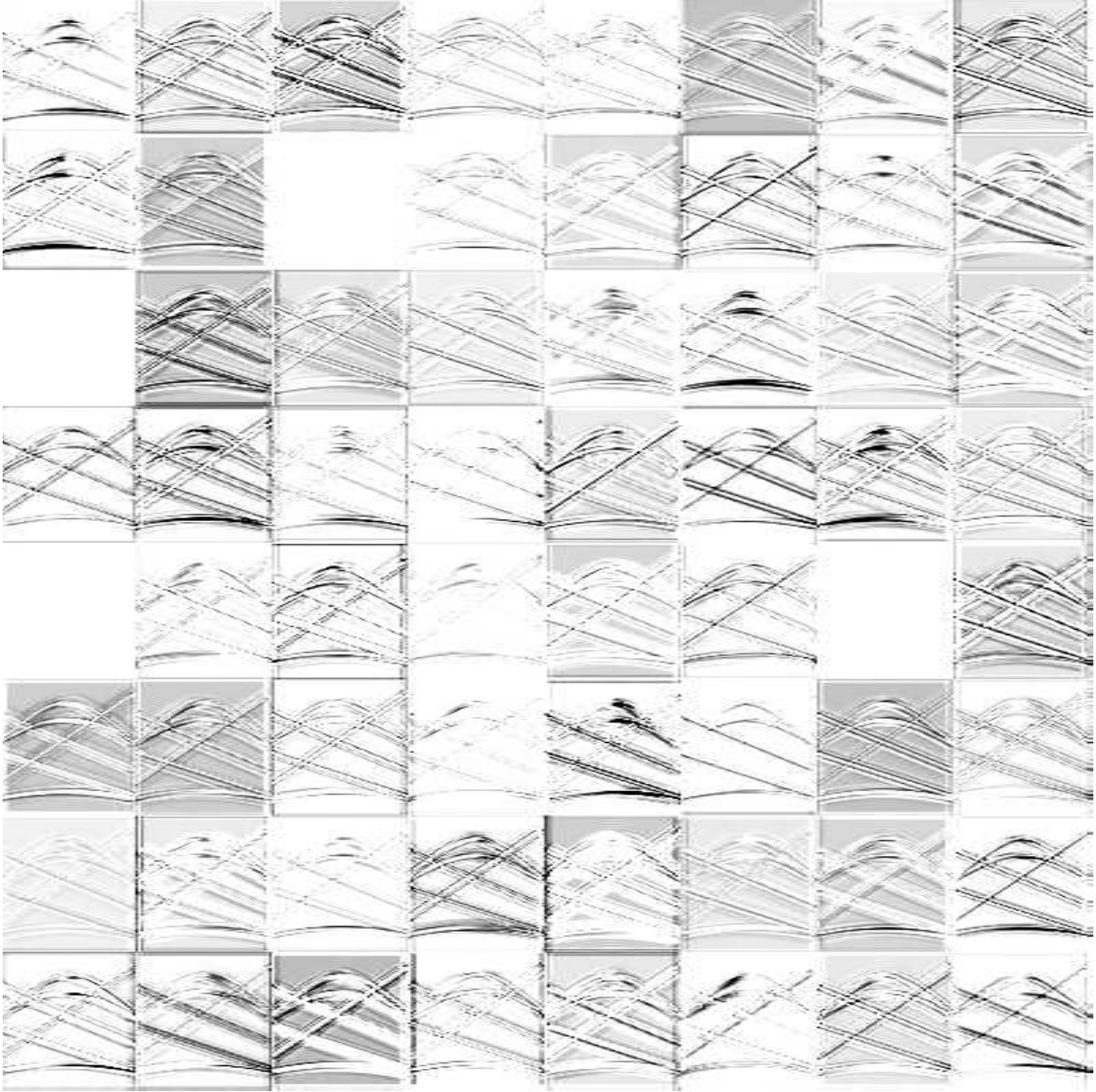}}
    \hfill
    \subfigure[]{\label{fig:07i-3}
    \includegraphics[width=0.3\textwidth]{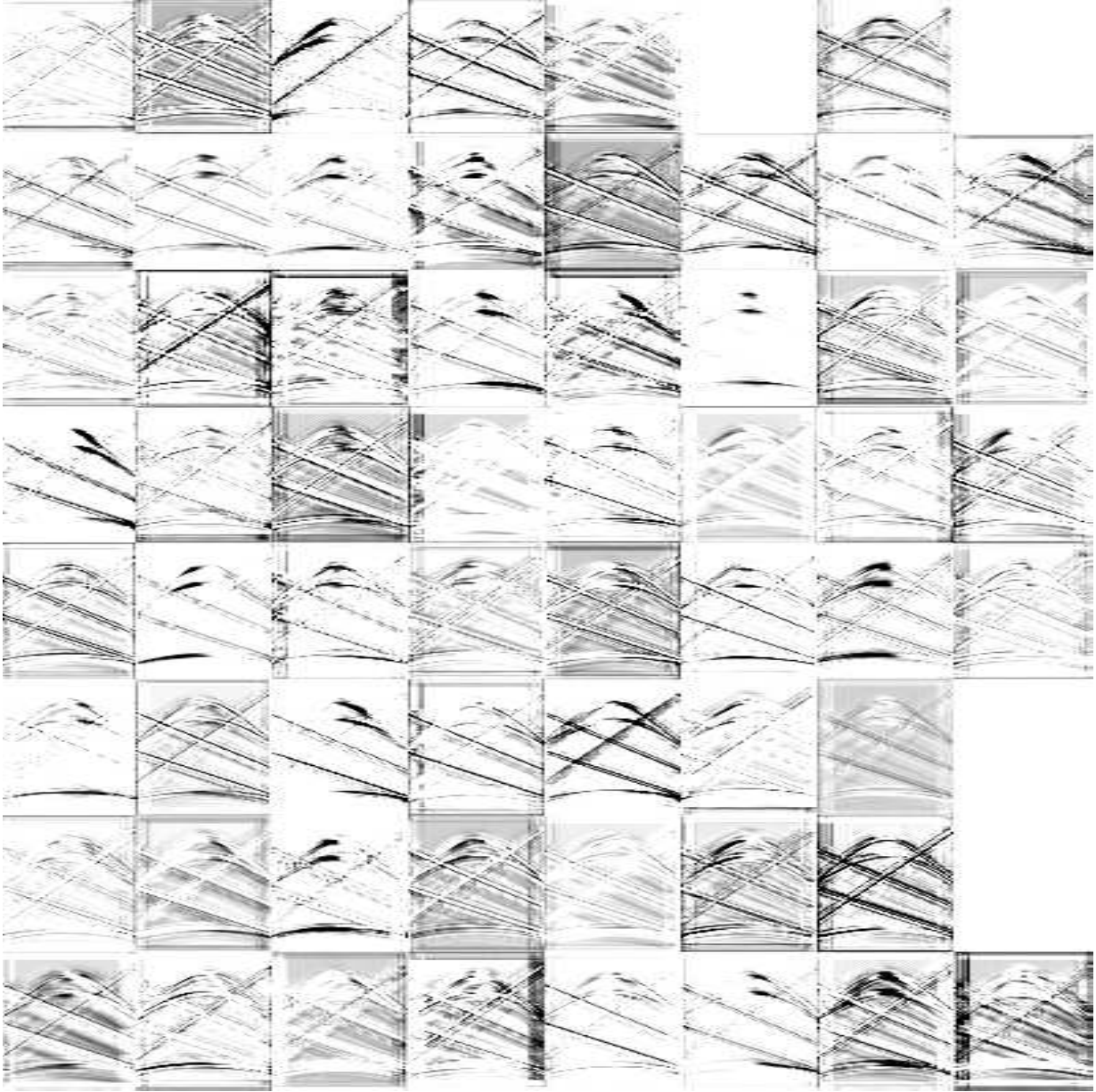}}

    \subfigure[]{\label{fig:07i-4}
    \includegraphics[width=0.3\textwidth]{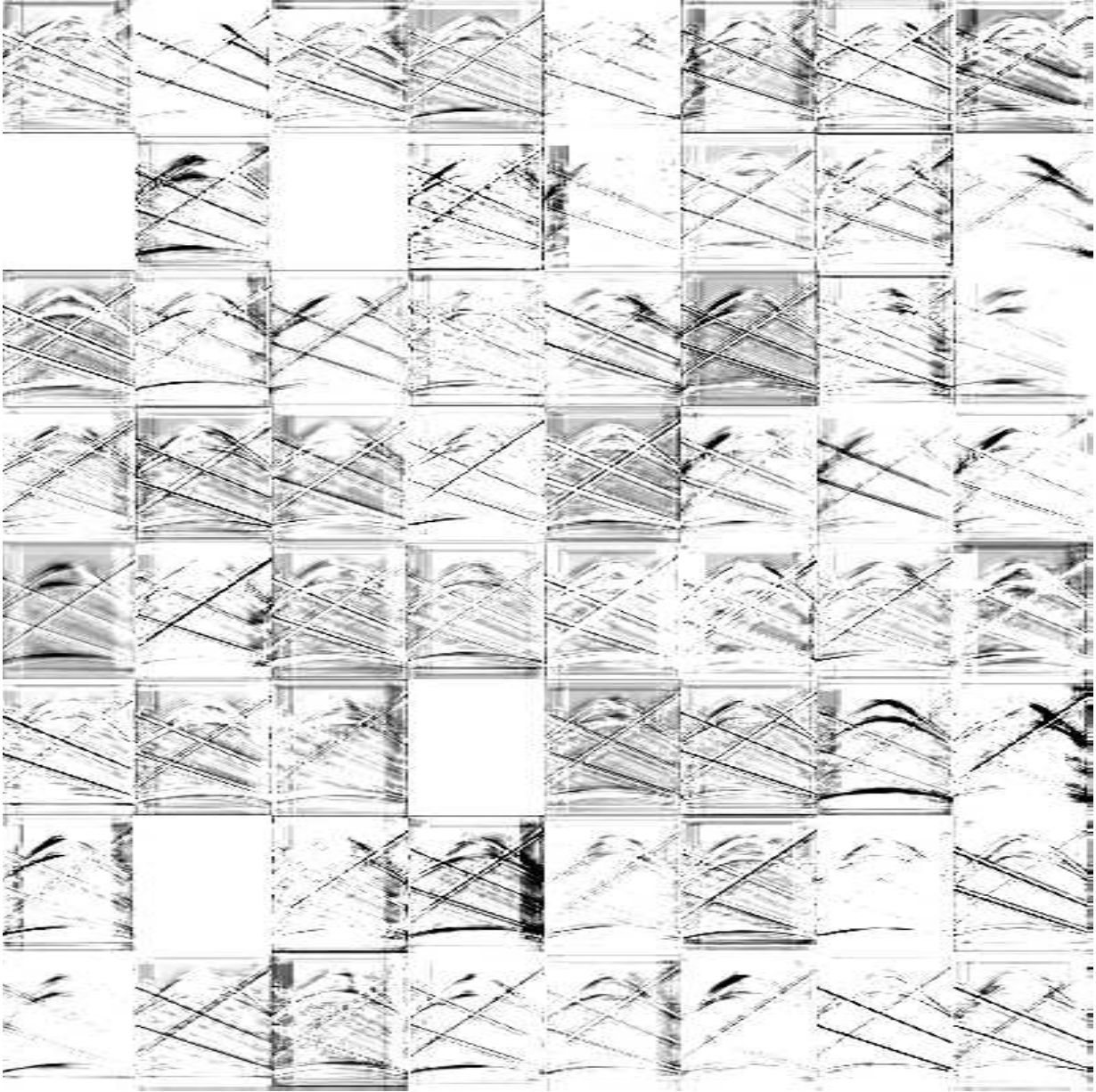}}
    \hfill
    \subfigure[]{\label{fig:07i-5}
    \includegraphics[width=0.3\textwidth]{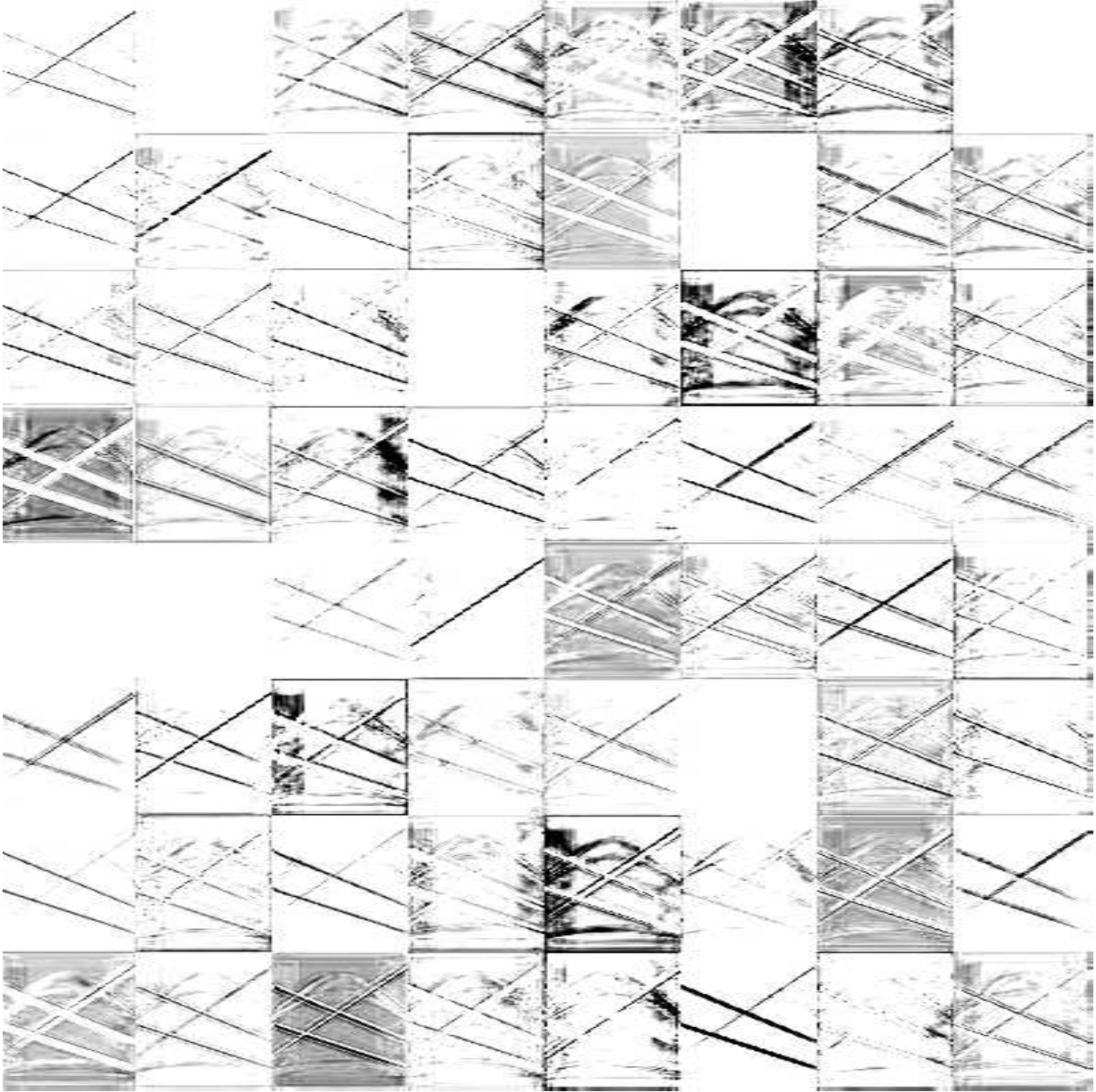}}
    \hfill
    \subfigure[]{\label{fig:07i-6}
    \includegraphics[width=0.3\textwidth]{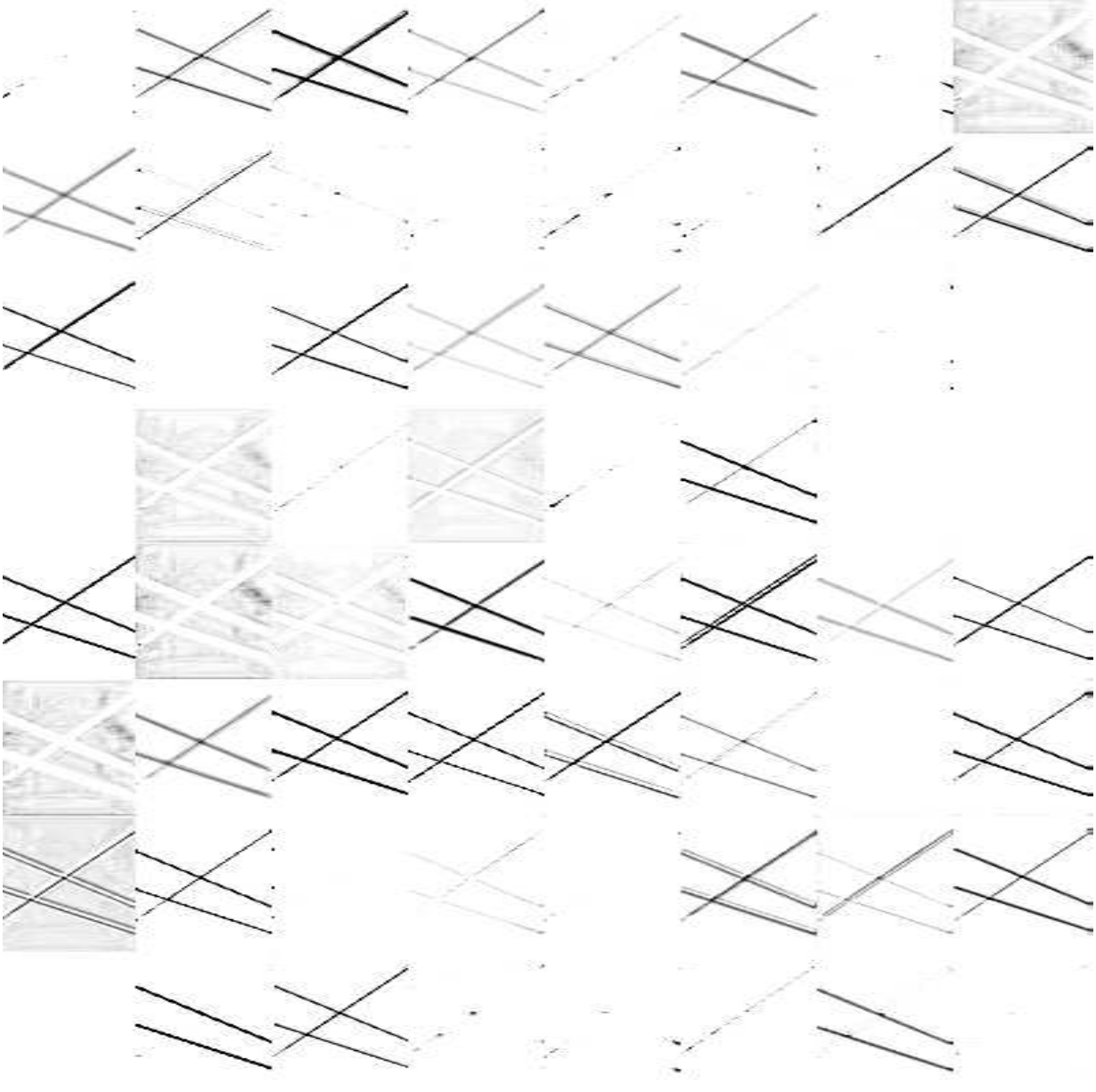}}
    \caption{ The output of six hidden layers after ReLU operation of CNN for linear noise attenuation. (a)--(f) are sorted according to the direction from input to output. The linear events are extracted in deeper layers.}
    \label{fig:linear_inside}
\end{figure*}

Figure \ref{fig:linear_inside} shows the outputs of six hidden layers. This test is performed on the data shown in Figure \ref{fig:07-1}. Each subfigure presents 64 outputs corresponding to the filters. It is clear that in the deeper layers, the hyperbolic events are removed gradually and linear events are preserved. 

\begin{figure*}
    \centering
    \subfigure[]{\label{fig:07b-1}
    \includegraphics[width=0.2\textwidth]{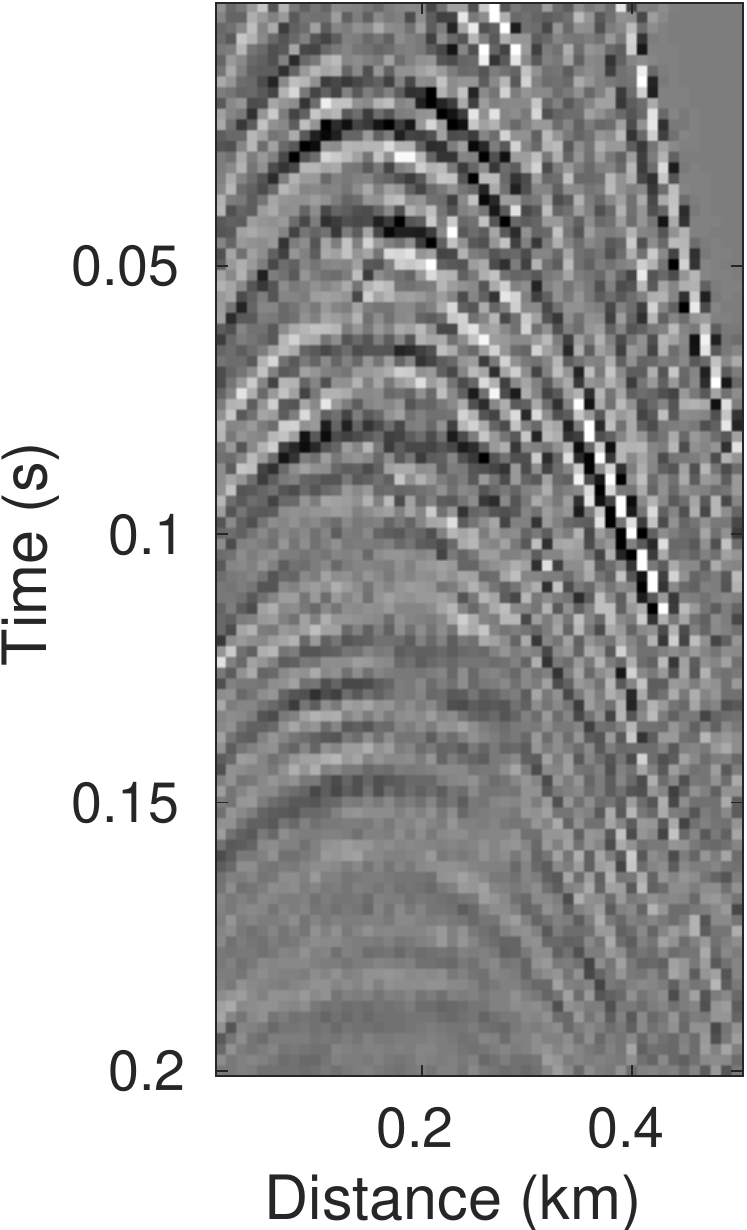}}
    \hfill
    \subfigure[]{\label{fig:07b-2}
    \includegraphics[width=0.2\textwidth]{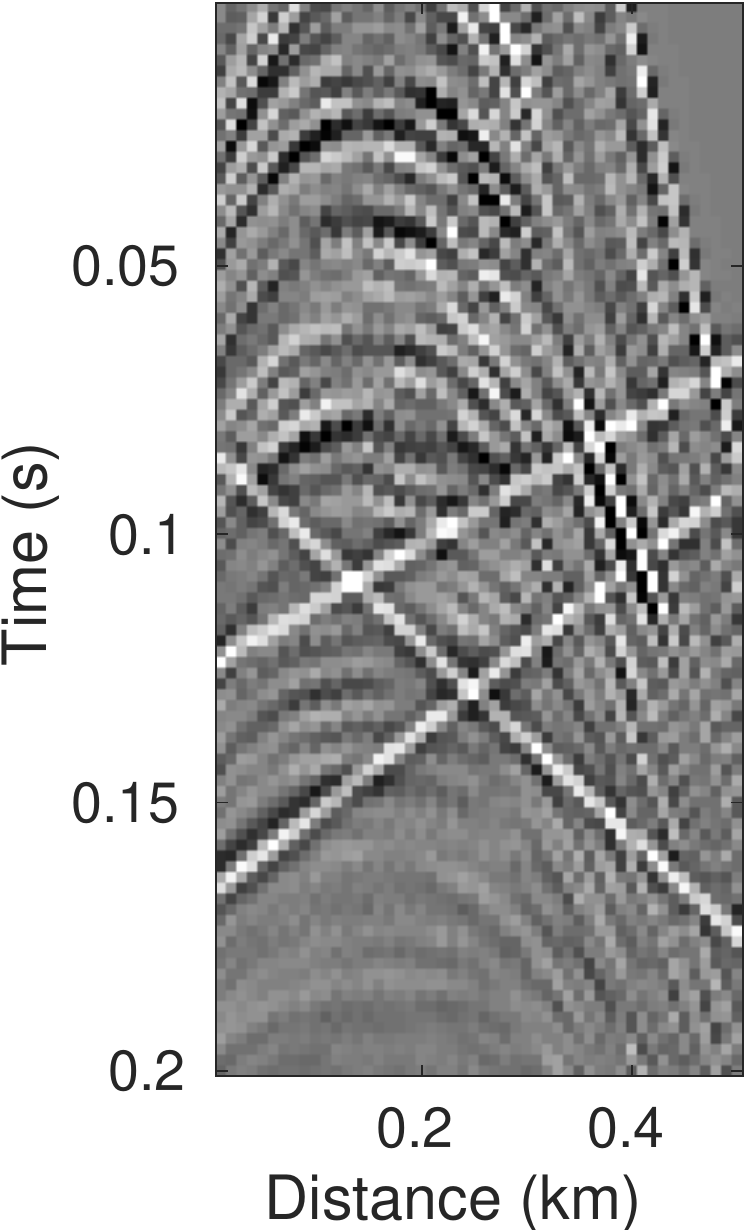}}
    \hfill
    \subfigure[]{\label{fig:07b-3}
    \includegraphics[width=0.2\textwidth]{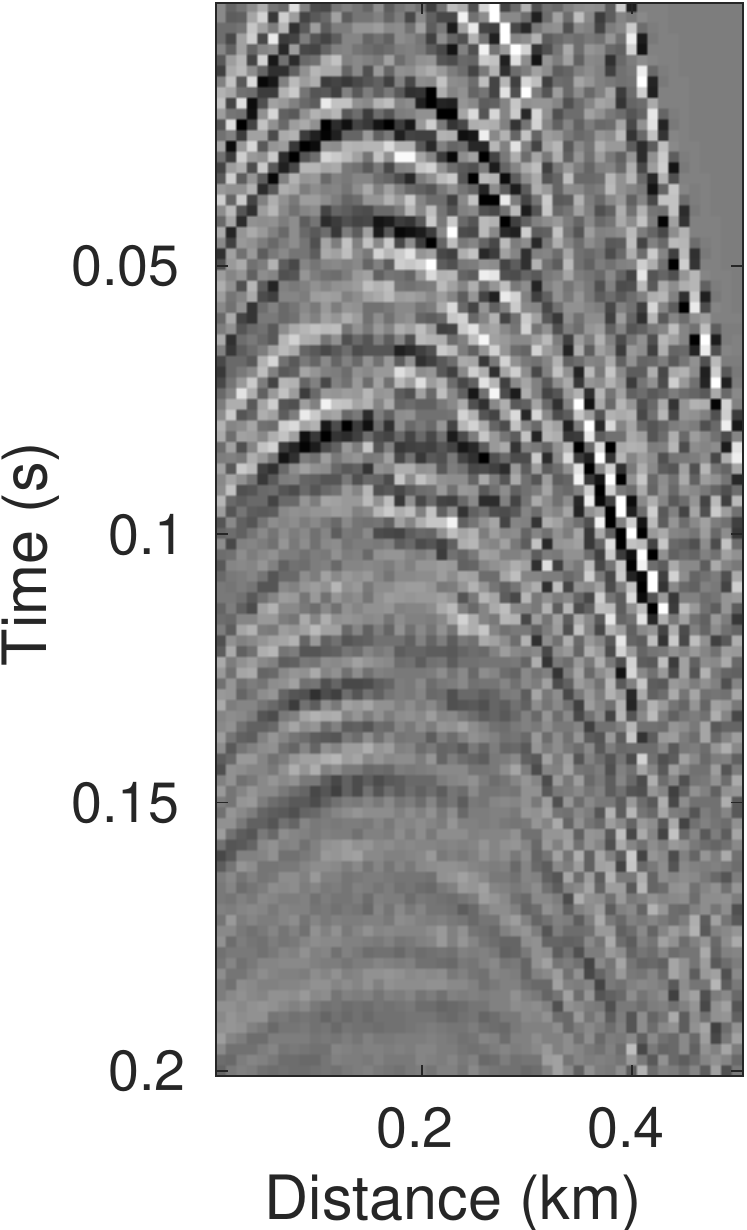}}
    \hfill
    \subfigure[]{\label{fig:07b-4}
    \includegraphics[width=0.2\textwidth]{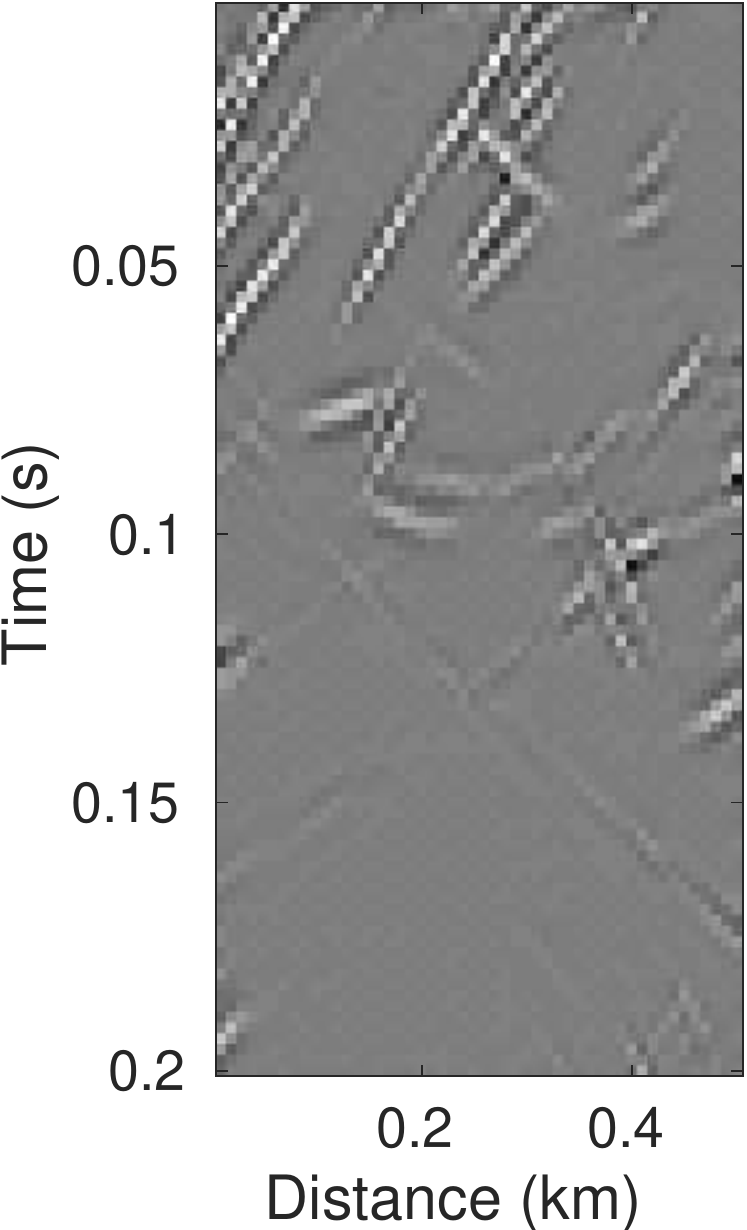}}

    \caption{{The previous trained CNN is applied on a `synthetic field dataset'. (a) A prestack dataset. (b) Linear noise is added to (a). (c) The denoised result obtained by CNN. (d) The difference between (a) and (c).}}
    \label{fig:linear_field}
\end{figure*}

A ``synthetic field dataset'' is also tested with the previous trained CNN. Figures \ref{fig:07b-1} and \ref{fig:07b-2} show a prestack dataset and the one contaminated with three linear events. Figures \ref{fig:07b-3} and \ref{fig:07b-4} show the denoised result with CNN and the difference map between the denoised result and the original dataset. The linear noise is successfully removed from Figure \ref{fig:07b-2}, which implies that the CNN can handle linear noise in the field dataset even if the CNN is trained with a synthetic training set. This can be explained by the fact that the `synthetic field dataset' in Figure \ref{fig:07b-1} has similar hyperbolical structures with the training set for Figure \ref{fig:linear_denoise}. However, in the difference map, obvious artifacts are left. 

\begin{figure*}
    \centering
    \subfigure[]{\label{fig:07c-1}
    \includegraphics[width=0.2\textwidth]{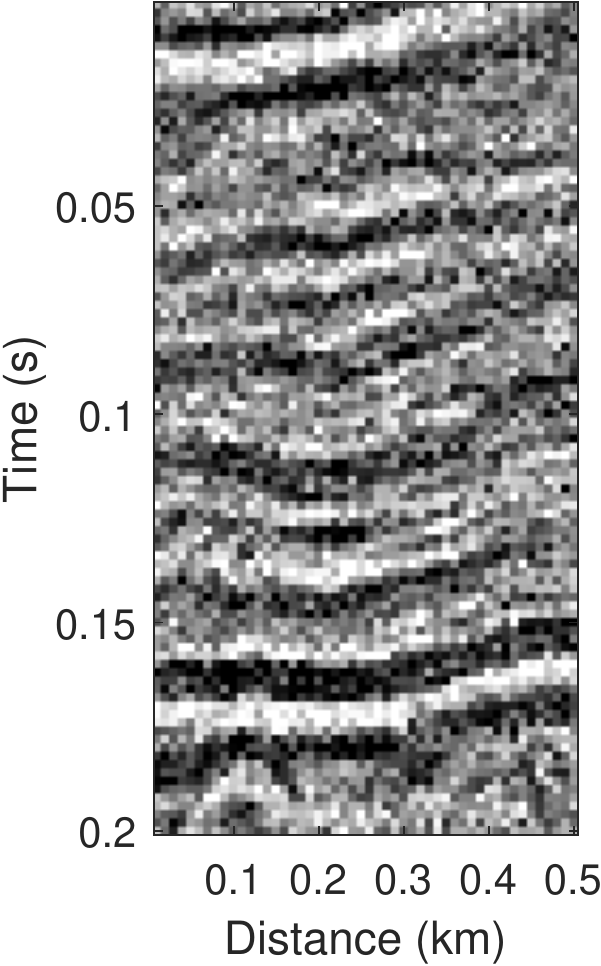}}
    \hfill
    \subfigure[]{\label{fig:07c-2}
    \includegraphics[width=0.2\textwidth]{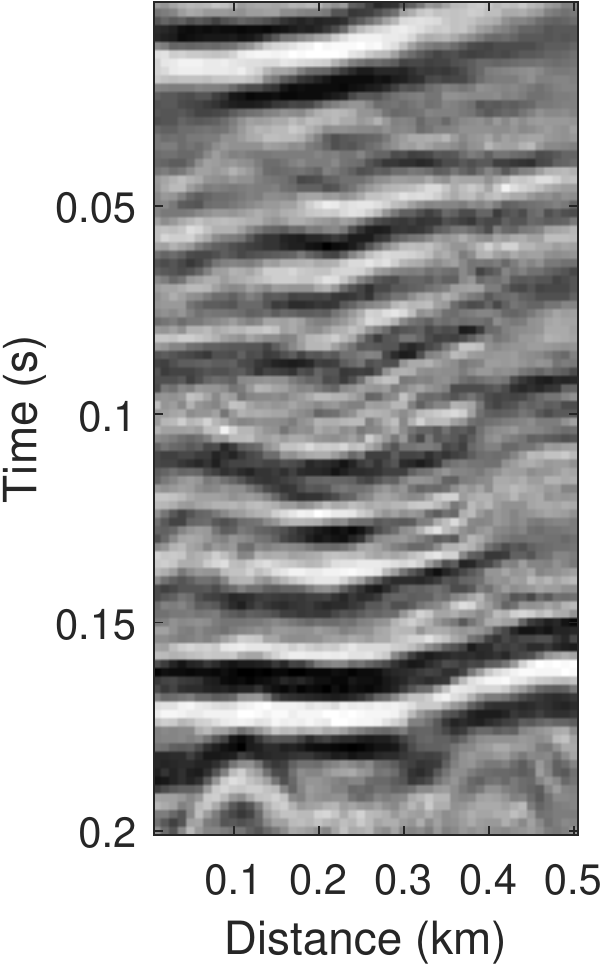}}
    \hfill
    \subfigure[]{\label{fig:07c-3}
    \includegraphics[width=0.2\textwidth]{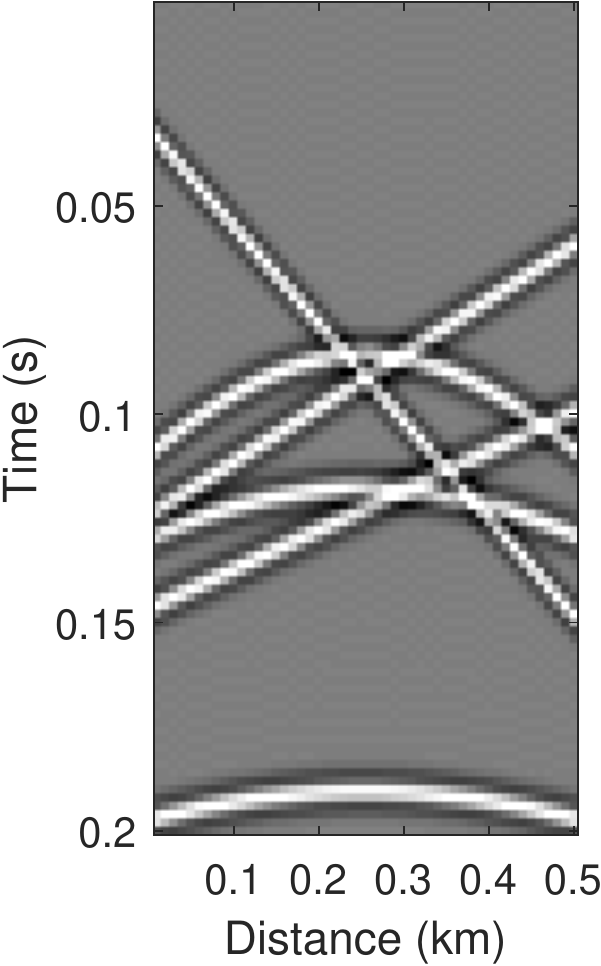}}
    \hfill
    \subfigure[]{\label{fig:07c-4}
    \includegraphics[width=0.2\textwidth]{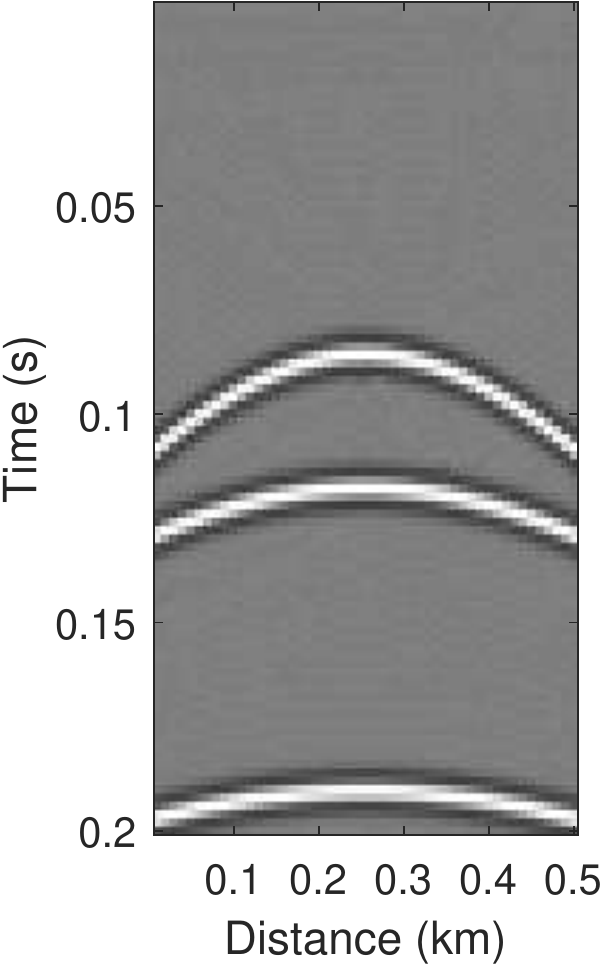}}

    \caption{{A single CNN is trained to attenuate both random and linear noise. (a) and (c) are the test datasets with random noise and linear noise respectively. (b) and (d) are denoised results of (a) and (c) with the trained CNN respectively.}}
    \label{fig:linear_random}
\end{figure*}

Can one model attenuate both random noise and linear noise with one dataset? A combined dataset with either random noise or linear noise is generated to test the performance of one single CNN on two different types of noise. Figure \ref{fig:07c-1} and \ref{fig:07c-3} show two datasets with random noise and linear noise from the test set separately. Figure \ref{fig:07c-2} and \ref{fig:07c-4} show the corresponding denoised results, which show that one single CNN is able to attenuate both random noise and linear noise. One merit of the deep learning method is that the designed network can handle more complex tasks if fed with enough training samples. 

\subsubsection{Multiple attenuation}

\begin{figure*}
    \centering
    \subfigure[]{\label{fig:08-1}
    \includegraphics[width=0.45\textwidth]{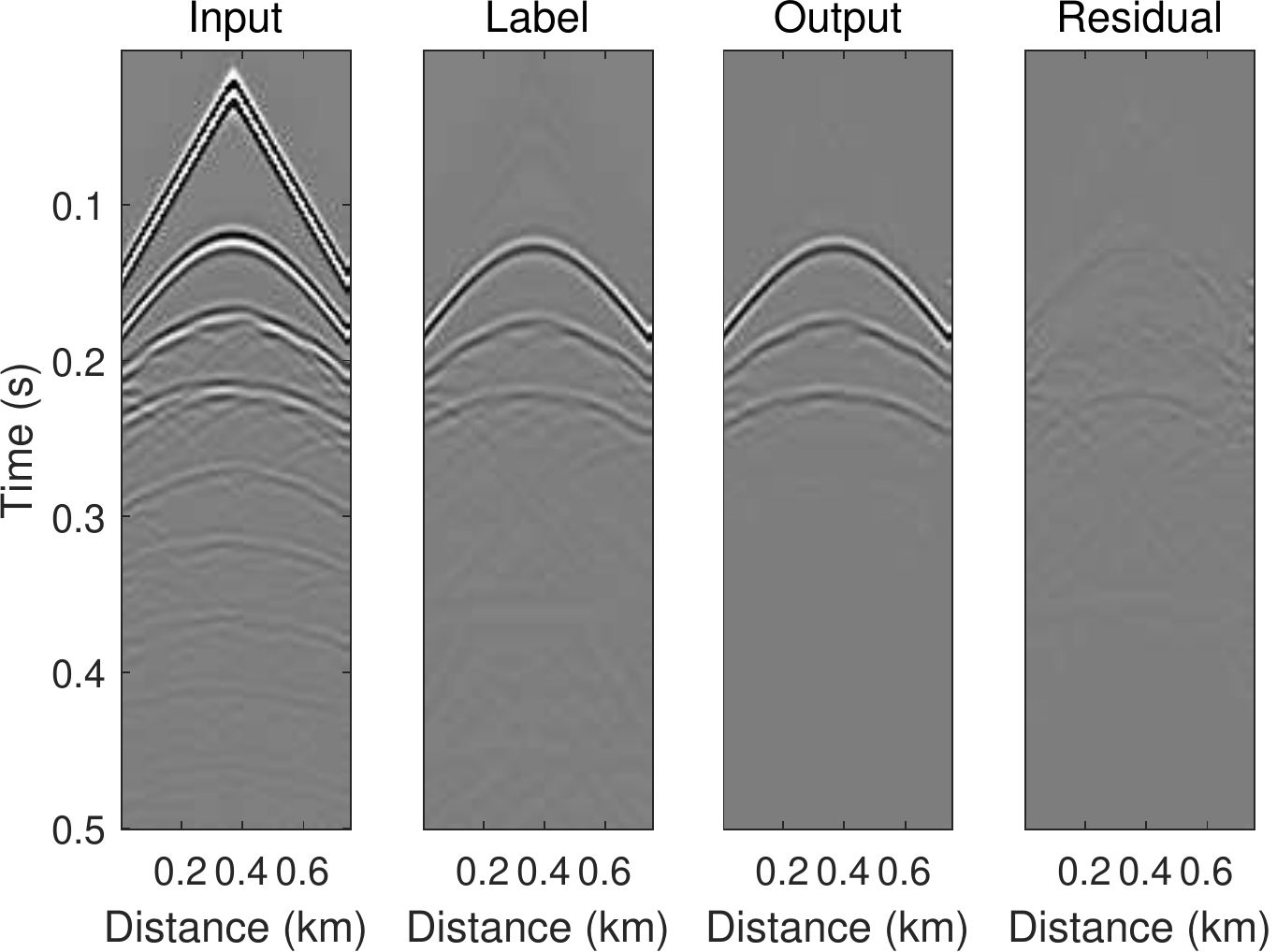}}
    \subfigure[]{\label{fig:08-2}
    \includegraphics[width=0.45\textwidth]{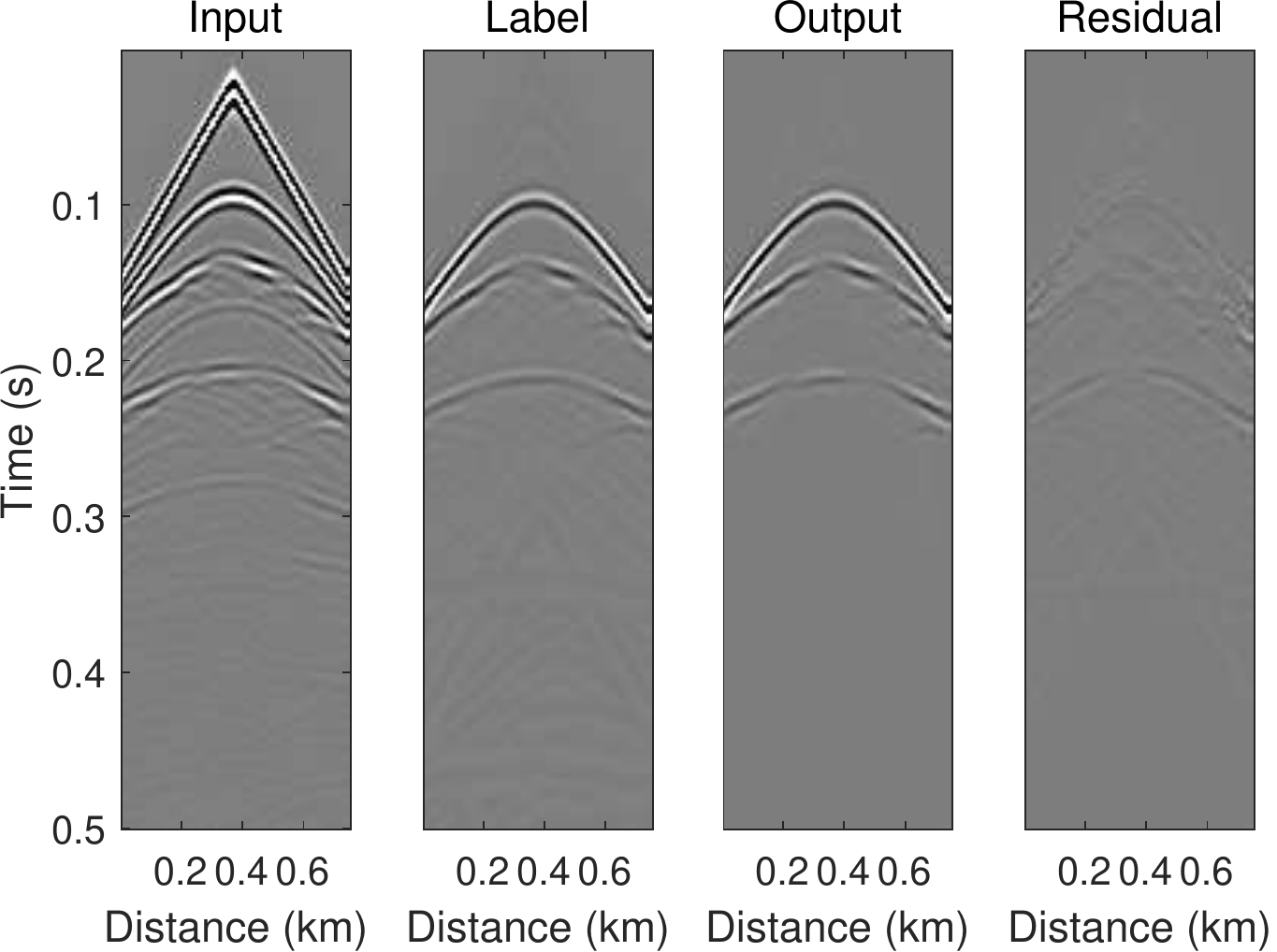}}
    \subfigure[]{\label{fig:08-3}
    \includegraphics[width=0.45\textwidth]{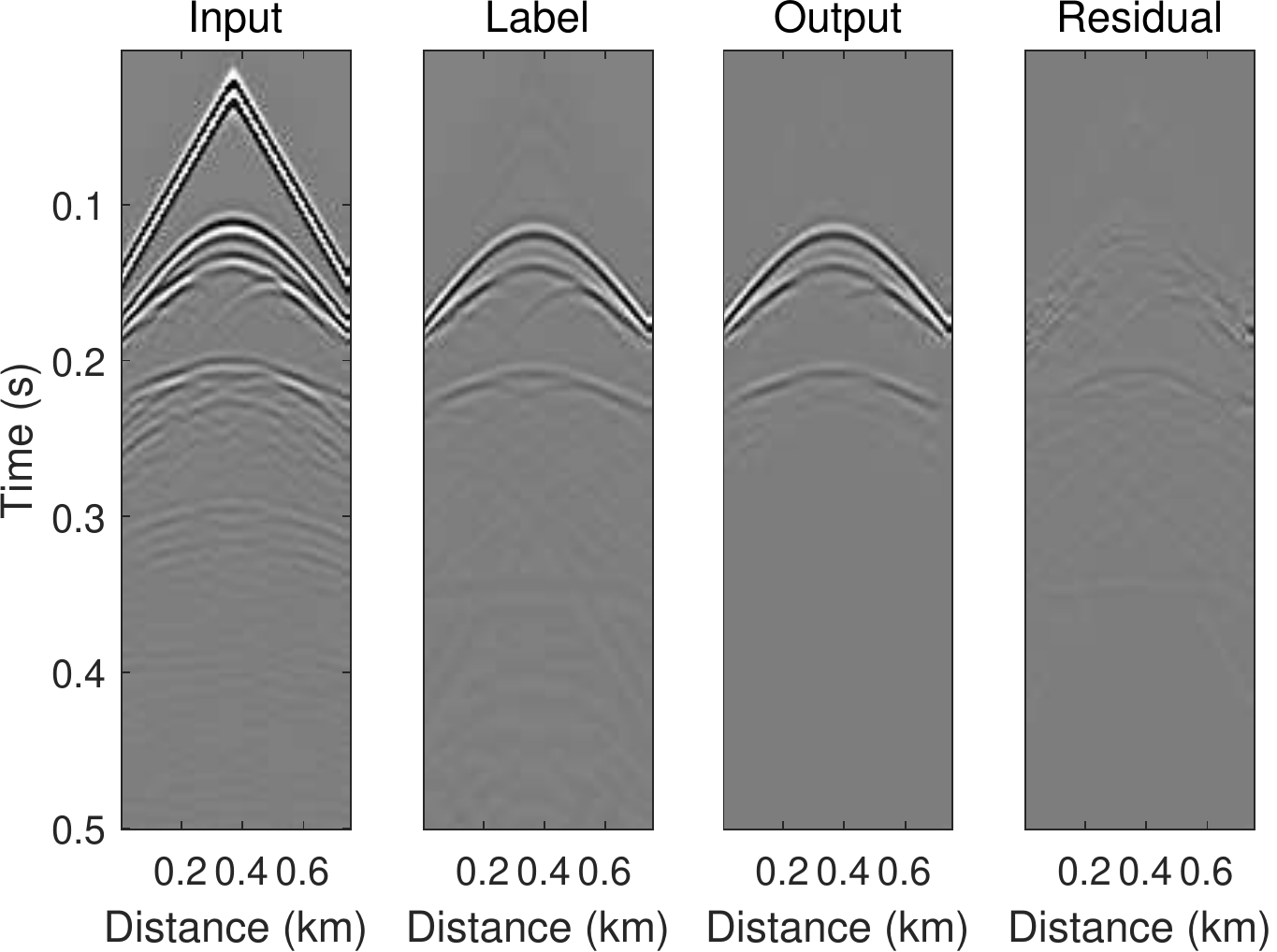}}
    \caption{ (a)--(c) are three multiple attenuation results from the testing set. In each subfigure, from left to right are the input, synthetics, output, and residual.}
    \label{fig:multiple_denoise}
\end{figure*}

The CNN architecture is the same as in the previous tests. The batch size 16. The number of epochs is 200. We tested the trained network with seismograms from the test set. The results are shown in Figure \ref{fig:multiple_denoise} with the direct waves removed. The events on the top are caused by the ghost source rather than the direct waves. In each subfigure, the left to right represents the following: input, synthetics, output, and residual. The results show the clean removal of the multiples without affecting the primaries. The residuals between the predicted seismograms and synthetics are acceptable.  The training time is approximately six hours, while the denoising time on the input of size $250 \times 75$ is 0.34 s on the CPU and 0.007 s on the GPU.

\begin{figure*}
    \centering
    \subfigure[]{\label{fig:data1-cost}
    \includegraphics[width=0.3\textwidth]{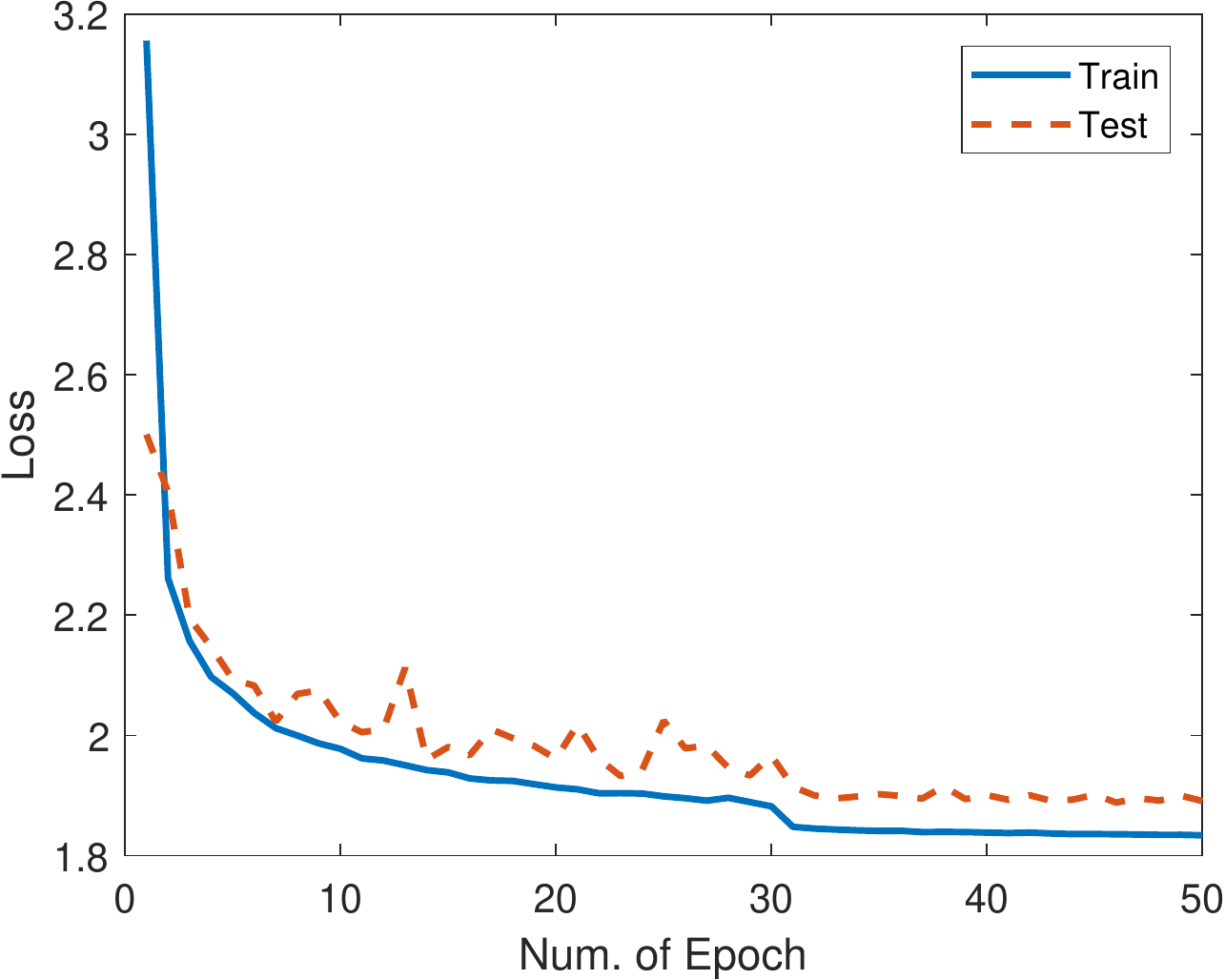}}
    \subfigure[]{\label{fig:07-cost}
    \includegraphics[width=0.3\textwidth]{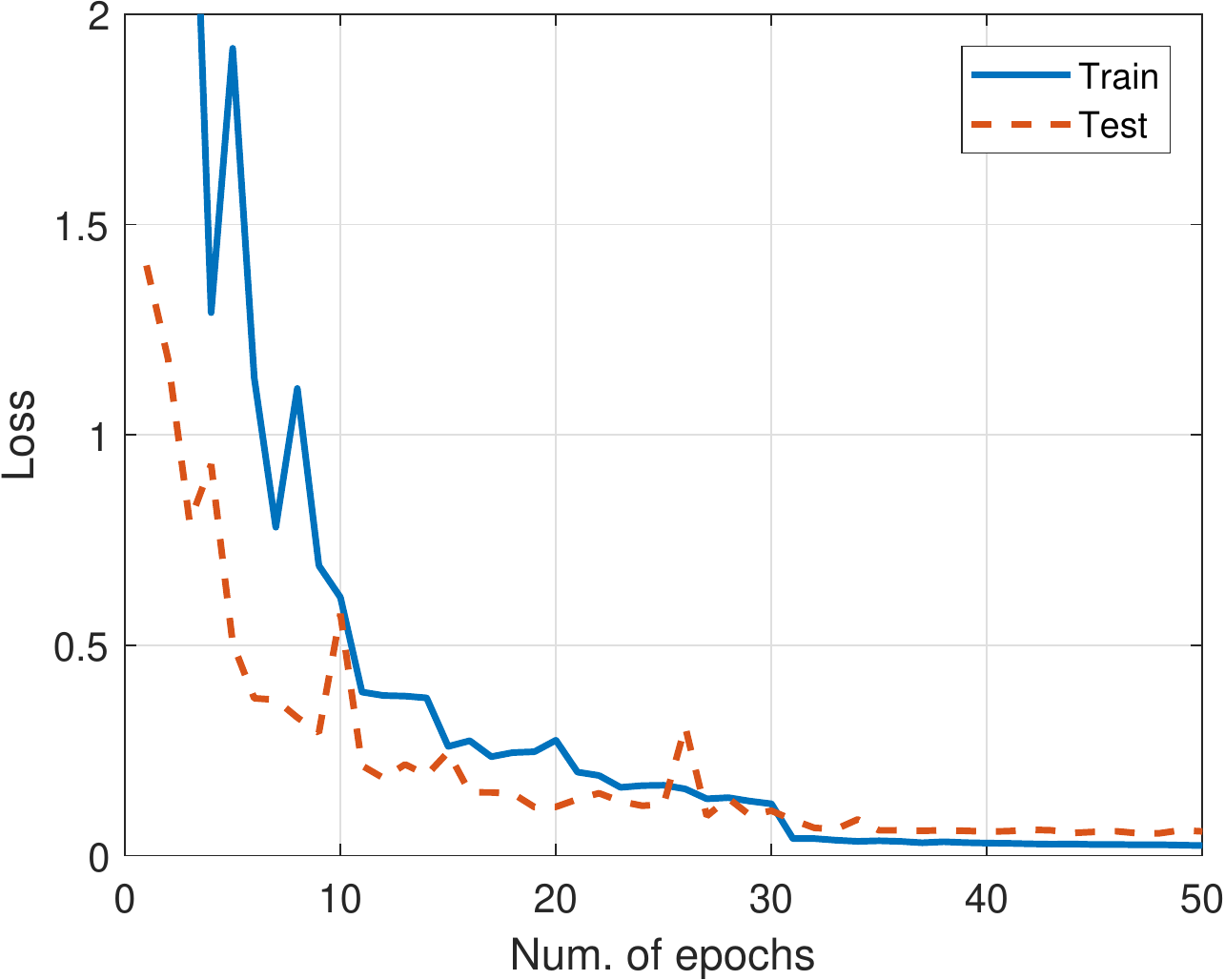}}
    \subfigure[]{\label{fig:08-cost}
    \includegraphics[width=0.3\textwidth]{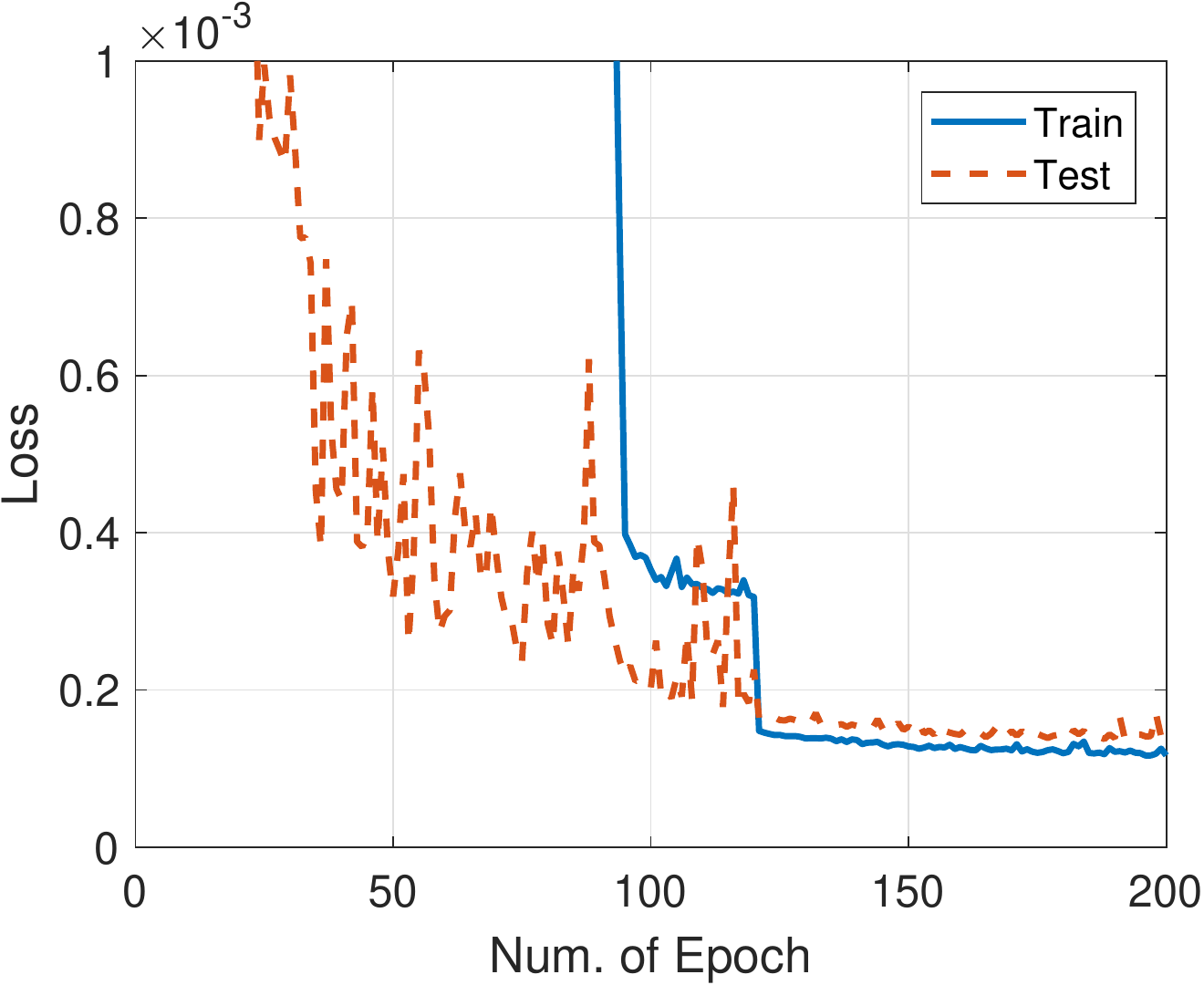}}
    \caption{{(a) --(c) are loss functions of the CNN versus the number of epochs in attenuation  of random noise, linear noise and multiple respectively.}}
    \label{fig:training_losses}
\end{figure*}

{We use the same network architecture for the previous tests, which indicates a potential generalization of the CNN for different kinds of noise attenuation. The test of different CNN architectures takes a long time, and we have already done such tests in random noise attenuation. In attenuation of linear noise and multiples, the borrowed network works well as expected. The reasons are: 1) The 17 layers CNN architecture contains enough parameters for generalization. 2) The training and testing samples are with relatively simple structures with respect to their sizes. 3) There are enough training samples in training set to avoid overfitting. Figure \ref{fig:training_losses} summarizes the average training losses and average validation losses per epoch versus the number of epochs while training CNN in attenuation of random noise, linear noise and multiple.}

\subsubsection{Real datasets random noise attenuation}

\begin{figure*}
    \centering
    \subfigure[]{\label{fig:24-1}
    \includegraphics[width=0.25\textwidth]{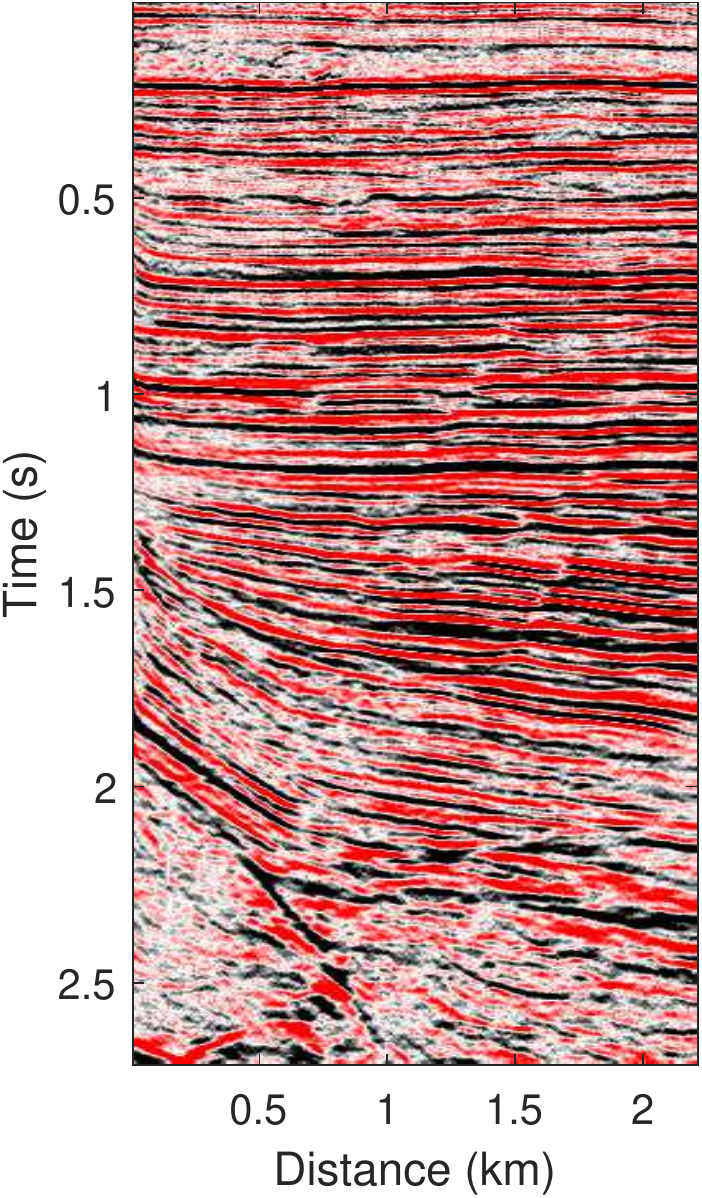}}
    \subfigure[]{\label{fig:24-2}
    \includegraphics[width=0.25\textwidth]{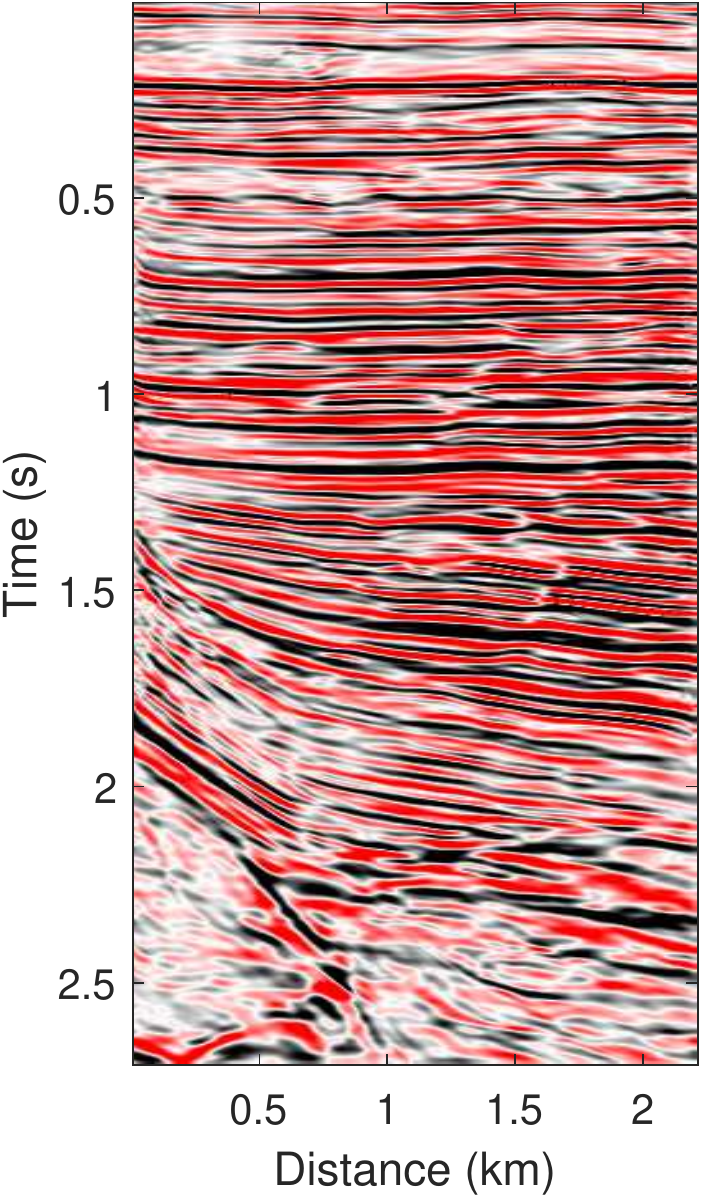}}
    \subfigure[]{\label{fig:24-test}
    \includegraphics[width=0.25\textwidth]{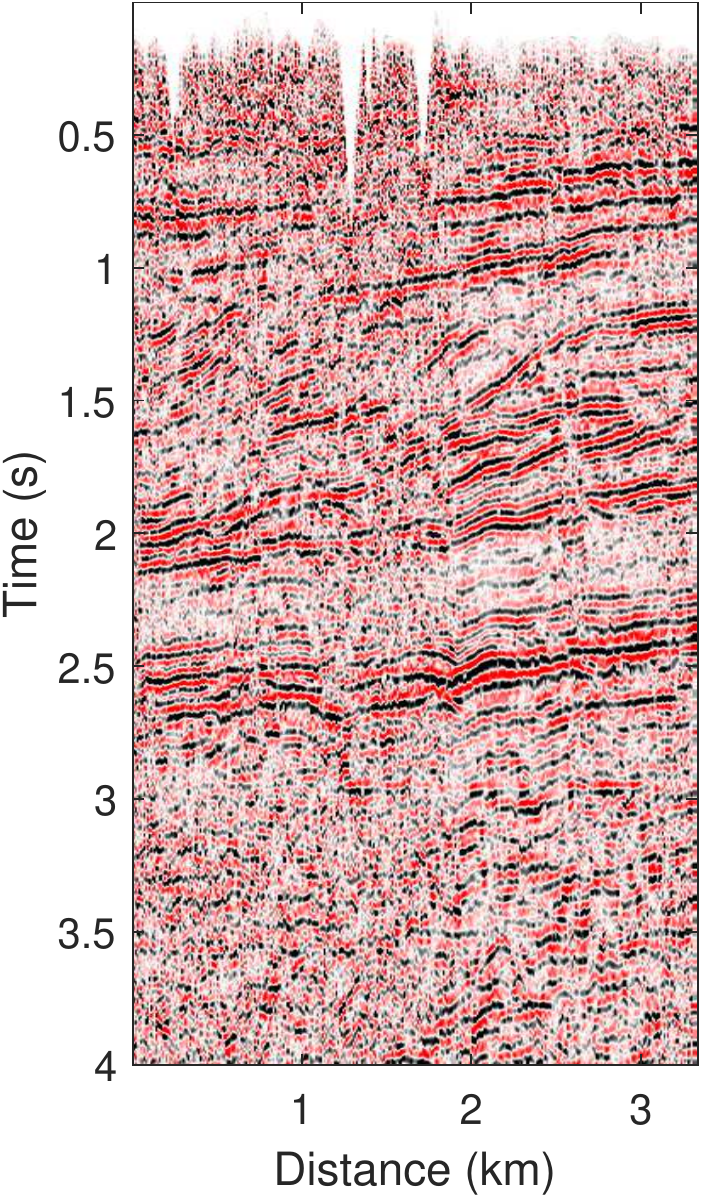}}
    \caption{ (a)(b) A pair of  input (noisy data) and label (with curvelet denoising) in a training set for real-data denoising. (c) A dataset for testing.}
    \label{fig:field_datasets}
\end{figure*}

\begin{figure*}
    \centering
    \includegraphics[width=1.0\textwidth]{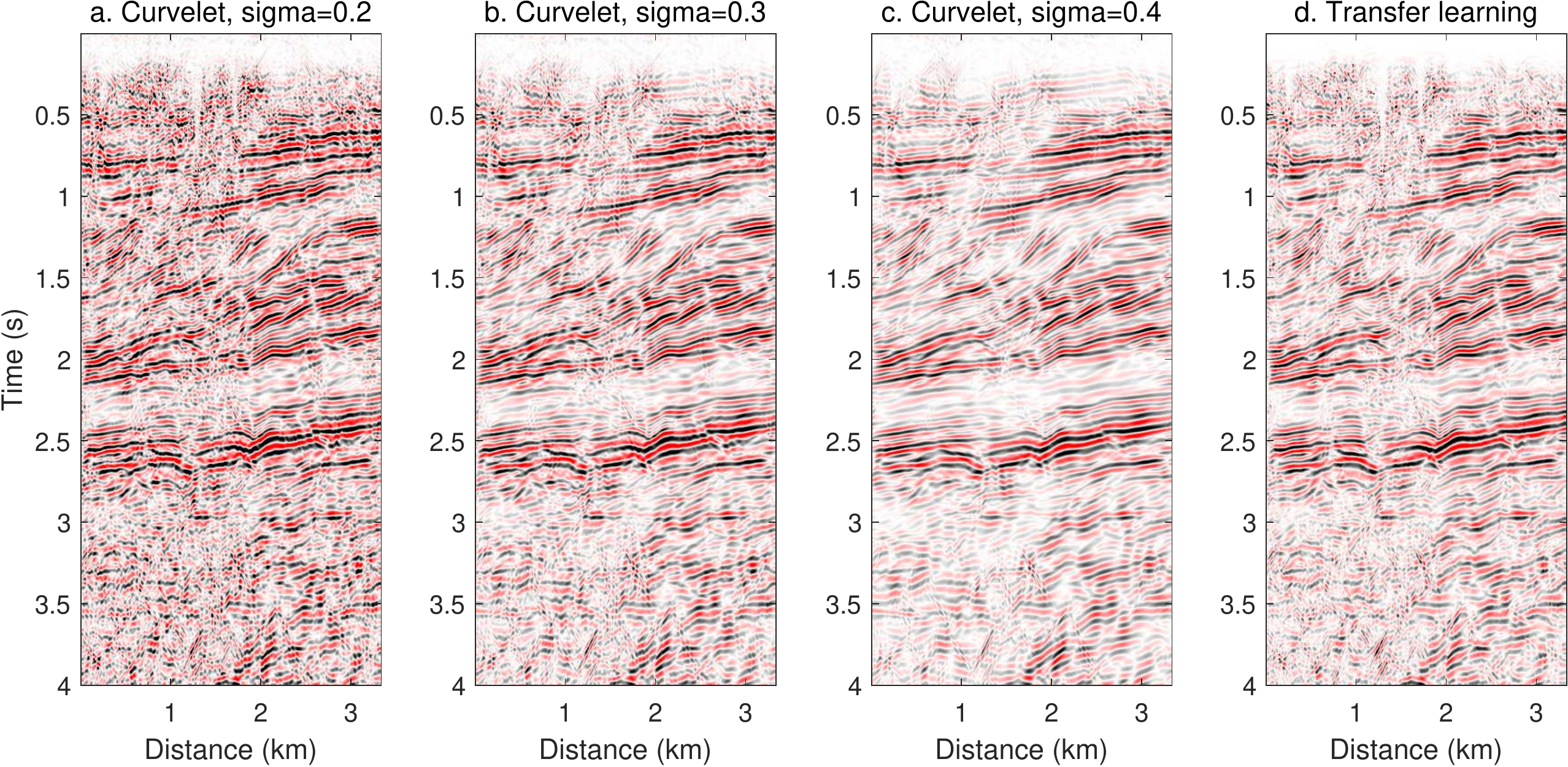}
    \caption{{Denoised sections. (a)--(c) Curvelet denoising with different thresholding parameter sigma. (d) CNN denoising with transfer learning.}}
    \label{fig:26-denoise}
\end{figure*}

\begin{figure*}
    \centering
    \includegraphics[width=1.0\textwidth]{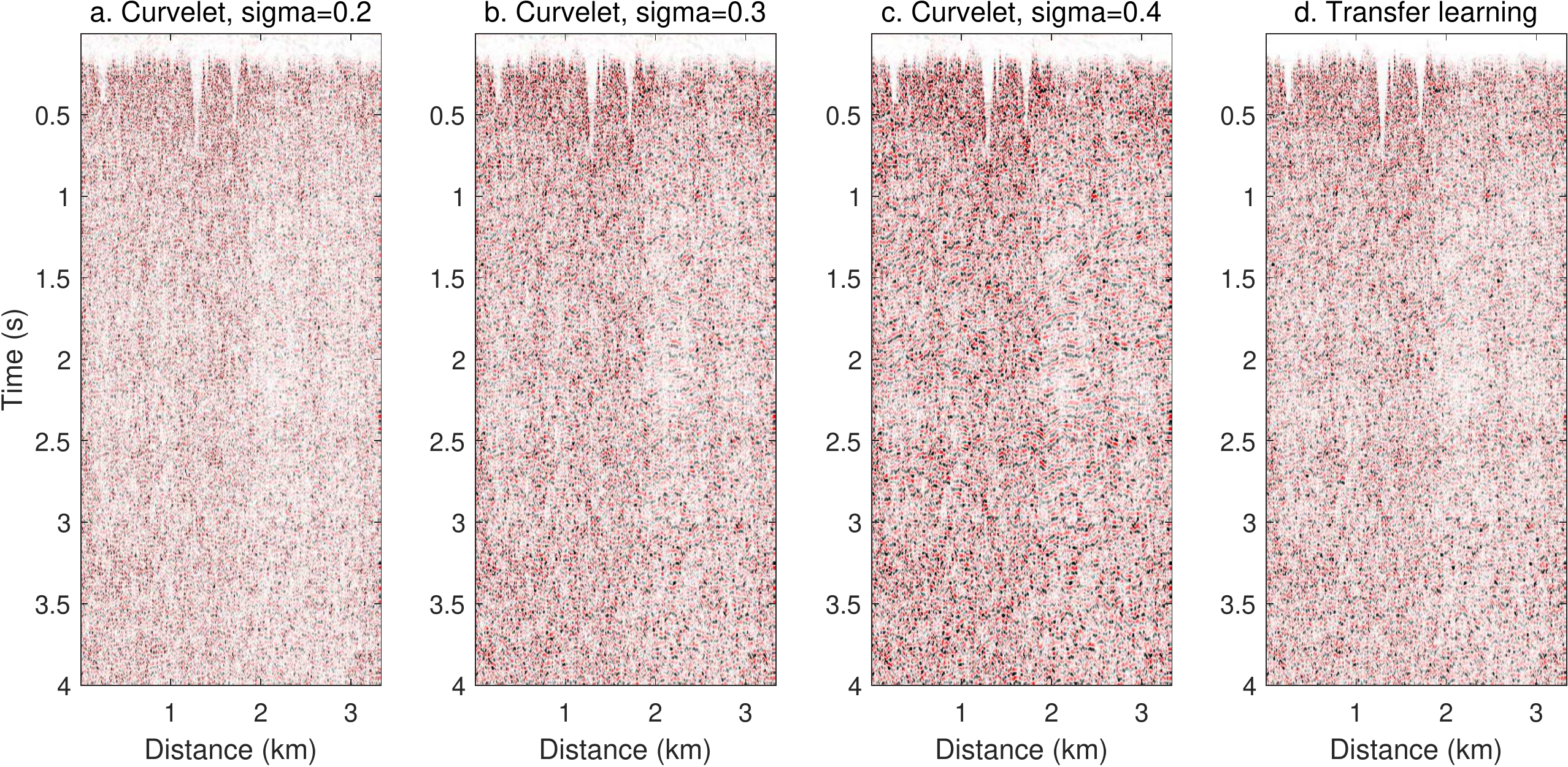}
    \caption{ Difference between Figure \ref{fig:field_datasets}c and Figure \ref{fig:26-denoise}. (a)-(c) Curvelet denoising with different thresholding parameter sigma. (d) CNN denoising with transfer learning.}
    \label{fig:27-noise}
\end{figure*}

The CNN trained from synthetic datasets in random noise attenuation cannot be applied on real datasets directly, because the real noises are not Gaussian distributed. However, training a new network with randomly initialized weights may be inefficient and the training set may be insufficient to train a valid network.  We assume that the field data exhibit local event structures similar to the synthetic data.  Therefore, transfer learning is used in real dataset training. The trained network with synthetic datasets is used to initialize the network for real datasets. The training set contains noisy datasets as inputs and denoised results with the curvelet method \citep{Hennenfent2006Seismic} as outputs. {For the curvelet method, we test different thresholding parameters and select the one with the best visual result by a human.}  One pair of training input and label is shown in Figure \ref{fig:field_datasets}a and b, respectively. Figure \ref{fig:field_datasets}c shows a dataset for testing. The training set contains four large sections, which are divided into 17152 subsections of the size of $40\times 40$. The network and training parameters are the same as in the synthetic situation. Figure \ref{fig:26-denoise} shows the denoised results with the curvelet and CNN methods, separately. Figure \ref{fig:27-noise} shows the corresponding noise sections. 

\begin{figure*}
    \centering
    \includegraphics[width=0.5\textwidth]{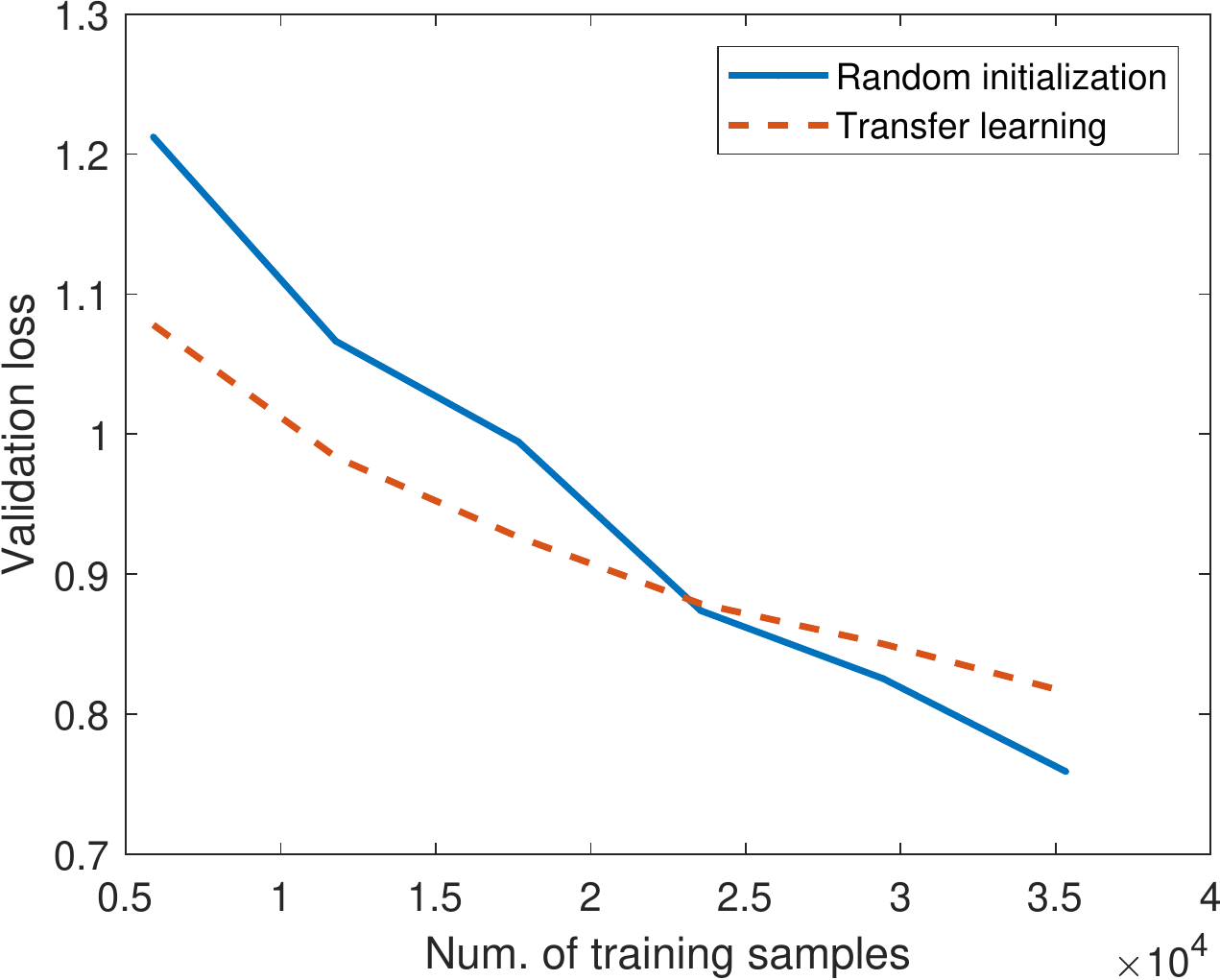}
    \caption{Validation loss of CNN versus the number of training samples, with randomly initialized weights (solid line) and transferred weights (dashed line) from the trained network in Figure \ref{fig:data1}f. }
    \label{fig:28-val-_loss_transfer}
\end{figure*}

In curvelet denoising, we tested three choices of the thresholding parameter sigma, which controls the harshness of denoising. No parameter required tuning in the CNN method, and the denoising result is similar to that in the curvelet method, with sigma = 0.3. When sigma = 0.2, the events appear under-smoothed. When sigma = 0.4, the events appear over-smoothed. The CNN method achieves noise attenuation adaptively without human intervention. The training time is approximately seven hours, the denoising time on the input of size $1501\times 333$ is 11.95 s on the CPU and 0.02 s on the GPU. Figure \ref{fig:28-val-_loss_transfer} shows the benefit of transfer learning on CNN training when the number of training samples is small. This test is performed on the same dataset generated previously. After the number of training samples reaches a certain point, a random initialization can provide better training.

\subsubsection{Real datasets scattered ground roll attenuation}

\begin{figure*}
    \centering
    \includegraphics[width=1.0\textwidth]{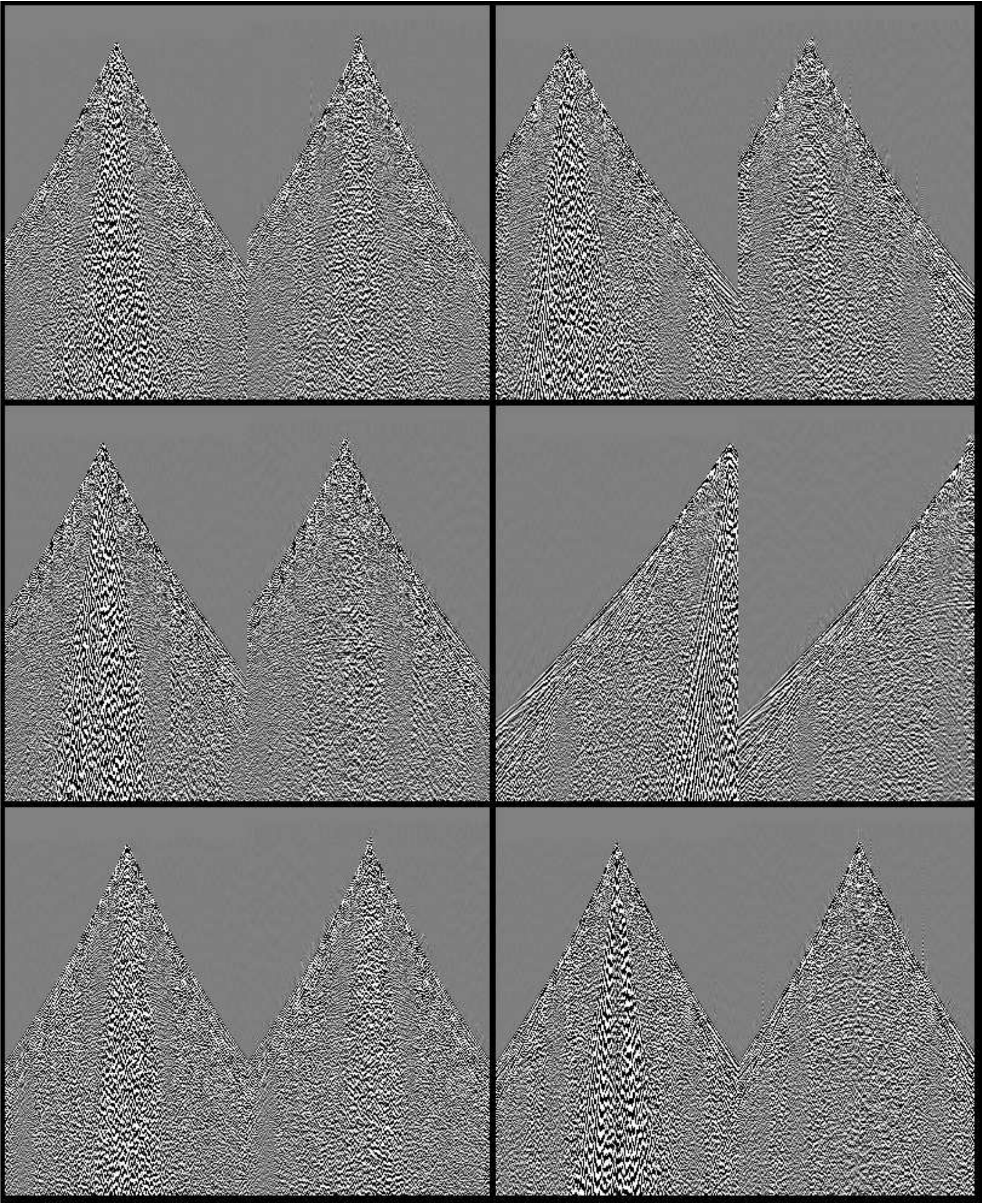}
    \caption{A subset of a training dataset for field ground roll attenuation.}
    \label{fig:sgr12}
\end{figure*}

\begin{figure*}
    \centering
    \subfigure[]{\label{fig:27-sgr-input}
    \includegraphics[width=0.3\textwidth]{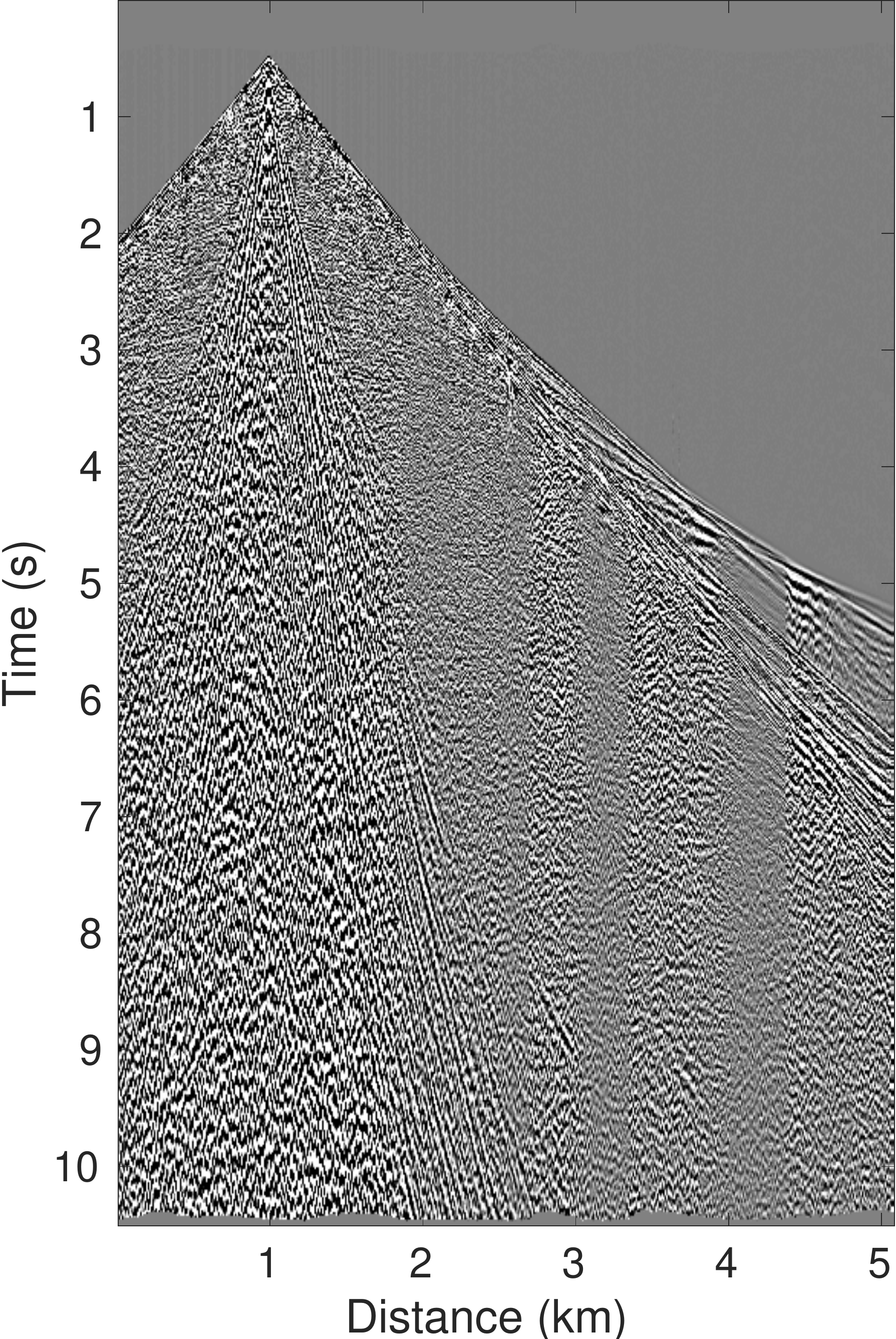}}
    \subfigure[]{\label{fig:27-sgr-output}
    \includegraphics[width=0.3\textwidth]{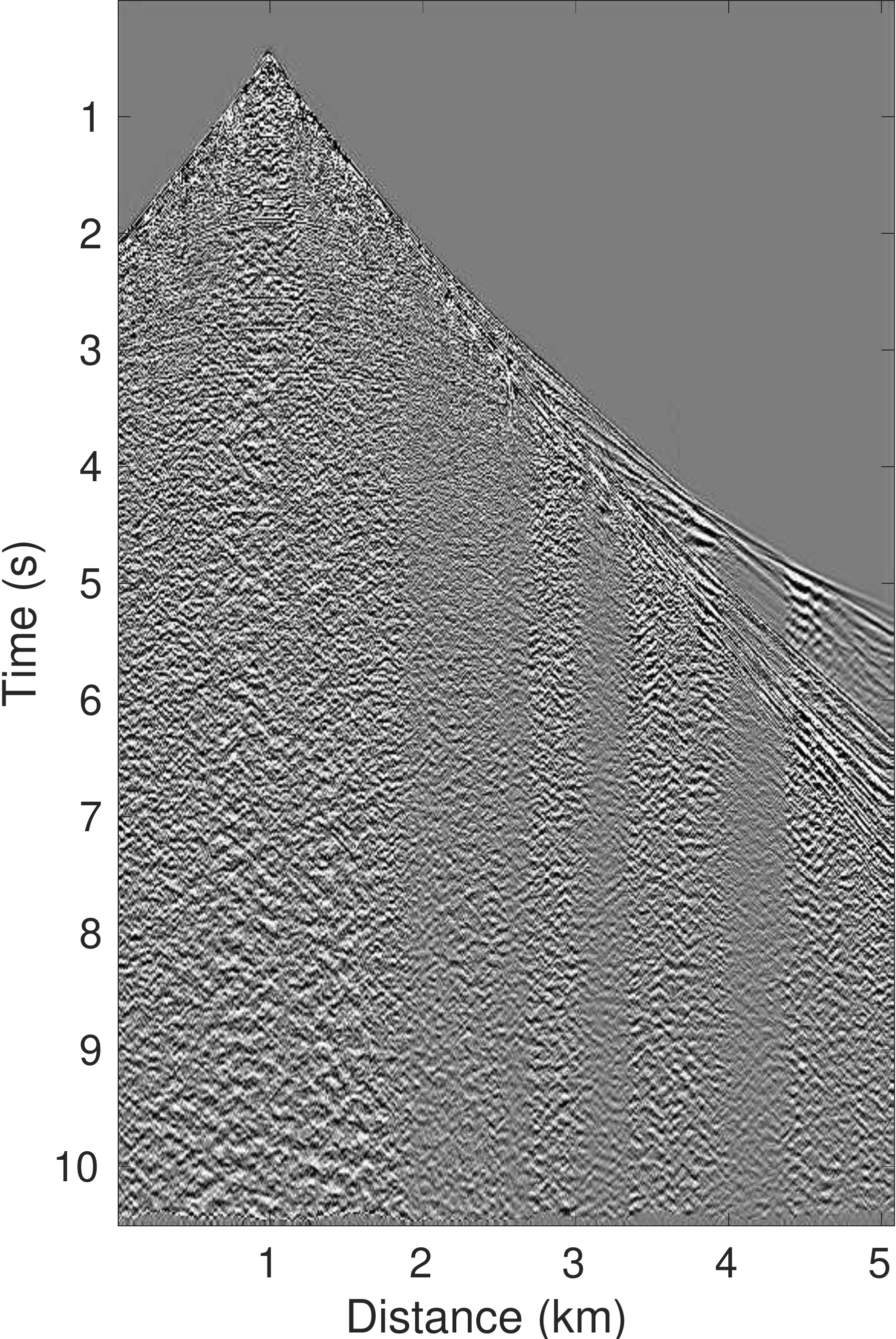}}
    \subfigure[]{\label{fig:27-sgr-label}
    \includegraphics[width=0.3\textwidth]{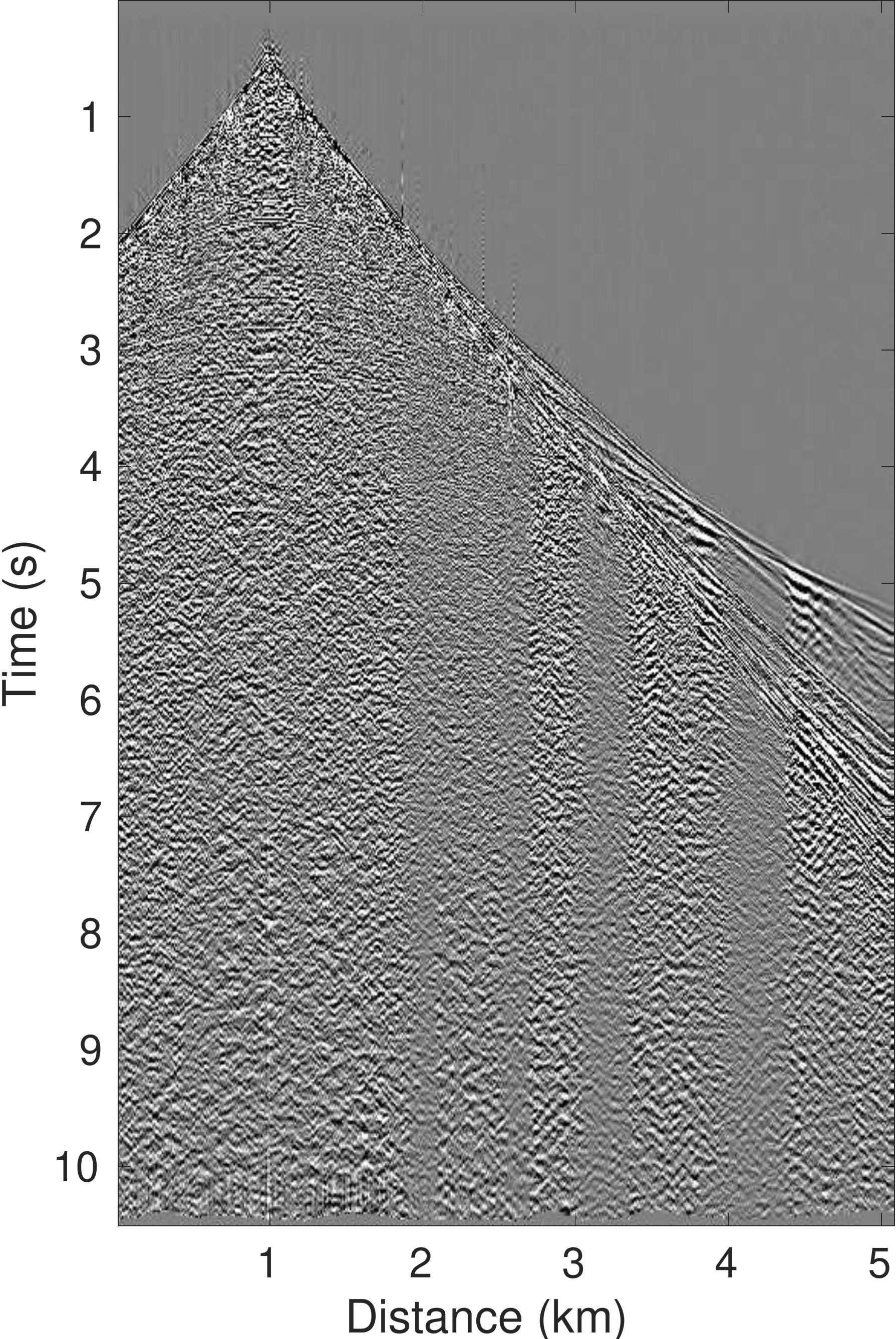}}

    \subfigure[]{\label{fig:27-sgr-error}
    \includegraphics[width=0.3\textwidth]{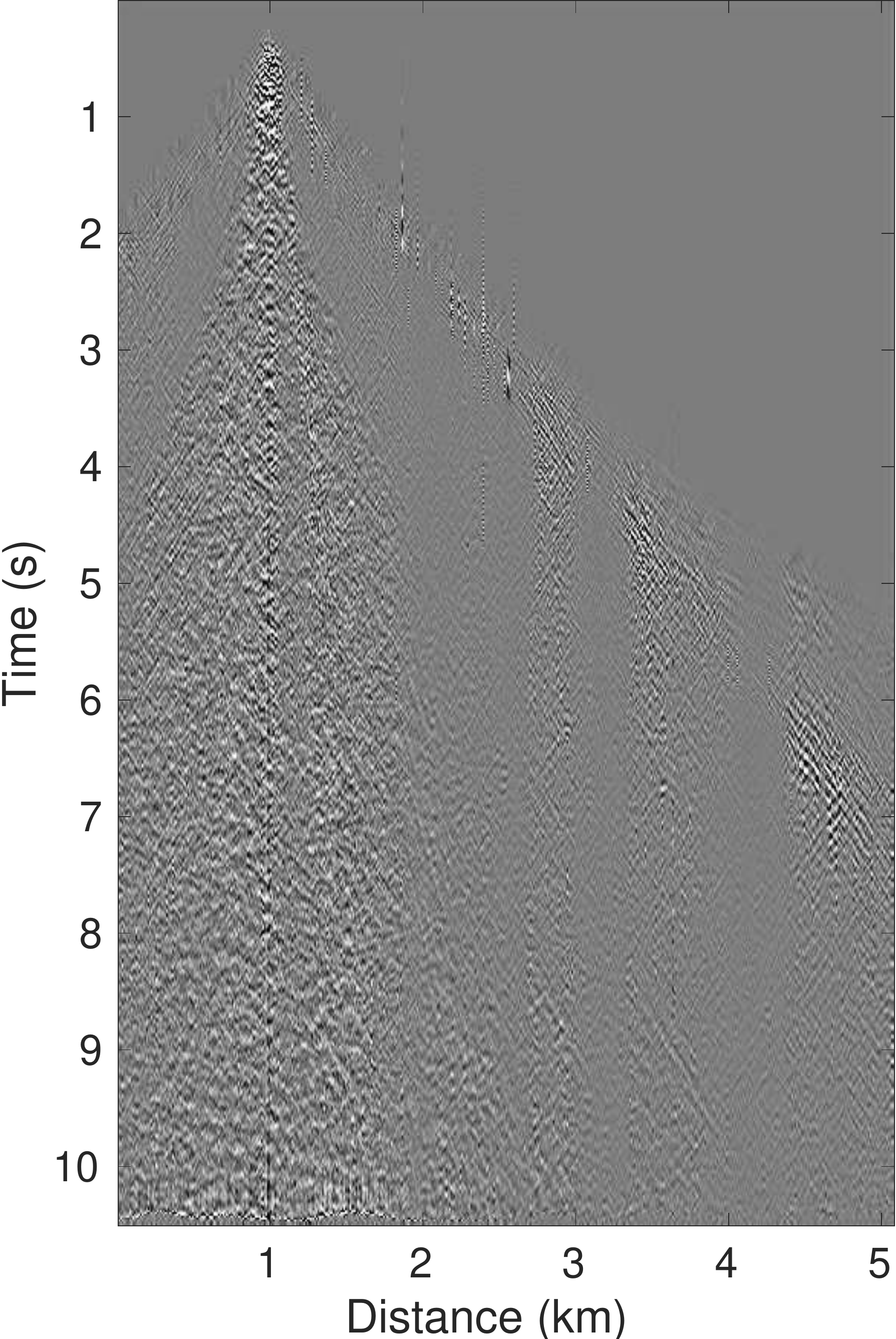}}
        \subfigure[]{\label{fig:sgr-spectrum}
    \includegraphics[width=0.45\textwidth]{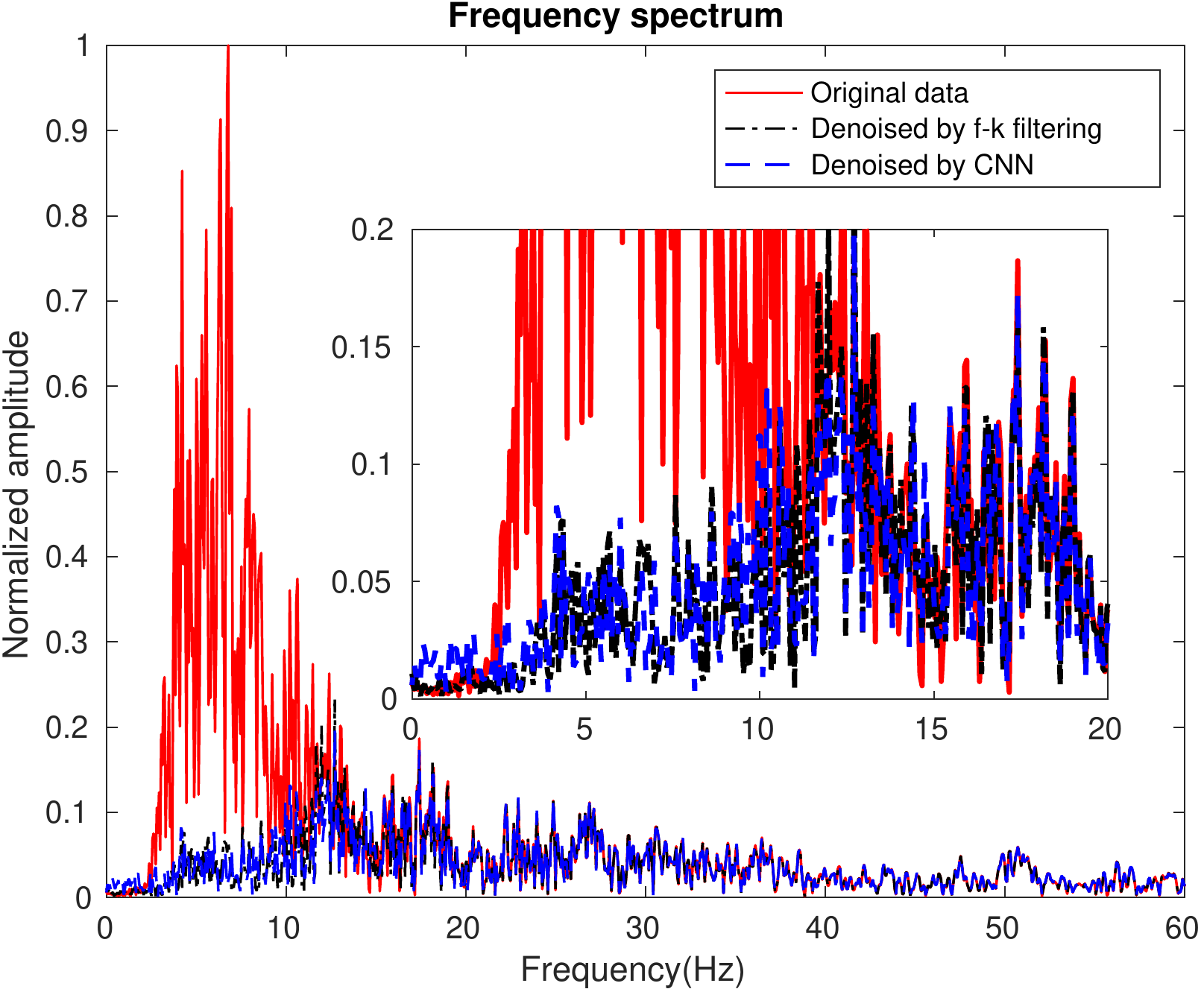}}
    \caption{Ground roll attenuation with CNN. (a) Original data. (b) Denoised by CNN. (c) Denoised data provided by the industry. {(d) The difference between (b) and (c). (e) Frequency spectrum of one trace (distance = 1.1 km) from the data  in (a)--(c)}}
    \label{fig:sgr}
\end{figure*}

Another field example is performed on scattered ground roll attenuation. Scattered ground roll is caused by the scattering of ground roll when the near surface is laterally heterogeneous. Scattered ground roll is difficult to remove as it occupies the same $f-k$ domain as reflected signals. Figure \ref{fig:sgr12} shows 6 from 160 pairs of training datasets. The training sample dataset (so-called clean dataset) is provided by industry. The datasets are split into $40\times 40$ patches to fit in the memory of the GPU. The network architecture is the same as in the previous tests. Figure \ref{fig:sgr}a-d show a test dataset, the denoised result of the CNN method, the result provided by the industry {and the difference map between Figure \ref{fig:sgr}b and Figure \ref{fig:sgr}c}. The CNN method removes the energy of the ground roll, implying that the ground roll is separable on the patch scale. The CNN method achieves similar results as the industry method since the network is trained from the training set whose clean data is obtained by the industry method. Figure \ref{fig:sgr-spectrum} shows the frequency spectrum of {one trace (distance = 1.1 km)} from the data in Figure \ref{fig:sgr}a-c. {The useful signals with frequencies higher than 15 Hz are successfully preserved.}

\section{Discussion}

Many unanswered questions arise in DL, such as the selection of hyperparameters. The introduction of DL may raise more problems than it solves. We discuss several interesting topics in DL in this section. 

\subsection{Can DL perform better than industrial standard tools?}

We discuss two situations: the clean dataset is generated by one existing method, and by more than one existing method. For the former, we use the random noise attenuation of real datasets as an example. If we use the industrial standard tools, such as  $f-x$ deconvolution, to obtain a clean dataset, can we obtain a CNN that performs better than $f-x$ deconvolution? For the clean dataset, we can tune the parameters manually, such as the length of the autoregression operator, to obtain an optimal denoising result. After the network is trained with these optimized inputs and {outputs}, the network can handle the denoising of new datasets adaptively without the tuning parameters. Meanwhile, if we use $f-x$ deconvolution for denoising the new datasets and tune the parameters manually, insufficient experience or time may lead to {poor} results. In the real datasets tests in Figure \ref{fig:26-denoise}, if the thresholding parameter is not correctly selected, the curvelet method fails to achieve satisfying denoising results. CNN can perform intelligent denoising without the tuning parameters. Such automation in DL is mostly missing in traditional methods.

For the second situation, it is natural to establish a more diverse training set with denoising results obtained from different methods suitable for different types of data. Subsequently, the trained network can benefit from different approaches and outperform any existing independent method. It is equivalent to training a network that can choose a suitable method adaptively to a specific type of dataset.

\subsection{How does the number of training samples contribute}

\begin{figure*}
    \centering
    \subfigure[]{\label{fig:28-train_loss}
    \includegraphics[width=0.45\textwidth]{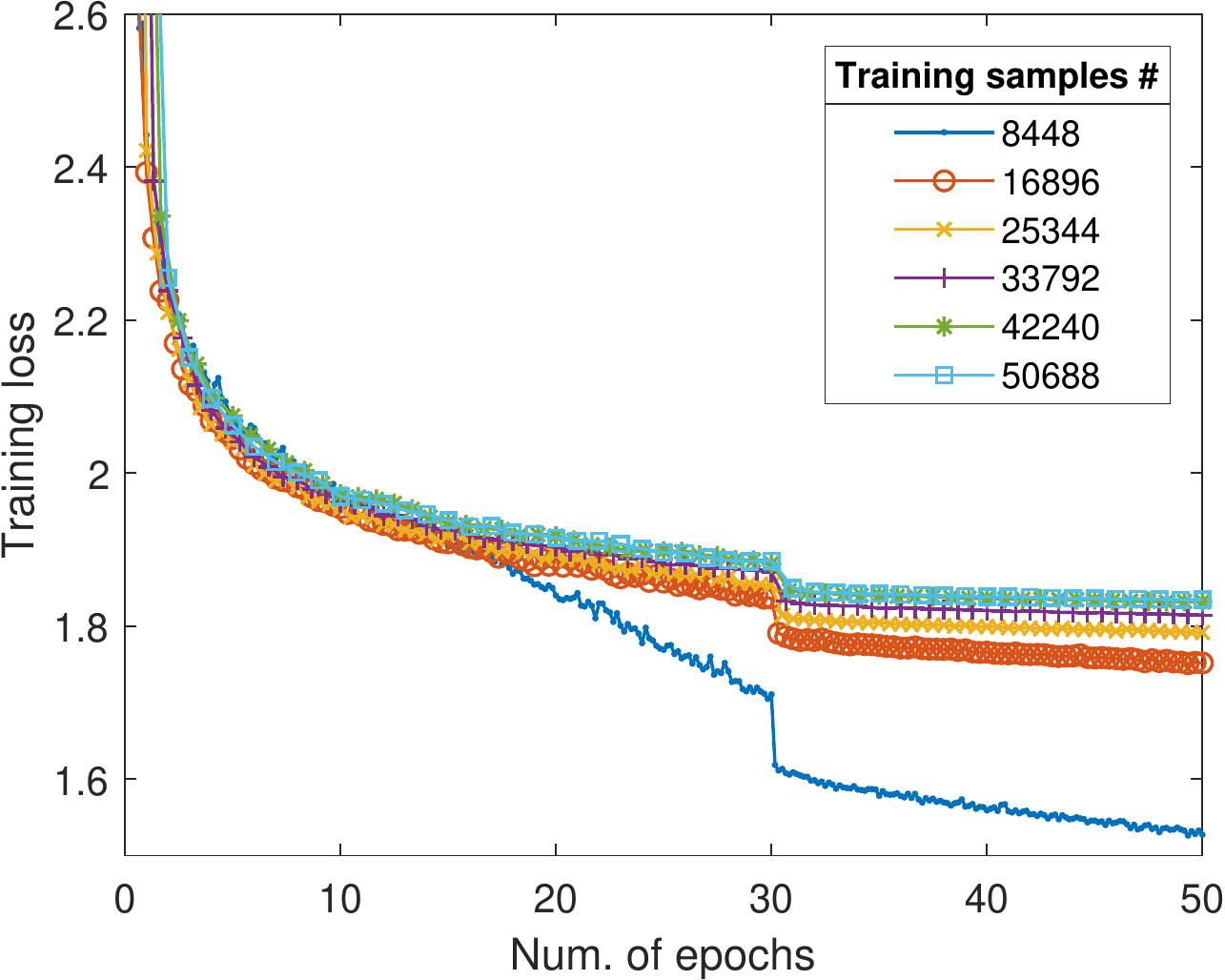}}
    \subfigure[]{\label{fig:28-test_loss}
    \includegraphics[width=0.45\textwidth]{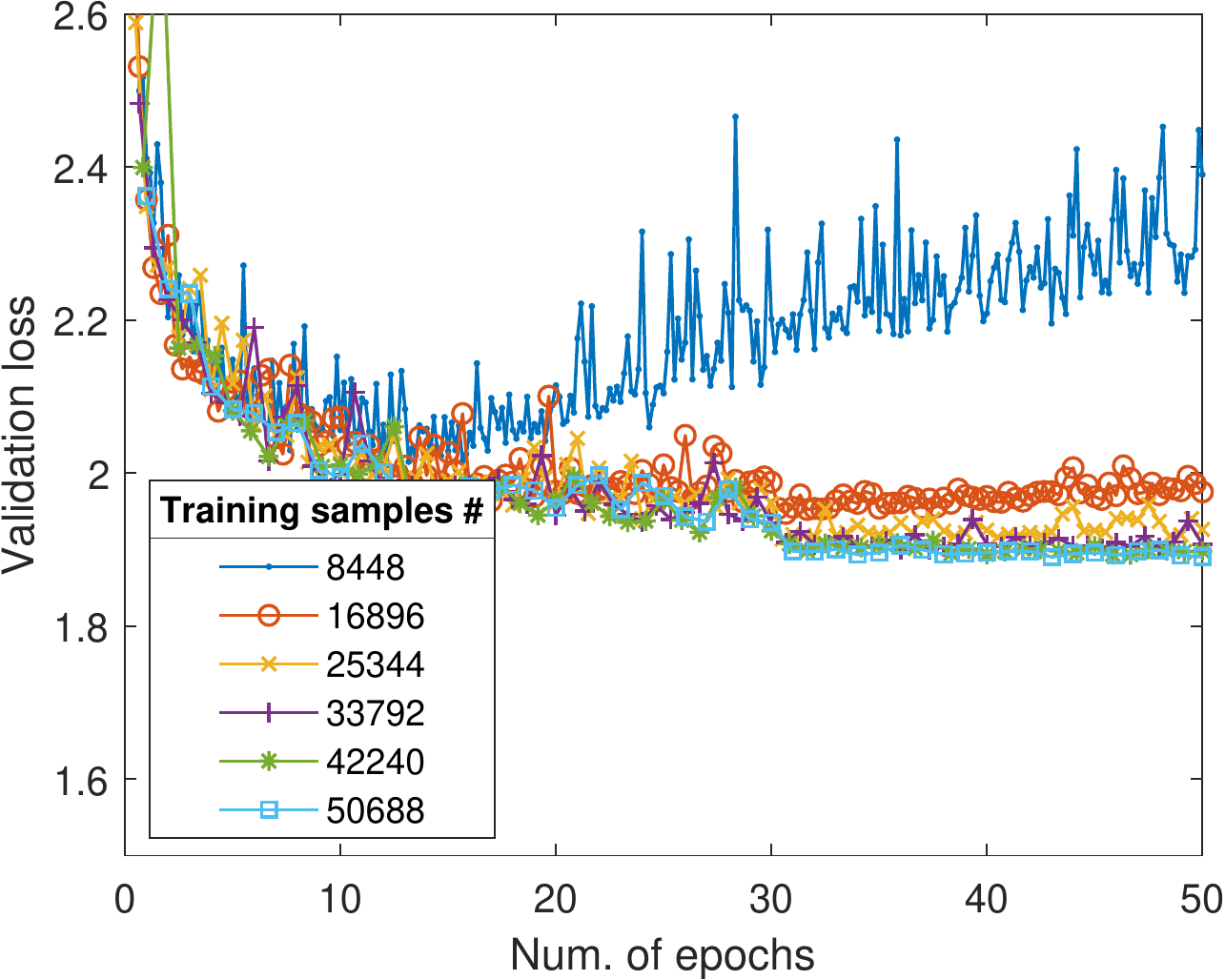}}
    \subfigure[]{\label{fig:deeploss}
    \includegraphics[width=0.45\textwidth]{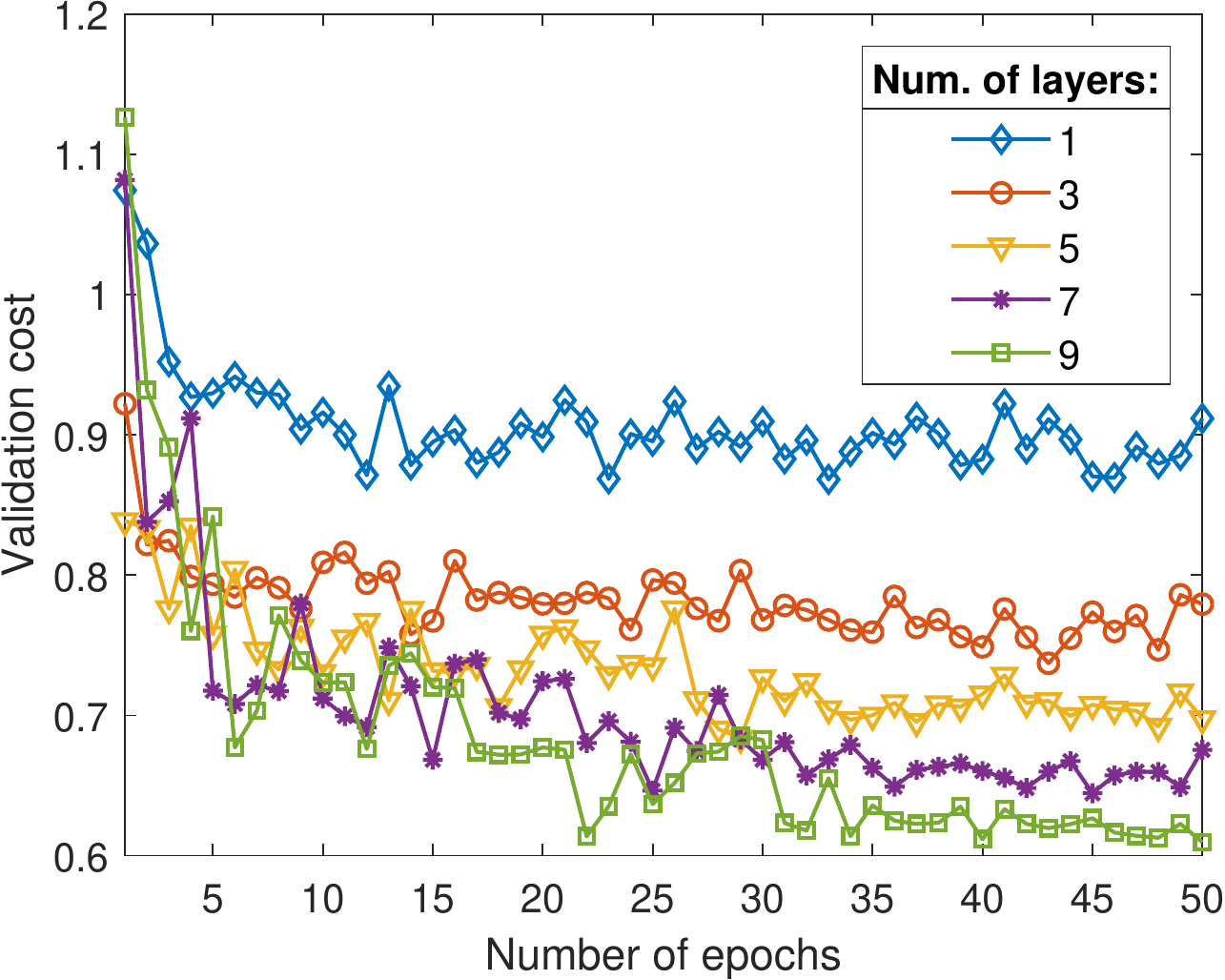}}
    \subfigure[]{\label{fig:snr_layers}
    \includegraphics[width=0.45\textwidth]{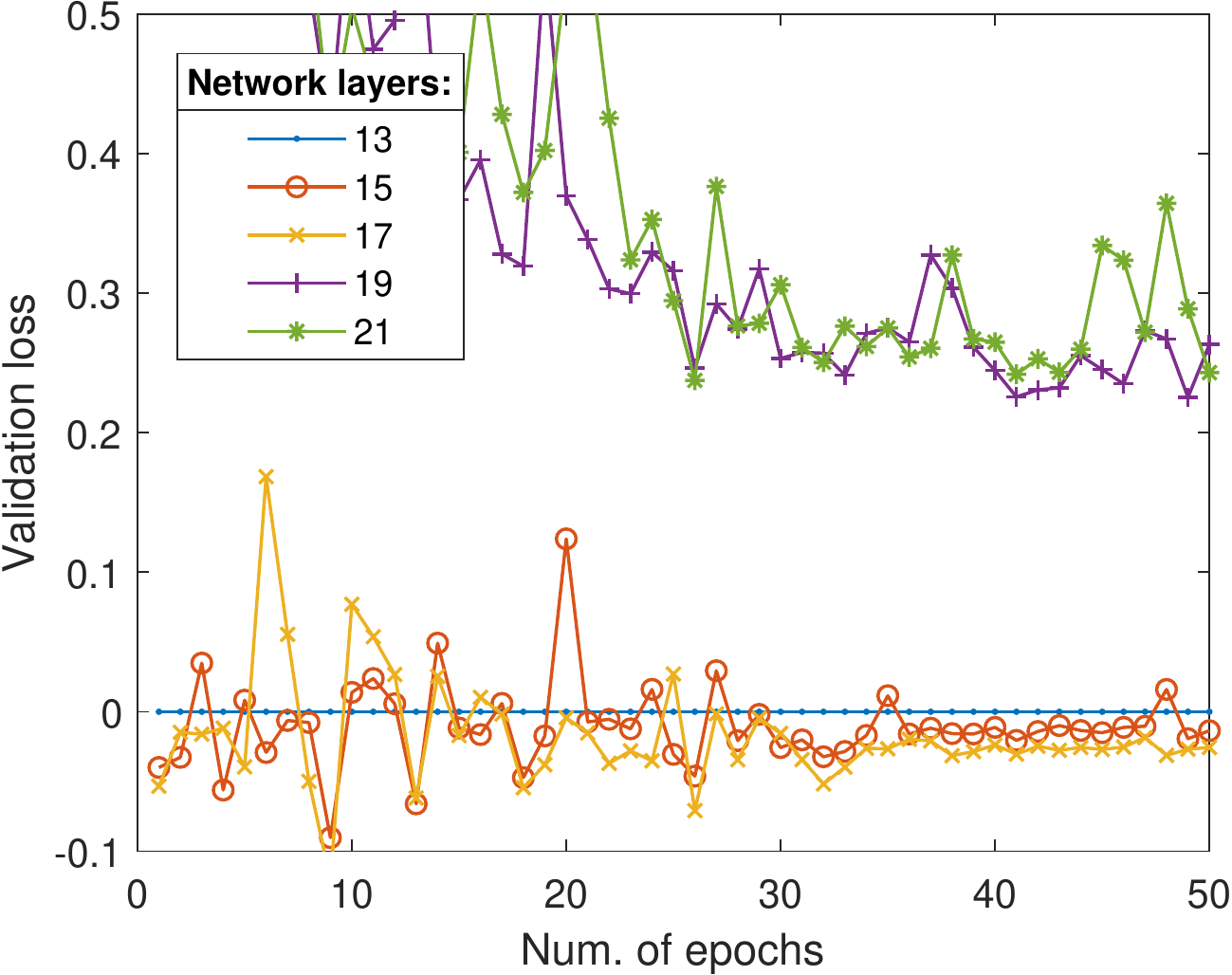}}
    \caption{{Tests of hyperparameters.} (a) Training loss  versus the number of epochs and the number of training samples. (b) Validation loss (the  loss function for the validation sets) versus the number of epochs and the number of training samples.  (c) Validation loss  versus the number of epochs for networks of different layers. (d) Comparisons of denoising results with CNN with different layers. Layer 13 is used as a baseline and $y$-axis indicates the differences in validation loss.}
    \label{fig:sample_num_test}
\end{figure*}

We are interested in how many training samples are needed to train a fully-specified neural network. In principle, the more the number of training samples, the more powerful is a network. However, a longer time is required to prepare the training set and train the network. We treat the number of training samples as a hyperparameter for training the network. This test is performed on the synthetic random noise training set. Figure \ref{fig:28-train_loss} and Figure \ref{fig:28-test_loss} shows the training loss and validation loss versus the number of epochs and the number of training samples. It is noteworthy that we used more epochs for small numbers of samples to ensure that the total numbers of samples used in each optimization are the same. The epochs are normalized to 50. We found that if the number of training samples is extremely small (8448), the training loss is small but the validation loss is large, thus causing overfitting. Early stopping and a larger number of training samples can avoid overfitting. The validation loss is almost the same for 44220 and 50688 training samples. After the number of training samples reaches a certain level, the extra training datasets contribute little to the validation loss. 

\subsection{Why “deep” matters}

Deeper layers incorporate a larger receptive field and more nonlinearities.  The receptive field represents the number of elements involved in one convolution operation. Larger receptive fields involve more information from the input layer. Figure \ref{fig:rf} shows that the receptive field size of a deeper layer is larger than that of a shallow layer in two dimensions. If the convolutional filter size is $3\times 3$, subsequently the receptive field of layer 2 with respect to layer 1 is $3\times 3$, while the receptive field of layer 3  with respect to layer 1 is $5\times 5$. {It is clear that more units  in the input are enrolled in computing the units in the deeper layers.} 

Meanwhile, if the other parameters are fixed, the CNN with deep layers incorporate more nonlinear layers, resulting in a more discriminative decision function  \citep{Simonyan2014Very}, and providing more parameters to fit the complex mappings. Figure \ref{fig:deeploss} shows how the different numbers of layers {(1, 3, 5, 7, 9)} affect the network performance on the test set in a random noise situation. The training and testing sets are the same in different setups. In each epoch, we calculate the validation loss.  The validation loss reduces as the layer increases. However, it is not always optimal to use extremely deep networks, as it may lead to low efficiency and overfitting. Hence, techniques for avoiding overfitting must be considered, such as dropout \citep{srivastava2014dropout}.

\subsection{Hyper-parameters}

{Hyperparameters are parameters that are set before the training begins. Other than the batch size and number of epochs, the CNN also contains numerous of hyperparameters in optimization, such as the learning rate, and momentum. Another type of hyperparameters pertains to the model architecture, such as the number of hidden layers, and the number of filters in each layer. Although parameter tuning is not required after training, hyperparameter tuning is required while designing network architecture and optimization. Even though guidance has been proposed for recommending the hyperparameters \citep{bengio2012practical}, it is difficult for readers who are not experts in DL and only wish to use DL rather than optimizing every hyperparameter. Fortunately, most of the hyperparameters in optimization, such as the batch size, typically impact the training time rather than the test performance \citep{bengio2012practical}. Therefore, except for the number of epochs, most hyperparameters in optimization can be borrowed from image denoising, such as \citet{Zhang2017Beyond}.}

If the number of model hyperparameters is small, such as 1 or 2, we can use the grid search method in the regular-domain or log-domain. Different configurations can be computed in parallel, and the one that achieves the minimum loss on the validation set is selected. The grid search method scales exponentially with the number of hyperparameters. When the number of model hyperparameter becomes more extensive, a practical solution is the random sampling of hyperparameters. More information on grid search and random sampling can refer to \citet{bengio2012practical}. In our work, the number of layers is determined with the grid search method. Other parameters, such as the number of filters, are determined by referring to \citet{Zhang2017Beyond}. 

In DnCNN, the authors used a network with 17 convolutional layers. Therefore, we tested  13, 15, 17, 19, and 21 convolutional layers and chose the one that achieved the least validation loss. This test was performed on the synthetic random noise training set. Figure \ref{fig:snr_layers} shows how the validation loss changes with the number of epochs and number of layers.  We used 13 layers as a baseline and show the differences between the other layers and 13 layers. The validation loss changes little for 13, 15, and 17 layers, but is much larger for 19 and 21 layers, which may be caused by overfitting when increasing the number of parameters in the network. When 17 layers were chosen, the least validation loss was shown in most cases.

\subsection{Distance between training and test set}

\begin{figure*}
    \centering
    \subfigure[]{\label{fig:04M-nf}
    \includegraphics[width=0.5\textwidth]{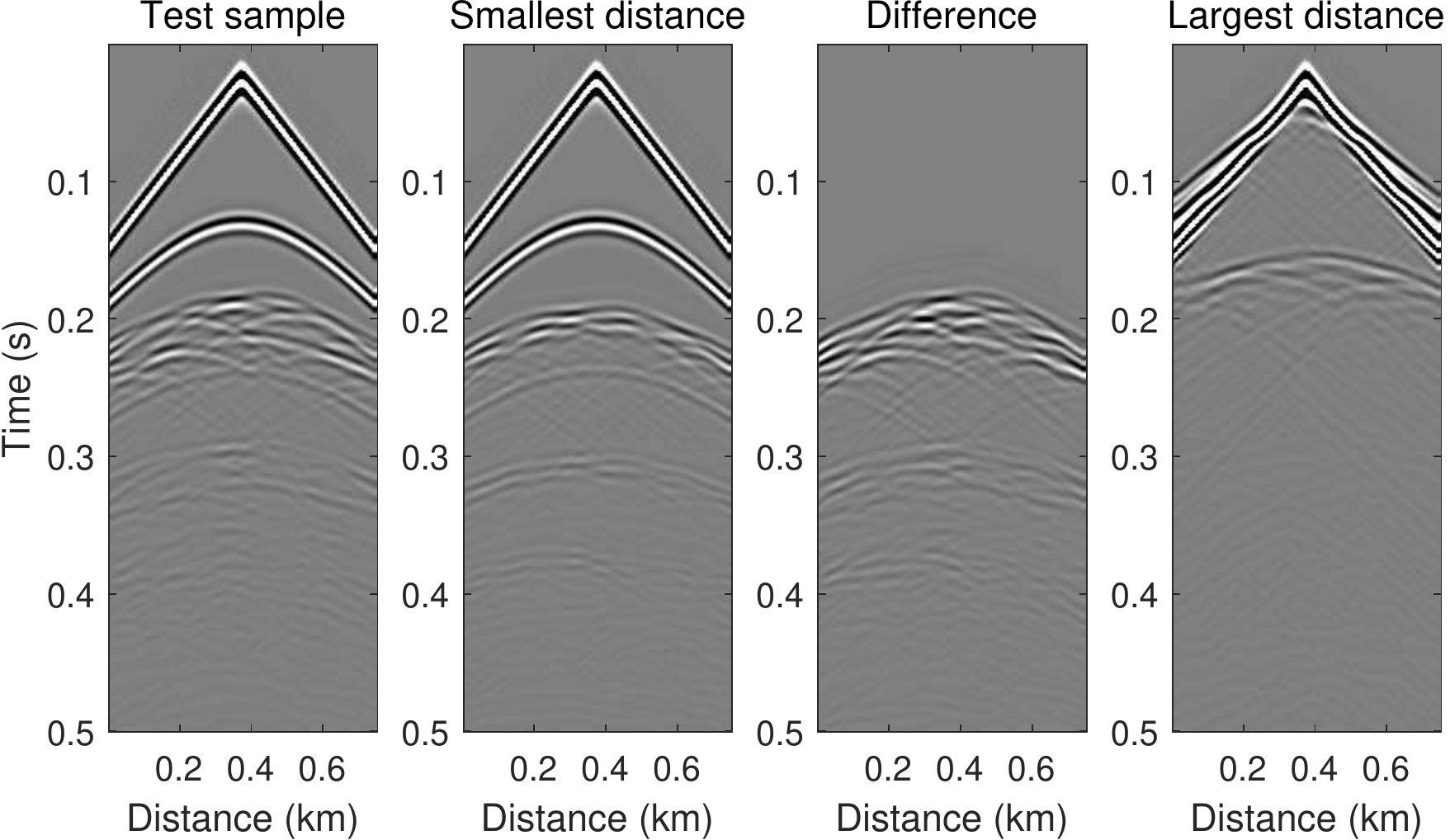}}
    \subfigure[]{\label{fig:04M-dist}
    \includegraphics[width=0.5\textwidth]{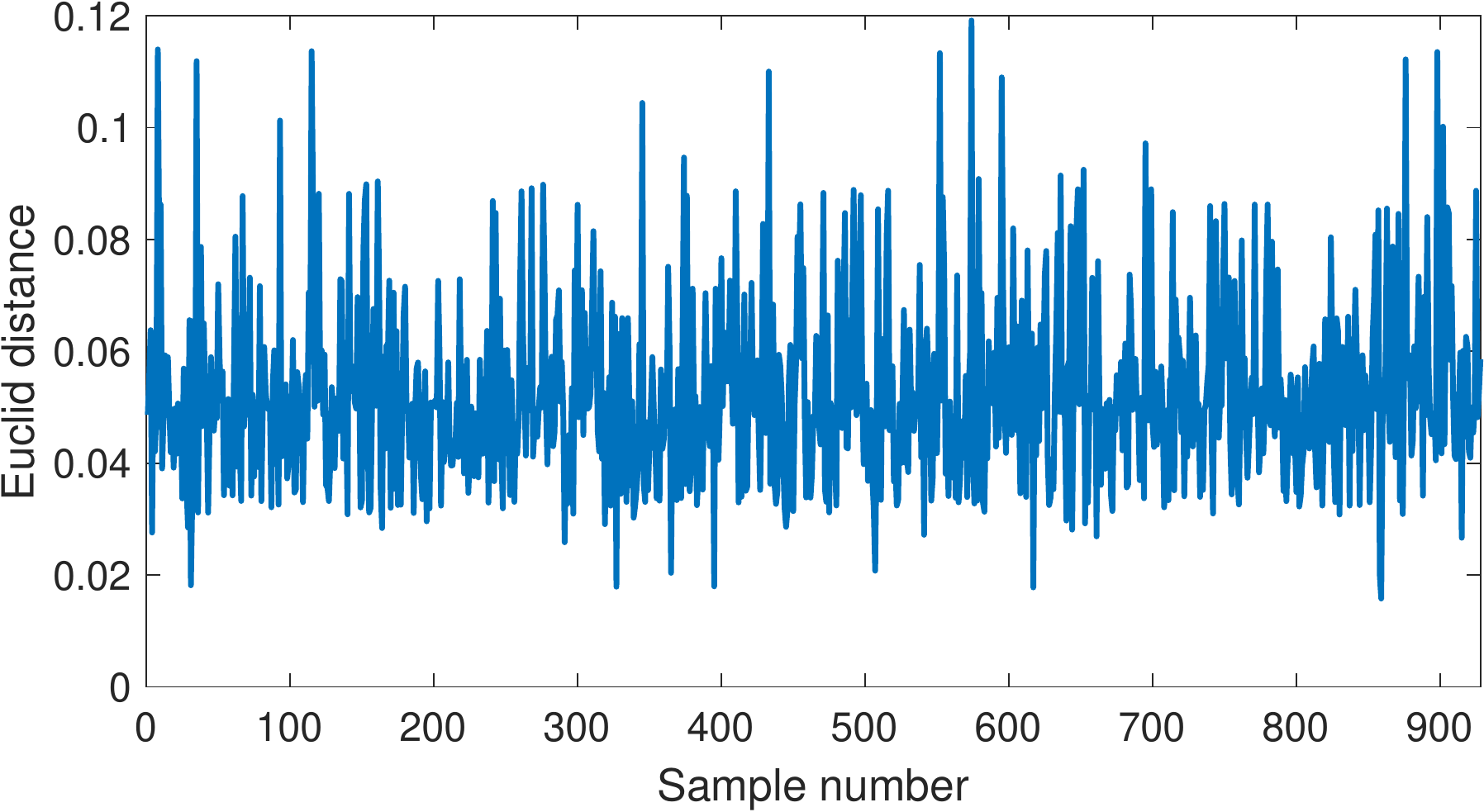}}
    \subfigure[]{\label{fig:linear-dist}
    \includegraphics[width=0.5\textwidth]{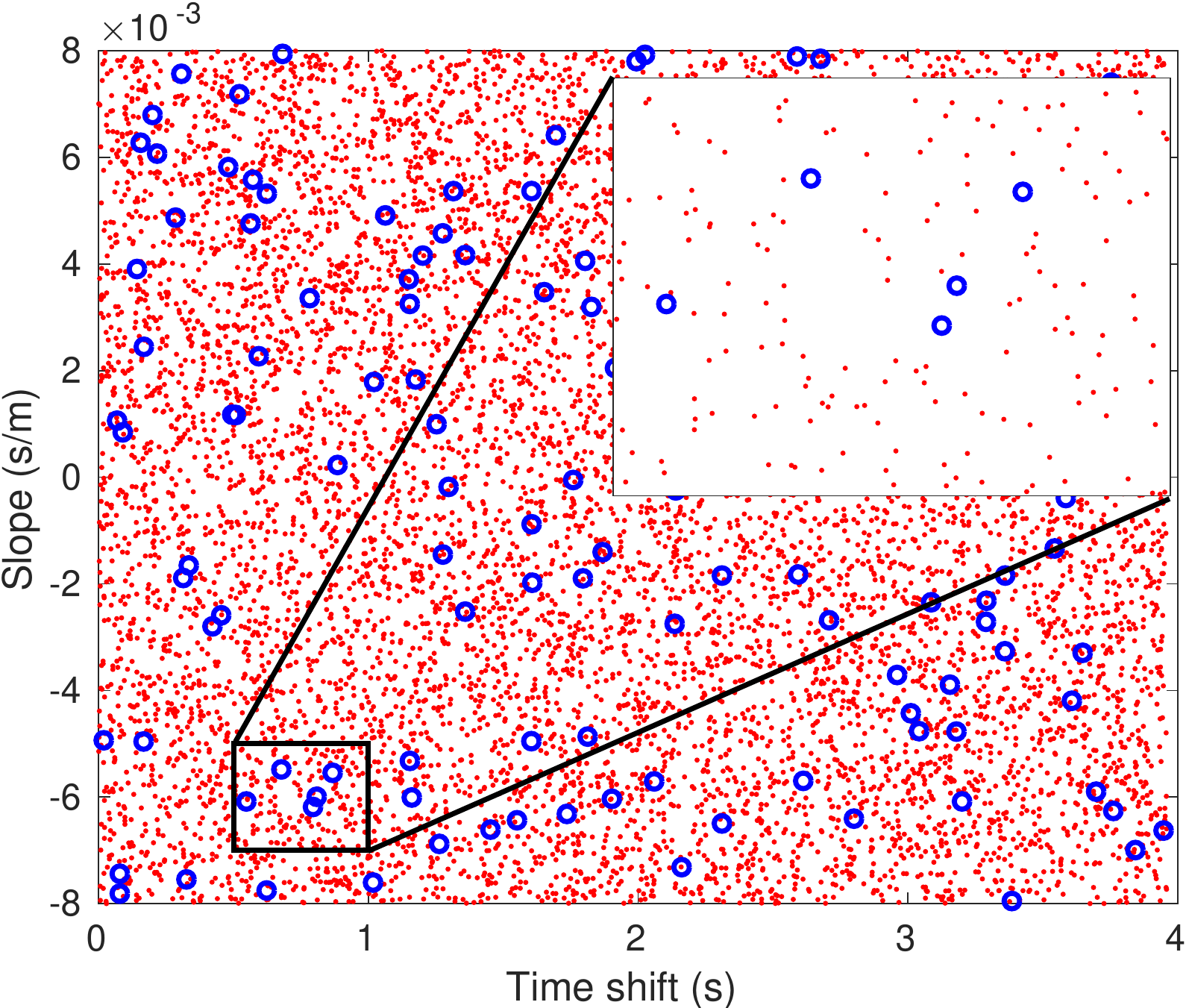}}
    \caption{  (a) From left to right are the testing sample,  training sample with the smallest distance to the testing sample,  difference between the training sample and the testing sample, and training sample with the largest distance to the testing sample. (b) The Euclidean distance between the testing sample in (a) and all training samples. (c) Distribution of slopes and translations of linear events on a two-dimensional plane. The red dots indicate the training samples and the blue circles indicate the testing samples.}
    \label{fig:samples-dist2}
\end{figure*}

The simplest method to measure the distances between two samples is using Euclidean distance. For the multiple datasets, we use a test sample and compute the distances between the test sample and all training samples. We present the training samples with the smallest and largest distance in Figure \ref{fig:04M-nf}. We found that the one that achieves the smallest distance contains the same first arrivals as the test sample but with different multiples. The distances are plotted in Figure \ref{fig:04M-dist}. The test sample does not copy any training sample.

\begin{figure*}
    \centering
    \includegraphics[width=1.0\textwidth]{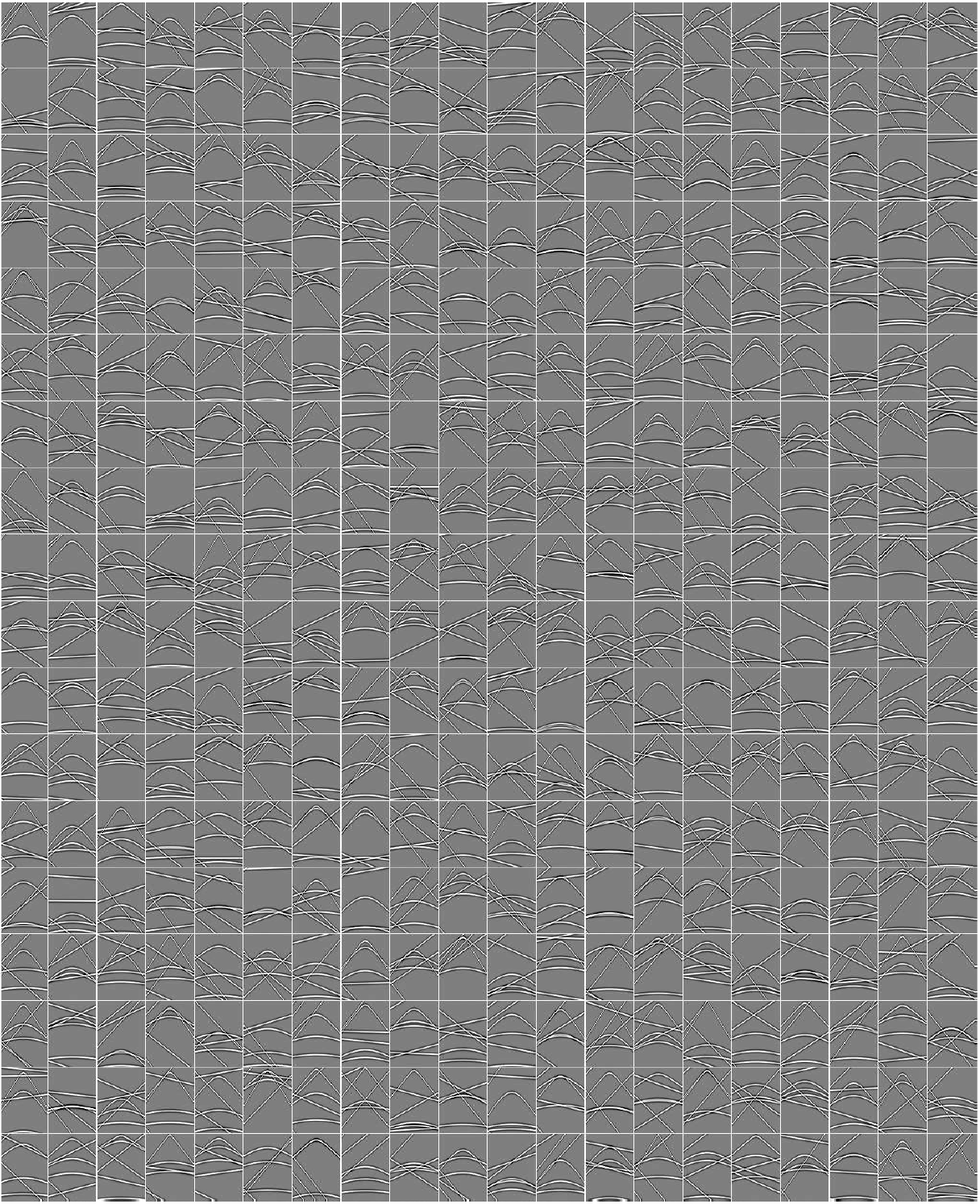}
    \caption{400 synthetic training samples from an 8000 sample dataset for linear noise attenuation.}
    \label{fig:linear400}
\end{figure*}

For the dataset containing linear noise, we plot the distribution of slopes and time shifts of the linear events on a two-dimensional plane in Figure \ref{fig:linear-dist}. Further, 400 of the training samples are plotted in Figure \ref{fig:linear400}. Theoretically, because we use a uniform distribution for generating the slope and time shift, different samples cannot be exactly the same. However, the plane may be crowded, especially when many samples are present. The red dots indicate the training samples and the blue circles indicate the testing samples. The blue circles locate among the red dots. If we memorize the red dots, we cannot obtain the blue circles by linear interpolation. Adding two events with slope $s_1$ and $s_2$ cannot yield an event with slope $s_1+s_2$. The training sample and test samples are generated independently by the same rule. Therefore, the network learns the relationship between the input and output rather than memorizing each training sample.

\subsection{{Multiple attenuation of field data}}

{No test on field multiple attenuation is performed for two reasons: the size of the training samples and the quality of the training dataset. First, the random noise is locally incoherent with the useful signals, such that a patching method is used to generate the training samples with small sizes (such as 35$\times$35). The linear noise is theoretically coherent with the useful signals on a local scale.  In practice, the numerical tests show that a larger patch size (such as 40$\times$40) also works. However, for multiple suppression, this patching strategy does not work. The whole dataset must be used as the training samples, which is currently not feasible in term of computational resources (such as the memory of the commonly available GPU's or the computational time). Future work on field-data multiple attenuation might start from training samples with relatively small sizes. Second, just like in random or linear noise attenuation for field data, we also need to construct a training set for multiple attenuation with the existing industry methods, which is not ideal. However, as the size of the training set grows larger and the hardware improved, we hope to obtain better predictions from the trained CNN.
}

\subsection{Failure situations}

\begin{figure*}
    \centering
    \includegraphics[width=0.3\textwidth]{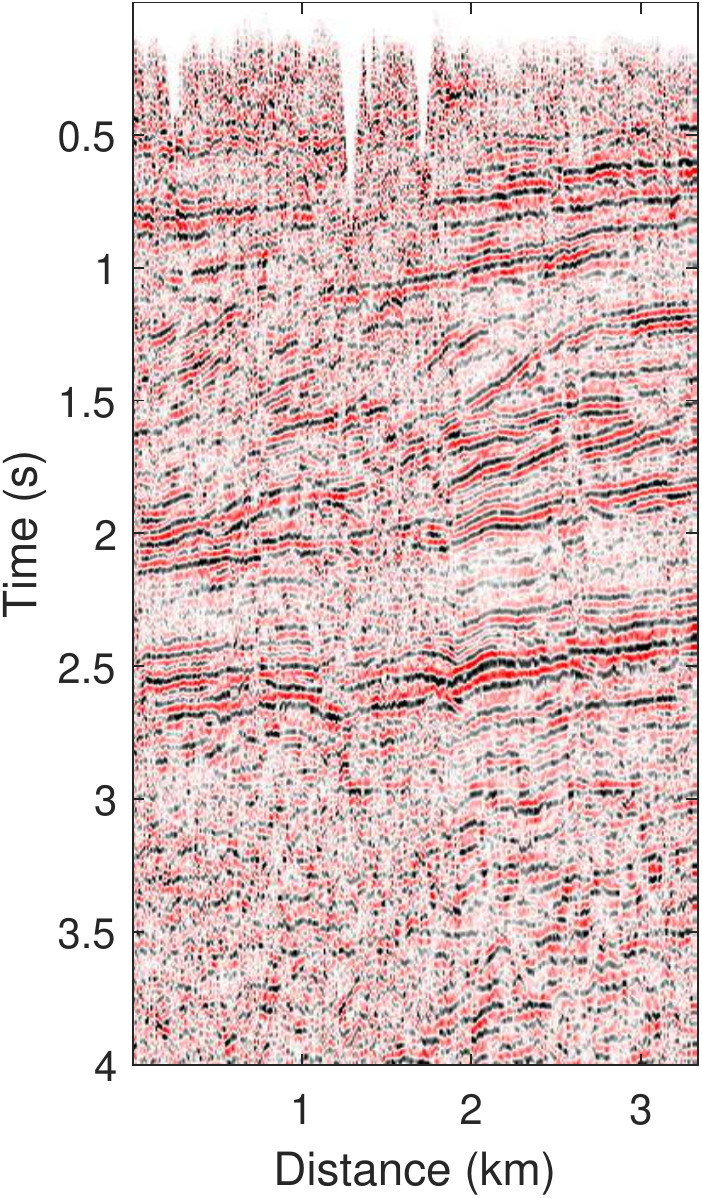}
    \caption{Field data random noise attenuation with CNN learned from synthetic training set.}
    \label{fig:field_syn}
\end{figure*}

\begin{figure*}
    \centering
    \subfigure[]{\label{fig:27-sgr-input}
    \includegraphics[width=0.3\textwidth]{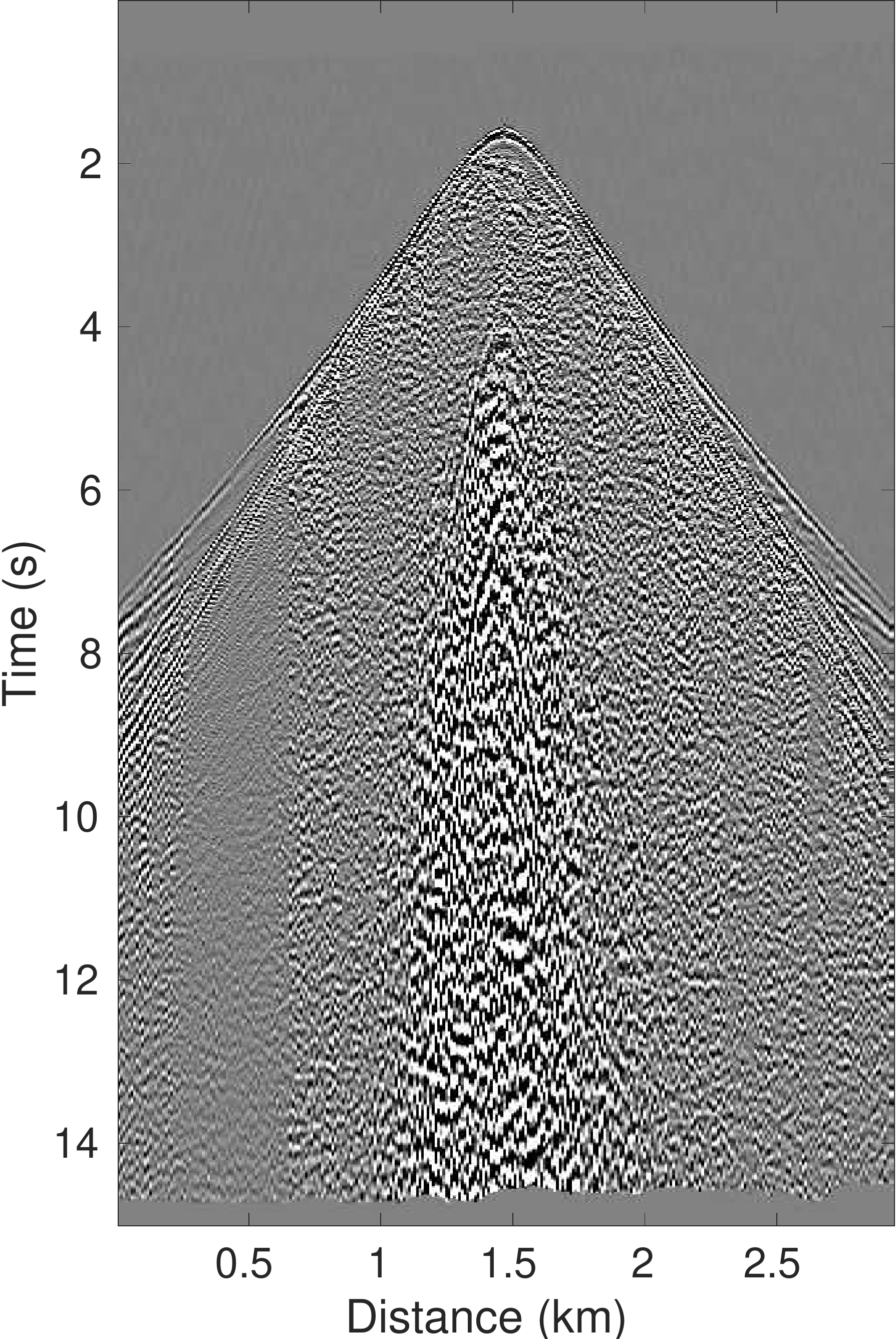}}
    \subfigure[]{\label{fig:27-sgr-output}
    \includegraphics[width=0.3\textwidth]{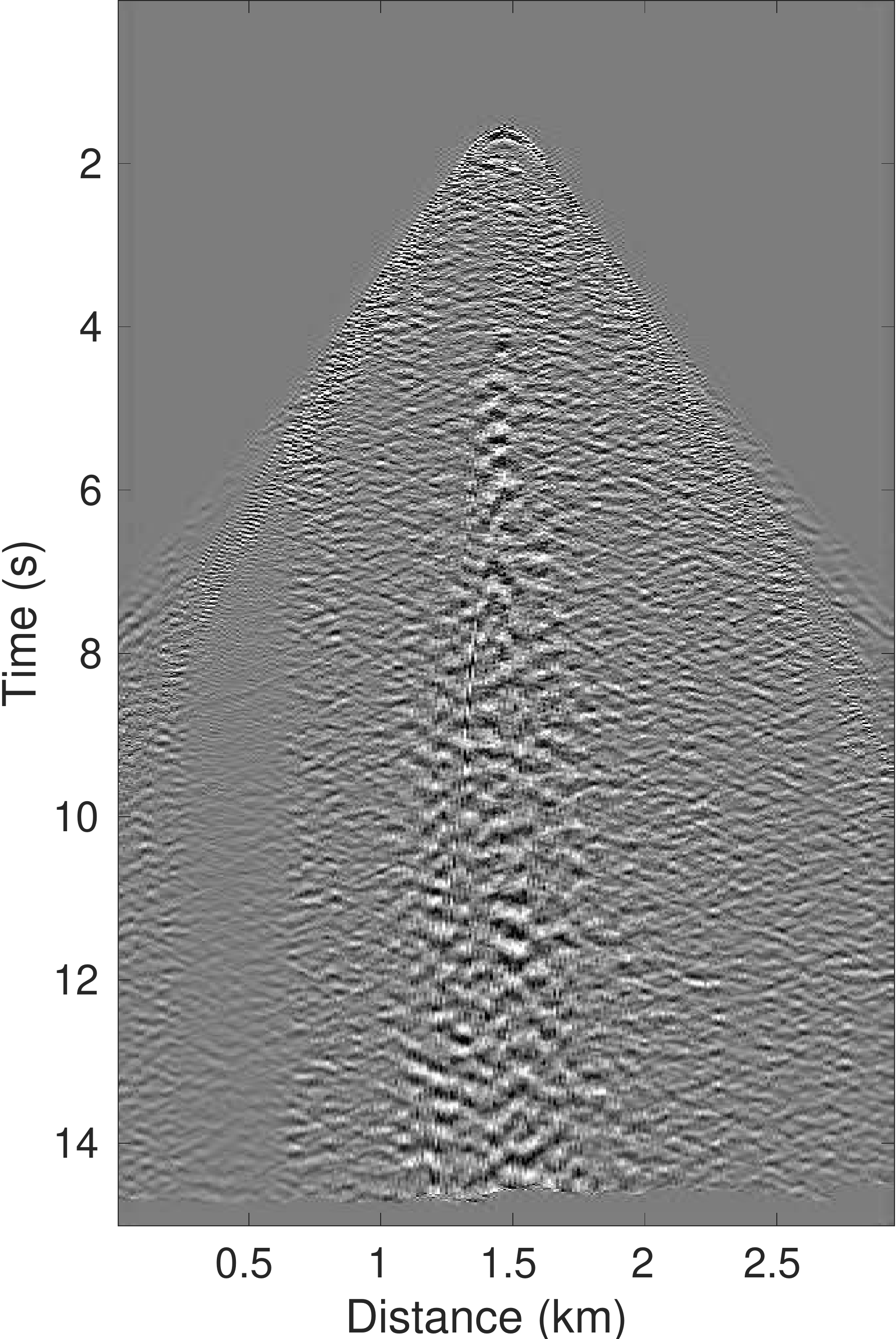}}
    \caption{{The previous trained CNN with the field ground roll datasets in Figure \ref{fig:sgr12} is applied on a different survey. (a) Original data. (b) Denoised by CNN.}}
    \label{fig:sgrb}
\end{figure*}

In general, DL will fail in the following scenarios: insufficient training samples (Figure \ref{fig:28-test_loss}), improper hyperparameter setting (Figure \ref{fig:snr_layers}), large input size (discussed in random noise dataset preparation section), or a test set that is entirely different from the training set. We use the trained network from the synthetic random noise situation and apply it on the field data in Figure \ref{fig:24-test}. The denoised result is shown in Figure \ref{fig:field_syn}, where the noise is hardly removed because the noise distribution in the field data is different from that in the synthetic dataset. {In the case of ground roll attenuation, the trained CNN with the datasets in Figure \ref{fig:sgr12} is applied to a different survey. Figure \ref{fig:sgrb} shows one noisy dataset and the denoised dataset, which is not acceptable. A possible solution is to make the training set as large and varied as possible. }

\section{Conclusion}

We show the applications of deep learning to seismic noise attenuation.  With DL, we achieve several advantages over traditional methods: (i) denoising quality in synthetic random noise attenuation, (ii) automation (no requirement of parameter tuning) and high efficiency on GPU's after training.

In real-dataset random noise attenuation, CNN and transfer learning enable automated/intelligent denoising without human intervention, where labor, ambiguity and time are saved. The training process, which in the cases presented here was about 6 hours, might be considered time-consuming. However, testing on a GPU is extremely fast (0.02 s on a 1501$\times$333 input), which makes it feasible for large-scale datasets. In field scattering ground roll attenuation, although the patching strategy is used, the ground roll is removed successfully. 

Large-scale and well-labeled field datasets are essential for real-data processing, which require intense human intervention. We proposed two possible directions for future studies. The first is to combine a large number of synthetic datasets generated by complex velocity models, e.g., the EAGE model, and a small number of labeled field datasets to form a new training set, which can be used to train a general network for seismic data processing. The second is unsupervised learning, where no clean signals are needed, and DL may separate the coherent noise and useful signals by exploiting the hidden information in the training set. Apart from insufficient training samples, DL also encounters several general problems, such as difficulties in understanding complicated networks, empirical hyperparameter settings, and intense computation while training. DL can also be used for other seismic data processing tasks, such as erratic noise attenuation, provided that a training set is properly constructed.

\section{ACKNOWLEDGEMENT}
The work was supported in part by the National Key Research and Development Program of China under Grant 2017YFB0202902 and Grant 2018YFC1503705, and NSFC under Grant 41625017, Grant 41804102 and Grant 41804108. We thank the Northeast Research Institute of Petroleum Exploration and Development for providing the scattered ground roll dataset.

\newpage

\bibliographystyle{seg}

\bibliography{deeplearning}{}

\newpage

\listoffigures

\newpage

\clearpage

\clearpage

\clearpage

\clearpage

\clearpage

\clearpage

\clearpage


\clearpage

\clearpage

\clearpage

\clearpage

\clearpage

\clearpage

\clearpage

\clearpage

\clearpage

\clearpage

\clearpage

\clearpage
 
\clearpage

\clearpage

\clearpage

\clearpage

\clearpage

\clearpage

\end{document}